\documentclass[aps,prd,twocolumn,nofootinbib,showpacs,preprintnumbers]{revtex4-1}
\usepackage{natbib}
\usepackage[pdftex]{graphicx}
\usepackage[usenames]{color}
\usepackage[dvips]{epsfig}

\usepackage{hyperref}
\usepackage{bbm}
\usepackage{marvosym}
\usepackage{verbatim}
\usepackage{psfrag}
\usepackage{slashed}

\usepackage{amsmath}
\usepackage{amssymb}
\usepackage{wasysym}
\usepackage{multirow}

\usepackage{diagbox}
\usepackage{siunitx} 
\usepackage{array} 
\usepackage{amssymb}	
\usepackage{booktabs,amsmath}
\usepackage{ bbold }
\usepackage[table,x11names]{xcolor}
\definecolor{maroon}{cmyk}{0,0.87,0.68,0.32}
\usepackage{amsfonts}

\definecolor{boxcolor}{HTML}{ffe6a1}

\def \Lat {{\Lambda}}
\def \LatB {{\Lambda^B}}
\def \LatM {{\Lambda^M}}
\def \LatSpat {{\Lambda_\sigma}}
\def \LatSpatB {{\Lambda_\sigma^B}}
\def \LatSpatM {{\Lambda_\sigma^M}}
\def \sumLatSpat {{\sum_{\{\LatSpatM,\LatSpatB\}}}}

\def \as {{a_{\sigma}}}
\def \at {{a_{\tau}}}
\def \Ns {{N_{\sigma}}}
\def \Nt {{N_{\tau}}}
\def \Nc {{N_{c}}}
\def \Nf {{N_{f}}}

\def \hmu {{\hat{\mu}}}

\newcommand{\expval}[1]{\langle #1 \rangle}

\def \QMat{\mathcal{M}}
\newcommand{\Exp}[1]{\exp \lr{#1}}

\newcommand{\calE}{\mathcal{E}}
\newcommand{\calA}{\mathcal{A}}
\newcommand{\calP}{\mathcal{P}}
\newcommand{\calQ}{\mathcal{Q}}
\newcommand{\calB}{\mathcal{B}}
\newcommand{\calH}{\mathcal{H}}
\newcommand{\calN}{\mathcal{N}}
\newcommand{\calZ}{\mathcal{Z}}

\newcommand{\Id}{{\mathbbm{1}}}
\newcommand{\lr}[1]{\left( #1 \right)}

\newcommand{\nn} {\nonumber\\}
\newcommand{\beqn} {\begin{equation}}
\newcommand{\eqn} {\end{equation}}

\def \beq{\begin{equation}}
\def \eeq{\end{equation}}
\def \bea{\begin{eqnarray}}
\def \eea{\end{eqnarray}}

\def \Tr {{\rm Tr}}
\def \tr {{\rm tr}}

\def \bet0{\beta_0}
\def \bet1{\beta_1}
\def \simgt{\,\rlap{\lower 7.5 pt\hbox{$\mathchar \sim$}}\raise 3 pt \hbox{$>$}\,}
\def \simlt{\,\rlap{\lower 7.5 pt\hbox{$\mathchar \sim$}}\raise 3 pt \hbox{$<$}\,}

\def\lsim{\raise0.3ex\hbox{$<$\kern-0.75em\raise-1.1ex\hbox{$\sim$}}}
\def\gsim{\raise0.3ex\hbox{$>$\kern-0.75em\raise-1.1ex\hbox{$\sim$}}}

\def \Zcal{\calZ}

\newcommand{\lsteel}{{\hspace{0.1em}\text{\Lsteel}\hspace{0.1em}}}
\newcommand{\tsteel}{{\hspace{0.1em}\text{\Tsteel}\hspace{0.1em}}}

\newcommand{\bareT}{\mathpzc{a}\hspace{-0.4mm}\mathcal{T}}
\newcommand{\bareMu}{\mathpzc{a\hspace{-0.3mm}\mu_B}}

\newcommand{\meson}{\mathfrak{m}}
\newcommand{\baryon}{\mathfrak{b}}
\newcommand{\hadron}{\mathfrak{h}}
\newcommand{\spin}{\mathfrak{s}}
\newcommand{\ka}{{n}}

\newcommand{\SU}{{\rm SU}}
\newcommand{\U}{{\rm U}}
\newcommand{\CT}{\rm CT}
\newcommand{\calG}{\mathcal{G}}
\newcommand{\acsch}{{\rm arcsch}}

\newcommand{\cbc}{\bar{\chi}\chi}

\DeclareMathAlphabet{\mathpzc}{OT1}{pzc}{m}{it}

\begin{document}

\title{Strong Coupling Lattice QCD in the Continuous Time Limit}
\author{
M.~Klegrewe$^{\rm a}$
}
\email{mklegrewe@physik.uni-bielefeld.de}
\author{
W.~Unger$^{\rm a}$,
}
\email{wunger@physik.uni-bielefeld.de}
\affiliation{$^{\rm a}$ Fakult\"at f\"ur Physik, Bielefeld University, D-33615 Bielefeld, Germany}

\begin{abstract}
We present results for lattice QCD with staggered fermions in the limit of infinite gauge coupling, obtained from a worm-type Monte Carlo algorithm on a discrete spatial lattice
but with continuous Euclidean time. This is obtained by sending both the anisotropy parameter $\xi= \as/\at$ and the number of time-slices $\Nt$ to infinity,
keeping the ratio $aT=\xi/\Nt $ fixed. The obvious gain is that no continuum extrapolation $\Nt \rightarrow \infty$ has to be carried out.
Moreover, the algorithm is faster and the sign problem disappears. We derive the continuous time partition function and the corresponding Hamiltonian formulation. We compare our computations with those on discrete lattices and study both zero and finite temperature properties of lattice QCD in this regime.
\end{abstract}

\maketitle
\tableofcontents

\section{Introduction}

The determination of the QCD phase diagram, in particular the location of the critical end point (CEP) is an important, long standing problem, requiring non-perturbative methods.
In lattice QCD, several approaches have been developed to investigate the phase transition from hadronic matter to the quark gluon plasma,
but either they are limited to rather small $\mu_B/T$, with $\mu_B$ the baryon chemical potential \cite{Fodor2001,deForcrand2003,Allton2005}, or they cannot yet address full QCD \cite{Aarts2013,Sexty2013,Langfeld2012} or study only low dimensional QCD-like toy models \cite{Alexandru2015,Schmidt2017,DiRenzo2017}.

The reason for this is the notorious sign problem \cite{Gattringer2016b}, which arises because the fermion determinant for finite $\mu_B$ becomes complex, and importance sampling is no longer applicable. 
In lattice QCD, the finite density sign problem is severe. There is however a limit where the sign problem can be made mild: this is the strong coupling limit, where a so-called dual representation in terms of color singlets is possible. In the strong coupling limit of lattice QCD (SC-LQCD) the sign problem is mild enough such that the full $(\mu_B,T)$ phase diagram can be measured via Monte Carlo methods based on the dual variables.
The method of dual variables has been successfully used in models with Abelian gauge symmetry \cite{Mercado2013,Gattringer2015}, and 
there have been attempts to dualize non-Abelian gauge theories \cite{Vairinhos2014,Gattringer2016a}, but it has not yet been possible to overcome the finite density sign problem. Our own approach discussed in \cite{Unger2017,Gagliardi2018,Gagliardi:2019cpa} is based on the strong coupling expansion, i.e.~an expansion in the inverse gauge coupling $\beta=\frac{2\Nc}{g^2}$. It is in principle possible to sample partition functions that include all orders via Monte Carlo, in a spirit of \cite{Wolff2008a,Wolff2008b}. In practice, the sign problem is reintroduced for large $\beta$.

In this paper, we will restrict to the strong coupling limit, since the focus is on deriving the Euclidean continuous time limit and apply the new formulation to Monte Carlo studies of QCD thermodynamics. Despite the fact that the strong coupling limit is the converse of the continuum limit, i.e.~the lattice is maximally coarse and it is not possible to set the scale, it nevertheless shares important features with lattice QCD on finer lattices: chiral symmetry breaking and its restauration at finite temperature as well as the nuclear liquid gas transition are also present in this model. We will extend the existing studies on SC-LQCD that are either based on mean field theory in the $1/d$ expansion \cite{Kawamoto1981,KlubergStern1983,Faldt1985,Bilic1991a,Bilic1992a,Kawamoto2005,Miura2016}
or on Monte Carlo
\cite{Rossi1984,Karsch1989,Forcrand2010}. 
In the past either the spectrum or the phase diagram and the nuclear properties \cite{Forcrand2010} have been studied. We investigate these phenomena in the continuous time limit, where the continuum limit of the temporal lattice spacing $\at\rightarrow 0$ is taken while leaving the spatial lattice spacing $\as$ finite. 
First simulations of SC-LQCD in continuous time have been performed by one of us in 
\cite{Unger2012}. Here, we improve upon the continuous time formulation and give many more results at zero and non-zero temperature.
The main advantage of the continuous time limit (CT) is that ambiguities arising from the anisotropy parameter $\gamma$ are circumvented. Also, the sign problem is absent, Quantum Monte Carlo (QMC) can be applied, and temporal correlation functions can be obtained with high resolution.\\

This paper is structured as follows: in Sec.~\ref{Formulation} we will derive the Quantum Hamiltonian formulation of strong coupling QCD and its generalization to an arbitrary number of colors. In Sec.~\ref{Algorithm} we will describe the worm algorithm operating in continuous time in detail and show that it indeed reproduces results consistent with the continuum extrapolation of simulations at finite $\Nt$. 
In Sec.~\ref{ZeroT} we apply SC-LQCD in the CT-limit to determine zero temperature observables. In Sec.~\ref{PhaseDiag} we investigate finite temperature properties, such as the grand-canonical phase diagram in the $\mu_B - T$ plane as well as the canonical phase diagram in the $n_B - T$ plane, with $n_B$ the baryon number density. 
In Sec.~\ref{CorrelationFunction} we discuss temporal correlation functions and how to extract pole masses. We provide both results at finite temperature and density.
In Sec.~\ref{TaylorExpansion} we show that the pressure at finite baryon density can also be reconstructed from Taylor coefficients. We conclude with remarks on the radius of convergence.
In the appendix, supplementary material for the various crosschecks of continuous time Monte Carlo and possible extensions such as for finite quark mass, more flavors and isospin chemical potential are discussed.\\ 

\section{Strong Coupling Lattice QCD in the Continuous Time Formulation}
\label{Formulation}

\subsection{Staggered Action of Strong Coupling QCD and its Dual Representation}

In SC-LQCD, based on the Euclidean lattice action, the gauge coupling is sent to infinity and thus the coefficient of the plaquette term $\beta=2\Nc/g^2$ is sent to zero. Hence the Yang Mills part $F_{\mu\nu} F_{\mu\nu}$ is absent.
Then, the gauge fields in the covariant derivative can be integrated out analytically. In fact, the order of integration is reversed compared to the standard representation of lattice QCD in terms of the fermion determinant: the gauge links $U_\mu(x)$ are integrated out before the Grassmann fields $\chi$, $\bar{\chi}$. Thus the final degrees of freedom of the partition function are color singlets composed of fermions: mesons and baryon.
However, as a consequence of the strong coupling limit, the lattice becomes maximally coarse and there is no way to set the scale: the lattice spacing $a$ cannot be specified in physical units. We will see however that specific dimensionless ratios can still be compared to continuum physics.

We shortly outline the procedure to obtain the dual representation for staggered fermions in the strong coupling limit where the action is only given by the fermionic part:
\begin{widetext}
\begin{align}
S[U,\chi,\bar{\chi}]&=\sum_x\Bigg[ 
\gamma\, \eta_0(x) \left( \bar{\chi}(x) e^{\at \mu_q} U_0(x)\chi(x+\hat{0})- \bar{\chi}(x+\hat{0}) e^{-\at\mu_q} U_0^\dagger(x)\chi(x)\right)\nn
&\quad\quad\quad+ \sum_{i=1}^d \eta_i(x) \left( \bar{\chi}(x) U_i(x)\chi(x+\hat{i})- \bar{\chi}(x+\hat{i}) U_i^\dagger(x)\chi(x)\right)+ 2am_q\bar{\chi}(x) \chi(x)\Bigg].
\label{SCQCDPF}
\end{align}
\end{widetext}
Here, $am_q$ is the quark mass and $\mu_q=\frac{1}{3}\mu_B$ the quark chemical potential. The bare anisotropy parameter $\gamma$ in the temporal Dirac coupling is introduced to vary the temperature continuously. 

Following the procedure discussed in detail in \cite{Karsch1989}, the gauge link integration over the Haar measure of \SU($\Nc$) can be performed analytically, as the integration factorizes in Eq.~(\ref{SCQCDPF}), i.~e.~the partition function can be written as a product of one-link integrals $z_\mu(x)$:
\begin{widetext}
\begin{align}
Z&=\int \prod_x \lr{d\bar{\chi}(x)d\chi(x)e^{2am_q\bar{\chi}(x)\chi(x)}\prod_\mu z_\mu(x)},\\
z_\mu(x)&\equiv \left.z(x,y)\right|_{y=x+\hmu}=\int_{\rm {SU(\Nc)}} dU_\mu(x) \exp\lr{\eta_\mu(x)\lr{\bar{\chi}(x) U_\mu(x)\chi(y) - \bar{\chi}(y)U_\mu^\dagger(x)\chi(x)}}\nn
&=\sum_{k=0}^{\Nc} \left\{ \frac{(\Nc - k)!}{\Nc! k!} 
\lr{\lr{\eta_\mu(x)\gamma^{\delta_{\mu 0}}}^2 M(x)M(y)}^k \right\} 
+  \lr{\rho(x,y)^{\Nc}\bar{B}(x)B(y)+(-\rho(y,x))^{\Nc}\bar{B}(y)B(x)},\\
M(x)&=\bar{\chi}(x)\chi(x), \qquad B(x)=\frac{1}{\Nc}\epsilon_{i_1\ldots i_{\Nc}}\chi_{i_1}(x)\ldots\chi_{i_{\Nc}}(x),\qquad\rho(x,y)=\eta_{\mu}(x)\lr{\gamma\exp(\pm \at\mu_q)\delta_{\mu 0}+(1-\delta_{\mu 0})}.
\end{align}
\end{widetext}
The new degrees of freedom after link integration on the right-hand side are the mesons $M(x)$ and the baryons $B(x)$.
The weight of the one-link integral is a sum over the so-called dimer number $k_\mu(x)=0,\ldots \Nc$ which corresponds to the number of (non-oriented) meson hoppings on that link, and on
$\rho(x,y)$ which is the weight for a baryon hopping $\bar{B}(x)B(y)$. 
The final partition function for the discrete system on a $\Ns^3\times \Nt$ lattice, after performing the Grassmann integrals analytically, is an exact rewriting from Eq.~(\ref{SCQCDPF}) and is given by:
\begin{widetext}
\begin{align}
\Zcal(m_q,\mu_q)&= \sum_{\{k,n,\ell\}}^{\text{GC}}\prod_{b=(x,\hat{\mu})}\frac{(\Nc-k_b)!}{\Nc!k_b!}\gamma^{2k_b\delta_{\hat{0}\hat{\mu}}}\prod_{x}\frac{\Nc!}{n_x!}(2am_q)^{n_x} \prod_\ell w(\ell)\nn
w(\ell)&=\prod_{x\in \ell}\frac{1}{\Nc!} \sigma(\ell)\gamma^{\Nc N_{\hat{0}}} \exp\lr{\Nc \Nt \omega(\ell) \at \mu_q}, \qquad \sigma(\ell)=(-1)^{\omega(\ell)+N_-(\ell)+1}\prod_{b=(x,\hmu)\in \ell}\eta_\hmu(x)
\label{SCPF}
\end{align}
\end{widetext}
The sum over all configurations $\{k,n,\ell\}$
is restricted to those that fulfill on each site $x$ the so-called Grassmann constraint (GC):
\begin{align}
&n_x+\sum_{\hat{\mu}=\pm\hat{0},\ldots \pm \hat{d}} \left(k_{\hat{\mu}}(x)+\frac{\Nc}{2}|b_\hmu(x)|\right)=\Nc
\label{GC} 
\end{align}
which expresses the fact that every Grassmann variables $\bar{\chi}_i(x)$, $\chi_i(x)$ ($i=1\ldots \Nc$) appear exactly once in the path integral.
After this exact rewriting of the strong coupling partition function the system can be described by 
confined, colorless, discrete degrees of freedom:
\begin{itemize} 
\item Mesonic degrees of freedom: $k_{\hat{\mu}}(x)\in \{0,\ldots \Nc\}$ (non-oriented meson hoppings called dimers) and
$n(x) \in \{0,\ldots \Nc\}$ (mesonic sites called monomers).
\item Baryonic degrees of freedom: they form oriented baryon loops $\ell$ and may wind $\omega(\ell)$ times in temporal direction, which results on its dependence on the chemical potential $\mu_q$.
The sign $\sigma(\ell)=\pm1$ of the loop $\ell$
depends on the loop geometry.
\item The baryonic loops are self-avoiding and do not touch the mesonic degrees of freedom, which follows from the Grassmann constraint Eq.~(\ref{GC}): for a given configuration, this gives rise to a decomposition of the lattice volume into mesonic sites and baryonic sites:
\begin{align}
\Lat=\Ns^3\times \Nt =\Lat_M\,\dot\cup\,\Lat_B.
\label{MBSites} 
\end{align}
\end{itemize}
It should be mentioned that this representation corresponds to unrooted staggered fermions. Due to the fermion doubling, one flavor of a staggered fermion comes in the multiplicity of four so-called tastes. However, in the strong coupling limit, the fermions are spinless and the taste breaking is maximal. Hence it is indeed a one-flavor theory with only one pseudoscalar meson as the Goldstone boson. To be more precise, in the chiral limit the action is invariant under the symmetry group $U_B(1)\times U_{55}(1)$:
\begin{align}
\chi(x) &\mapsto e^{i\theta_{B}+i\epsilon(x)\theta_{55}}\chi(x), &\epsilon(x)&=(-1)^{\sum_\mu x_\mu}
\label{ChiralTrafo}
\end{align}
which is due to the even-odd decomposition of the bipartite lattice for staggered fermions, i.~e.~even and odd sites can be transformed independently.
The symmetry $e^{i\theta_{B}}\in U_B(1)$ corresponds to baryon conservation and 
$e^{i\theta_{55}}\in U(1)_{55}$ is a subgroup of the full $SU_L(4)_L\times SU_R(4)$ chiral symmetry for unrooted staggered fermions. 
In the spin-taste basis this corresponds to the channel $\gamma_5\otimes \xi_5$. 
At finite quark mass $U(1)_{55}$ is explicitly broken, and in the dual representation this is due to the presence of monomers: the number of monomers on even sites equals its number on odd sites.
In the chiral limit we expect O(2) critical exponents for the chiral phase transition. This is also the case away from the strong coupling limit, as long as the lattice spacing is finite.
In this work we will restrict to the chiral limit, $m_q=0$, where monomers are absent: $n_x=0$. We discuss the prospects of the continuous time formulation at finite quark mass in the appendix \ref{FiniteMass}.

\subsection{SC-LQCD at Finite Temperature and the Continuous Time Limit}
\label{FiniteTCTLimit}

In the staggered action Eq.~(\ref{SCQCDPF}) we have introduced a bare anisotropy $\gamma$ in order to vary the temperature continuously. Hence also in the dual representation 
the weights for temporal meson or baryon hoppings in Eq.~(\ref{SCPF}) contain the anisotropy parameter $\gamma$. 
We will now explain why this is necessary and why it is also a key step to derive the continuous time limit.\\

The main objective of SC-LQCD is to study thermodynamic properties. Since $\beta=0$, we cannot vary the temperature $T=1/(\Nt a(\beta))$ continuously via the lattice spacing, but only with the lattice extent $\Nt$. The chiral transition is however at temperatures much higher than $1/2$, such that for temperatures $1/\Nt$ we are always in the chirally broken phase. The solution is to introduce an anisotropy in the Dirac Operator to favor fermion propagation in temporal direction. In contrast to the chemical potential, the bare anisotropy does not distinguish between forward and backward temporal direction. 
The temperature on an anisotropic lattice is given by the inverse of the lattice extend in temporal direction
\begin{equation}
 T=\frac{1}{\at \Nt}=\frac{\xi(\gamma)}{\as \Nt} \qquad \text{with} \qquad \xi(\gamma)\equiv\frac{\as}{\at}
\end{equation}
but the functional dependence $\xi(\gamma)$ of the ratio of the spatial and temporal lattice spacings on the bare anisotropy  is not known a priori.
Hence also the dependence of $T$ on $\gamma$ is unknown. The main motivation for this study is to overcome this difficulty.

The weak coupling analysis of Eq.~(\ref{SCQCDPF}) suggests that $\xi(\gamma)=\gamma$, but this does not carry over to strong coupling, where quarks are confined on links to color singlets.
In the mean-field approximation of SC-LQCD \cite{Bilic1992a} based on $1/d$-expansion (with $d$ the spatial dimension) the critical temperature is given by
\begin{equation}
 \gamma_c^2=\Nt \frac{d (\Nc+1)(\Nc+2)}{6(\Nc+3)}
\label{MFT}
\end{equation}
suggesting that $\as T_c\propto\frac{\gamma_c^2}{\Nt}$ is the sensible $\Nt$-independent identification in leading and next-to leading order in $d$. 

\begin{figure}[h!]
\includegraphics[width=0.5\textwidth]{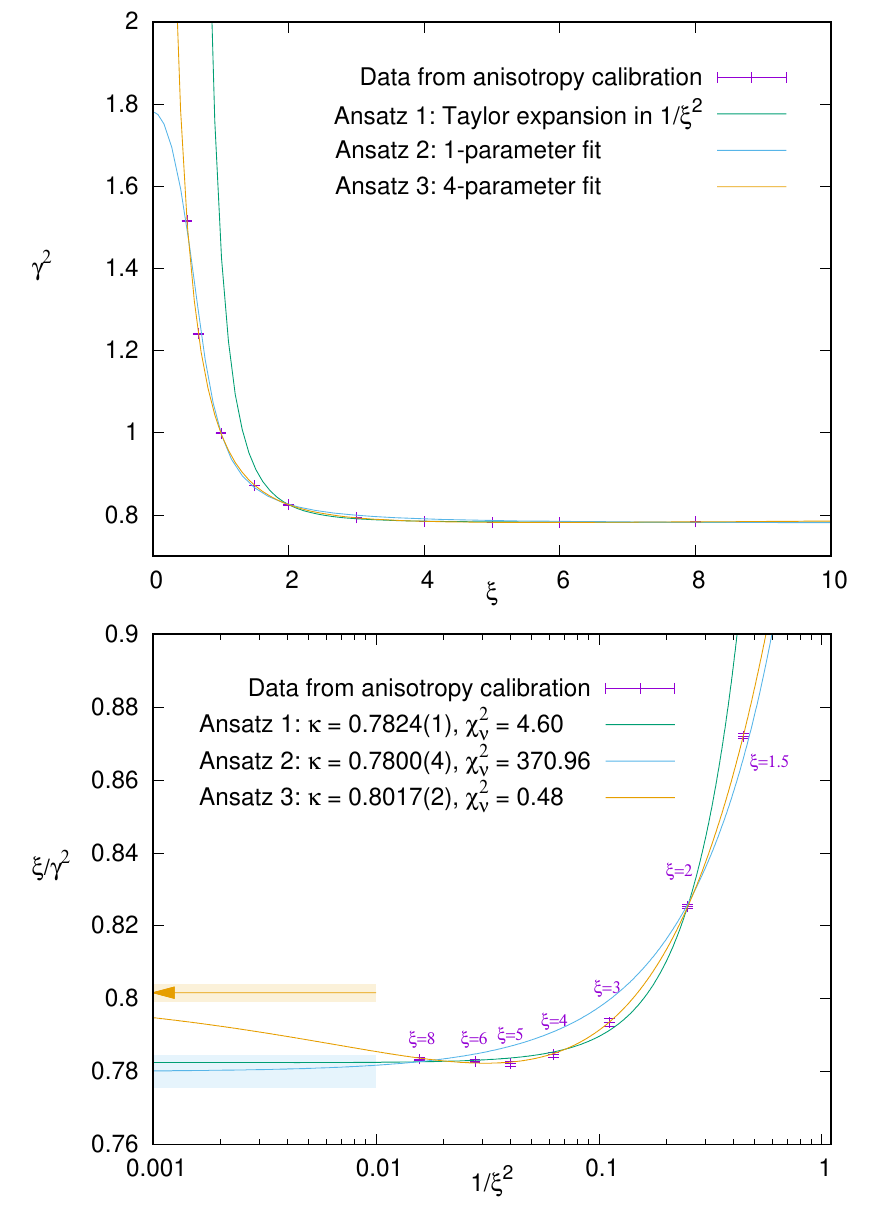}
\caption{
\emph{Top}: Interpolation of $\gamma^2$ as a function of $\xi$ with Ansatz 1: Eq.~(\ref{AnsatzGamma01}) for $\xi\geq2$, Ansatz 2: Eq.~(\ref{AnsatzGamma02}) and Ansatz 3: Eq.~(\ref{AnsatzGamma03}), both for all $\xi$. \emph{Bottom:} Extrapolation of the anisotropy correction factor $\frac{\xi}{\gamma^2}$ towards  $1/\xi^2\sim a_\tau^2\rightarrow 0$, extracting $\kappa$. It is evident that the extrapolated results depend on the Ansatz. Clearly, Ansatz 3 has the smallest reduced chi-squared.
}\label{AnisotropExtrapolation}
\end{figure}

It is however possible to determine the function $\xi(\gamma)$ non-perturbatively on anisotropic lattices with 
\begin{align}
\Ns\as&=\Nt\at,& \xi&=\frac{\Nt}{\Ns} 
\end{align}
by a bare anisotropy calibration $\gamma_0(\xi)$ via conserved currents in both spatial and temporal direction \cite{deForcrand2017}.
For large $\Nt$ (implying large $\xi$ and $\gamma$), it turns out numerically that $\xi$ diverges as
\begin{align}
\lim_{\Nt\rightarrow \infty}\xi(\gamma)&=\kappa\gamma^2.
\label{DefAnisoCorrFactor}
\end{align}
The precise value of $\kappa$ can only be determined non-perturbatively and a posteriori, based on the values $\gamma_0(\xi)$ measured via the anisotropy calibration (see also Sec.~\ref{MeasureKappa}) and has been extrapolated for SU(3) from the set $\xi=\{0.5,1,1.5, 2,3,4,5,6,8\}$ to $\xi\rightarrow \infty$.
The function $\xi/\gamma^2$, which we call the anisotropy correction factor, can be either parameterized by $\xi$ or $\gamma$ and is well described by the Ans\"atze
\begin{align}
\frac{\xi}{{\gamma^2}}(\xi)&\simeq \kappa + \frac{a}{\xi^2}+ \frac{b}{\xi^4}, \label{AnsatzGamma01}\\    
\frac{\xi}{\gamma^2}(\gamma)&\simeq \kappa + \frac{1-\kappa}{1+\kappa(\gamma^4-1)}, \label{AnsatzGamma02}\\
\frac{\xi}{{\gamma^2}}(\xi)&\simeq 
\frac{
\xi}{
\left(
\frac{\xi}{\kappa+A\xi^{Q_1}} + 
\frac{
\xi^{1/2}}{{\kappa}^{1/2}+B\xi^{Q_2}}
\right)^2}.
\label{AnsatzGamma03}
\end{align}
Clearly, the extrapolated value for $\kappa$ based on $\xi\rightarrow \infty$ will depend on the Ansatz, as shown in Fig.~\ref{AnisotropExtrapolation}.
The Taylor expansion in $1/\xi^2\sim \at^2$ in Eq.~(\ref{AnsatzGamma01}) is limited to the fit range $\xi\geq 2$ and results in the value $\kappa=0.7824(1)$, which is consistent with the already determined value in \cite{deForcrand2017}. The second Ansatz Eq.~(\ref{AnsatzGamma02}) has only $\kappa$ as a free parameter, and interpolates the data for all $\xi$ surprising well, although there are deviations. By construction, $\xi/\gamma^2=1$ for $\gamma=1$. In order to improve on this one-parameter fit, the third fit Ansatz Eq.~(\ref{AnsatzGamma03}) introduces 3 additional independent fit parameters to connect the regime $\xi> 1$ with the opposite regime $\xi<1$; with $Q_1>0$ and $Q_2<0$:
\begin{align}
\lim_{\xi\rightarrow \infty}\frac{\xi}{\gamma^2}&=\kappa, &
\lim_{\xi\rightarrow 0}\frac{\xi}{\gamma^2}&=\frac{\kappa^2}{\xi}, & \left.\frac{\xi}{\gamma^2}\right|_{\xi=1}=1.
\end{align}
The fit parameters $A$ and $B$ are thus not independent.
This fit results in a non-monotonic behavior, which reflects the fact that 
$\left.\frac{\xi}{\gamma^2}\right|_{\xi=8}=0.7834(2)$ is larger than $\left.\frac{\xi}{\gamma^2}￼￼\right|_{\xi=6}=0.7828(2)$. Also, it has the smallest reduced chi-squared. Thus we think that the extrapolated result $\kappa=0.8017(2)$ is more trustworthy. The error is purely statistical, and the systematic error due to the choice of the fit ansatz is unknown.
In Sec.~\ref{MeasureKappa}) we will overcome the ambiguities of the extrapolation $a_\tau\rightarrow 0$ by measuring $\kappa$ directly in the continuous time limit.\\

We will see in Sec.~\ref{Crosschecks} that many observables and the phase diagram have a strong $\Nt$-dependence, which can even be non-monotonic. This requires large $\Nt$ to have control over the extrapolation.
Hence we want to eliminate $\gamma$ and $\Nt$ all together from the partition function Eq.~(\ref{SCPF}) and replace them by the temperature $aT$. 
The continuous time definition of the temperature in lattice units is
\begin{align}
a T_{CT}&=\kappa  \lim_{\substack{\Nt\rightarrow \infty\\\gamma\rightarrow \infty}} \bareT(\gamma,\Nt),& \bareT=\frac{\gamma^2}{\Nt}=const.
\label{CTLimit}
\end{align}
where we have dropped the subscript, $a\equiv \as$. The limit $\Nt\rightarrow \infty$, $\gamma\rightarrow \infty$ is a joint limit and the second condition implies that $\gamma$ diverges as $\gamma=\sqrt{\bareT \Nt}$ for $\Nt\rightarrow \infty$.
Likewise we can define unambiguously the continuous time chemical potential to replace the chemical potential $\at\mu_q$ in Eq. (\ref{SCPF}):
\begin{align}
 a\mu_{B,CT}&=\kappa  \lim_{\substack{\Nt\rightarrow \infty\\\gamma\rightarrow \infty}} \bareMu(\gamma,\Nt),& \bareMu= \Nc \gamma^2 \at\mu_q =const.
\label{CTLimitMu}
\end{align}
which is also consistent with the $\gamma$-dependence of the mean-field critical chemical potential $\mu_c(T=0)$ obtained via $1/d$-expansion \cite{Bilic1991a}, similar to Eq.~(\ref{MFT}):
\begin{align}
 \at \mu_{q,c} &= \frac{d}{4\gamma^2}+\mathcal{O}(\gamma^6).
 \end{align}
Now all discretization errors from finite $\at$ are removed.
The new partition function in continuous Euclidean time will be derived in the next section. We will then have to check numerically that the above limits are well defined for the typical observables. We will present a worm-type Monte Carlo algorithm which samples the partition function efficiently. 
We denote $\bareT$, $\bareMu$ as the bare temperature and bare chemical potential which are then renormalized by $\kappa$.
We will see that we can determine $\kappa$ non-perturbatively directly by Monte Carlo simulations in the continuous time limit.

\newcommand{\Ord}[1]{\mathcal{O}\lr{#1}}
\newcommand{\arcsinh}{\rm arcsinh}
\newcommand{\GC}{\delta_{\sum\limits_\mu k_\mu(x),\Nc}}

\subsection{Continuous Time Partition Function}
\label{ContTimeParFunc}

We will now explain in detail how to derive the continuous time partition function from the discrete time partition function Eq.~(\ref{SCPF}) by tracing the $\gamma$-dependence and neglecting subleading terms that vanish in the limit $\Nt\rightarrow \infty$. 
The first step to obtain these results is to factorize Eq. (\ref{SCPF}) into the temporal and spatial part:
\begin{widetext}
\begin{align}
\calZ(\gamma, \at \mu_q,\Nt) 
&= \gamma^{\Nc \Lat}  \sum_{\{k,\ell\}} \left\{\left(\prod_{x\in \Lat_M}\GC \frac{(\Nc-k_0(x))!}{k_0(x)!}
\prod_{i=1}^{d}\frac{(\Nc-k_i(x))!}{\Nc!k_i(x)!}\gamma^{-2k_i(x)}
\right)\right.\nn
&\qquad\qquad\qquad\quad \times \left. \left(\prod_{\ell\subset \Lat_B} \sigma(\ell) \prod_{(x,\mu)\in \ell} \Exp{(\delta_{\hmu,+\hat{0}}-\delta_{\hmu,-\hat{0}})\Nc \at\mu}\right)\prod_{i=1}^d\lr{\gamma^{-\Nc \delta_{\mu i}}}  \right\}
\end{align}
\end{widetext}
where the factor $\Nc!$ from the site weights for zero monomer number cancels the $1/\Nc!$ in the temporal gauge link, and a prefactor $\gamma^{N_c \Lat}$ was pulled out such that spatial links are now suppressed by $1/\gamma^2$ for mesons and $1/\gamma^\Nc$ for baryons. Also we have put the Grassmann constraint Eq.~(\ref{GC}) into the above equation via a Kronecker delta and the decomposition Eq.~(\ref{MBSites}).
We will now neglect the sub-leading terms, i.e.~we will only keep terms that survive in the limit Eq.~(\ref{CTLimit}). For any temperature, the average contribution per time location is $1/\gamma^2$. 
This will have drastic consequences, as spatial baryons for $\Nc\geq 3$ and spatial dimer occupation numbers $k_i>1$ will vanish.
We will later see how to interpret this outcome and also show numerically that this is well justified.
For now we note that the average dimer density will depend on the temperature, and (anti-) baryons are static for $\Nc\leq 3$ for all temperatures and chemical potential. For large $\gamma, \Nt$ the partition function becomes
\begin{widetext}
\begin{align}
\tilde{\calZ}(\gamma, \at\mu_q,\Nt)
&=  \sum_{
\substack{
\left.\{k\}\right|_{\LatM}\\
\left.\{\omega\}\right|_{\LatSpatB}
}} \left\{\left(\prod_{x\in \LatM}\GC
\frac{(\Nc-k_0(x))!}{k_0(x)!}
\left(\delta_{k_i(x),0}
+\delta_{k_i(x),1} \frac{1}{\Nc}\gamma^{-2}
\right)
\right)\left(\prod_{\vec{x}\in \LatSpatB}  e^{\omega(\vec{x})\Nc \at \mu_q \Nt}\right)\right\}\nn
&= \sumLatSpat\sum^{\rm GC}_{
\left.\left\{
\substack{k_0\in\{0,\ldots \Nc\}\\k_i\in\{0,1\}}
\right\}\right|_{\LatM}
} \left\{\left(\prod_{(\vec{x},\tau)\in \{ x|k_i(x)=1\}  }\frac{
v(k_0^{-}|k_0^{+})_{(\vec{x},\tau)}
v(k_0^{-}|k_0^{+})_{(\vec{x}+\hat{i},\tau)}
}{\gamma^{2}}
\right)\left( 2 \cosh(\mu_B/T)\right)^{|\LatSpatB|}\right\}
\label{PFLargeGamma}
\end{align}
\end{widetext}
where we have dropped the overall prefactor $\gamma^{\Nc \Lat}$ and we have used 
\begin{align}
 \mu_B/T&= \bareMu/\bareT=  \Nc \at\mu_q \Nt.
\end{align}
We have resummed static baryons  and anti-baryons \linebreak $\omega=\pm1$ in the second line, with $|\LatSpatB|$ the number of spatial sites occupied by (anti-) baryons with $\LatB=\LatSpatB\times \Nt$, $\LatM=\LatSpatM\times \Nt$. 
The sum over configurations contains all possible partitions of the spatial lattice $\{\LatSpatM,\LatSpatB\}$ with
\begin{align}
\LatSpatB\,\dot\cup\,\LatSpatM=\LatSpat\equiv \Ns^3.
\end{align}
The vertex weights introduced in the second line $v_{(\vec{x},\tau)}$ depend on the dimers $k_0^{-}=k_0(\vec{x},\tau-1)$ and $k_0^{+}=k_0(\vec{x},\tau)$ $k_0^{-}=k_0(\vec{x},\tau-1)$ and simplify due to the Grassmann constraint, $k_0^{-}
+k_0^{+}+1=\Nc$:
\begin{align}
 v(k_0^{-}|k_0^{+})&= \sqrt{\frac{1}{\Nc}
 \frac{(N_c-k_0^{-})!}{k_0^{-}!}\frac{(\Nc-k_0^{+})!}{k_0^{+}!}}\nn
&= \sqrt{
 \frac{(N_c-
 k_0^{-}
 )(1+k_0^{-})}{
 \Nc}}
 \label{Vertices}
\end{align}
They come in pairs of adjacent spatial sites \linebreak $(\vec{x},\tau)$, $(\vec{x}+\hat{i},\tau)$ at both ends of a spatial dimer on a bond $b=(\vec{x},\tau;i)$.
There are $\Nc$ types of vertices since $k_0^{-}$ can take the values from $0$ to $\Nc-1$. 
For $\Nc=3$, there are only three types of vertices, 
\begin{align}
 v(0|2)&= v(2|0)=1, &
v(1|1)&= \frac{2}{\sqrt{3}}.
\label{DimerBasedVertices}
\end{align}
The important observation is that the mesonic part of the partition function only depends on the number of vertices, and not on the precise temporal position.
The temporal intervals between vertices attached to spatial dimers have a trivial weight: due to the Grassmann constraint on every site where no spatial dimer is attached, $k_0^{-}
+k_0^{+}=\Nc$ implies that  
the dimer numbers form alternating chains as shown in Fig.~\ref{DimerChains} and cancel in weight:
\begin{align}
 \sqrt{
 \frac{(N_c-k_0^{-})!}{k_0^{-}!}\frac{(\Nc-k_0^{+})!}{k_0^{+}!}}
&= 1
\end{align}
Only the relative order of the vertices $v$ is important, but not the length of the intervals between them.
The partition function of SC-LQCD with $\Nc=3$ can be written in terms of these vertices as follows:
\begin{widetext}
\begin{align}
\tilde{\calZ}(\gamma, \mu_B/T,\Nt)  
&= \sumLatSpat\sum_{\ka}\sum_{
\left\{\left.\substack{
k_0(\vec{x},0),\\
n_\lsteel, n_\tsteel}
\right\}\right|_{\LatM}
}^{\text{GC}}
\left\{\left( 
\prod_{(\langle \vec{x},\vec{y}\rangle_j,\tau_j,),j=1}^{\ka} \left(\frac{v_\lsteel}{\gamma}\right)^{n_\lsteel(\vec{x},\tau_j)} \left(\frac{v_\tsteel}{\gamma}\right)^{n_\tsteel(\vec{y},\tau_j)}
\right)
\left( 2 \cosh(\mu_B/T)\right)^{ |\LatSpatB|}\right\}.
\label{ParFuncDTG}
\end{align}
\end{widetext}
The temporal dimers in the first time  slice \linebreak $k_0(\vec{x},\tau=0)$ are now dynamic variables in the partition sum.
The $\lsteel$-vertices and $\tsteel$-vertices at sites $x\in \LatM$ are defined in terms of the previous vertices,\linebreak
$v_\lsteel=v(0|2)=v(2|0)$, $v_\tsteel=v(1|1)=\frac{2}{\sqrt{3}}$, and the order in the high temperature expansion is given by the number spatial dimers:
\begin{align}
 n&=\frac{1}{2}\sum_{\vec{x}\in\LatSpatM}\int_0^{1/\bareT} \hspace{-2mm}d\tau \lr{n_\lsteel(\vec{x},\tau) + n_\tsteel(\vec{x},\tau)}\equiv N_{Ds}.
 \end{align}
In the partition sum, not all temporal positions of the vertices are possible due to the Grassmann constraint (GC).
We still need to replace $\gamma$ by the temperature $aT$, which requires bookkeeping of possible locations for spatial dimers. We will provide the details in the appendix Sec.~\ref{DerivationCT}. A simplified argument that allows to understand the temperature dependence is that for the first spatial dimer there are up to $\Nt$ possible locations between two adjacent spatial sites $\langle \vec{x},\vec{y} \rangle$, but due to the even-odd decomposition there are only $\Nt/2$ possible locations for the second spatial dimer, and likewise for all other dimers, as long as $\Nt$ is large.
Hence every spatial dimer, after summing over possible locations, has weight $\frac{\Nt}{2\gamma^2}=1/(2 \bareT)$.
The final result is
\begin{widetext}
\begin{align}
\calZ_{\rm CT}(\bareT,\bareMu)&=\sumLatSpat
\sum_{
\left. 
\left\{\omega\right\}
\right|_{\LatSpatB}}
e^{\omega_{\vec{x}} \mu_B/T}
\sum_{\ka\in 2\mathbb{N}}\frac{1}{\ka!}\frac{1}{(2\bareT)^{\ka}}
\sum_{\calG\in \Gamma_\ka^{\LatSpatM}} v_\tsteel^{N_\tsteel(\calG)},& 
N_\tsteel&=\sum_{\vec{x}\in \LatSpatM} \int_0^{1/\bareT} d\tau\, n_\tsteel(\vec{x},\tau),
\label{ParFuncCT}
\end{align}
\end{widetext}
where $\Gamma_\ka=\{n_\lsteel(\vec{x},\tau),n_\tsteel(\vec{x},\tau)\}$ is the set of all valid configurations on the mesonic sublattice $\LatSpatM$ with $\ka\equiv N_{Ds}$ spatial dimers and $N_{\tsteel}\leq 2\ka$ is the total number of $\tsteel$-vertices, integrated over the compact temporal direction. Since $v_\lsteel=1$, we do not need to include them into the weight. The prefactor $1/\ka!$ is due to time-ordering. In the next section we will simplify this result further by a Hamiltonian formulation, where we obtain a meaningful expression for $\Gamma_\ka$\\

We now want to discuss the interpretation of the final partition function: as illustrated in Fig.~\ref{absorptionemission}, as the temporal lattice spacing $\at\simeq a/\xi(\gamma) \rightarrow 0$,
multiple spatial dimers become resolved into single dimers. 
The overall number of spatial dimers remains finite in the CT-limit, as the sum over ${\cal O}(\gamma^2)$ sites compensates the $1/\gamma^2$ from spatial dimers. Its number is a function of the temperature and will signal spontaneous chiral symmetry breaking, see Sec.~\ref{ChiralTransition}.
As shown in Fig.~\ref{Suppression} it takes large $\Nt$ such that double dimers vanish, but it does not require large $\Nt$ to make baryons static. The sign problem has completely vanished as $\sigma(\ell)=1$ for static baryons loops $\ell$.
The set of all baryonic sites coincides then with the fermion bags that have been discussed in \cite{Gattringer2018}.
The expansion in $\ka$ is an all order high temperature expansion. It will also hold at very low temperatures and we will be able to address zero-temperature phenomena.

\subsection{Hamiltonian Formulation}

In order to rewrite the partition function further, we make use of a diagrammatic expansion. These methods, giving rise to Quantum Monte Carlo, are nowadays widely used in condensed matter \cite{Gull2011,Pollet_2012}. The general idea is to decompose the Hamiltonian $\calH=
\calH_0+\calH_i$ and express the partition function in terms of an expansion parameter $\ka$ which keeps track of the number of interactions described by $\calH_i$.
After summing over all configurations of a given order in $\ka$, one integrates over all possible times at which interaction events may take place
\newpage

\begin{figure}[h!]
\includegraphics[width=0.5\textwidth]{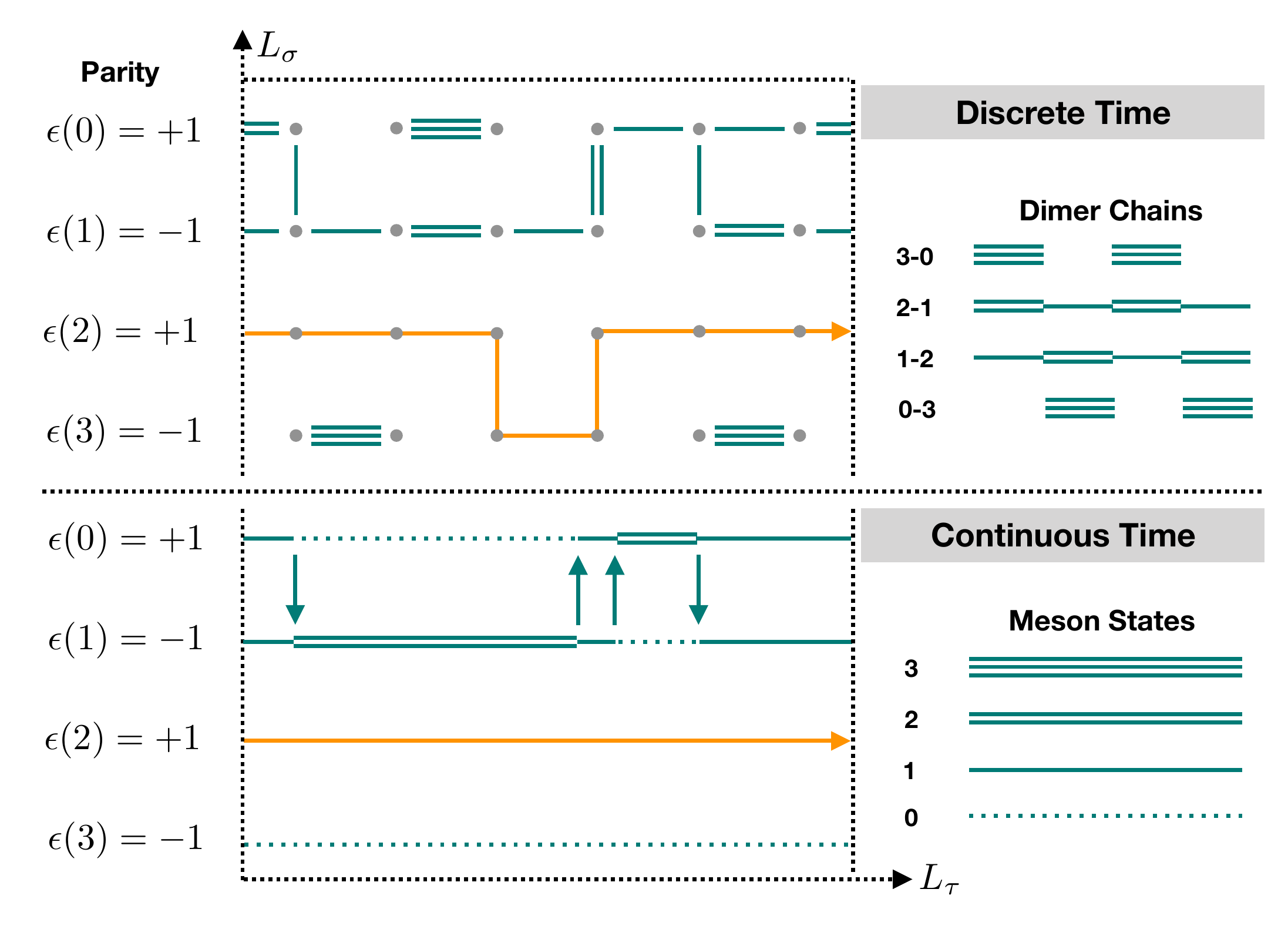}
\caption{
Correspondence between discrete time configurations in terms of dimer coverings and baryon world lines (top) and in terms of hadron occupation numbers in continuous time (bottom). 
Multiple spatial dimers become resolved in single spatial dimers (which can be oriented consistently from emission sites $\calE$ to absorption sites $\calA$, indicated by the arrow), baryons become static and only vertices of $\lsteel$-shape or $\tsteel$-shape survive as $\at\rightarrow 0$.
}
\label{DimerChains}
\end{figure}

We will take a step back and reformulate Eq.~(\ref{PFLargeGamma}) in new degrees of freedom: the temporal dimers $k_0(x)$ are replaced by an occupation number $\meson(x)$
by the following assignment:
\begin{align}
k_0(x)   &\mapsto  \meson(x)=\epsilon(x) \left(k_0(x)-\frac{\Nc}{2}\right)+\frac{\Nc}{2}\nn
\meson(x)&\in \left\{0,1,\ldots \Nc \right\}
\label{MesonState}
\end{align}
with $\epsilon(x)=\pm1$ the parity of a site introduced in Eq.~(\ref{ChiralTrafo}). As a consequence, the alternating dimer chains will be replaced by meson occupation numbers $\meson(x)$ which is constant on the interval between attached spatial dimers (see Fig.~\ref{DimerChains}), and the dimer-based vertices $v(0|2)$, $v(1|1)$, $v(0|2)$ in Eq.~(\ref{DimerBasedVertices}) are replaced by occupation number-based vertices $\tilde{v}(\meson|\meson')$, which change the meson state by one unit: $\meson(x) \mapsto \meson'(x)=\meson(x)\pm 1$:
\begin{widetext}
\begin{align}
\mathcal{\tilde{Z}}(\gamma,  \mu_B/T,\Nt)
&= \sum_{\ka=0}^{\infty}\sum_{\left.\{\meson,(\tau,l))\}\right|_{\LatM}} 
 \left\{\left(\prod_{  (\vec{x},i)\in l_j }
\frac{\hat{v}(\meson^-|\meson^+)_{(\vec{x},\tau)}}{\gamma}
\frac{\hat{v}(\meson^-|\meson^+)_{(\vec{x}+\hat{i},\tau)}}{\gamma}
\right)\left( 2 \cosh(\mu_B/T)\right)^{|\LatSpatB|}\right\}
\end{align}
\end{widetext}
In fact there is a conservation law: if a quantum number $\meson(x)$ is raised/lowered by a spatial dimer, then at the site connected by the spatial dimer the quantum number is lowered/raised. This is a direct consequence of its definition Eq.~(\ref{MesonState}): the parity of the two sites connected by a spatial dimer is opposite.
We therefore can replace the vertices by raising and lowering operators:
\begin{widetext}
\begin{align}
\calZ_{\rm CT}(\bareT,\bareMu)&= \sumLatSpat\left\{
\Tr_{\meson|\LatSpatM} \left[
\exp\left(\frac{1}{2\bareT}
\sum_{\langle \vec{x},\vec{y}\rangle
} 
\left(
\hat{v}(\vec{x})_{\meson,\meson+1}\hat{v}(\vec{y})_{\meson,\meson-1}
+
\hat{v}(\vec{x})_{\meson,\meson-1}\hat{v}(\vec{y})_{\meson,\meson+1}
\right)
\right)
\right]
\Tr_{r|_{\LatSpatB}}\left[
 e^{\hat{\omega} \mu_B/T}\right]
\right\}\nn
&=\Tr_\hadron\left[e^{(\hat{\calH}+\hat{\calN}\bareMu)/\bareT}\right],\qquad
\hat{\calH}=\hat{\calH}_0+\hat{\calH}_i,\qquad \hat{\calH}_0=0,\qquad
\hat{\calH}_i=\frac{1}{2}\sum_{
\langle\vec{x},\vec{y}\rangle}
\left(
\hat{J}^{+}_{\vec{x}} \hat{J}^{-}_{\vec{y}}+
\hat{J}^{-}_{\vec{x}} \hat{J}^{+}_{\vec{y}}
\right),\qquad
\hat{\calN}=\sum_{\vec{x}}\hat{\omega}_x,\nn
\hat{J}^+&=\left(
\begin{array}{cccc|cc}
0   & 0   & 0   & 0 &  & \\
\hat{v}_\lsteel & 0   & 0   & 0 &  & \\
0   & \hat{v}_\tsteel & 0   & 0 &  & \\
0   & 0   & \hat{v}_\lsteel & 0 &  & \\
\hline
 &  &  &  & 0 & 0\\
 &  &  &  & 0 & 0\\
\end{array}
\right),\qquad
\hat{J}^{-}=(\hat{J}^{+})^T, \qquad
\hat{\omega}=\left(
\begin{array}{cccc|cc}
0 & 0 & 0 & 0 &  & \\
0 & 0 & 0 & 0 &  & \\
0 & 0 & 0 & 0 &  & \\
0 & 0 & 0 & 0 &  & \\
\hline
 &  &  &  & 1 & 0\\
 &  &  &  & 0 & -1\\
\end{array}
\right),\qquad
| \hadron\rangle=| \meson,\baryon\rangle =\left(
\begin{array}{c}
0 \\
\pi \\
2\pi \\
3\pi \\
\hline
B^+ \\
B^- \\
\end{array}
\right).
\label{ParFuncHam}
\end{align}
\end{widetext}
This result is valid for $\Nc=3$, $\Nf=1$.
A corresponding result for $N_f=2$ is given in the appendix Sec.~\ref{ParFuncNf2}.
In the second line we have included the baryonic sites into the trace, 
and introduced the mesonic raising and lowering operators $\hat{J}^+$,  $\hat{J}^{-}$ (which contain the vertices), 
and the baryon number operator $\hat{\calN}$. 
The block-diagonal structure expresses the fact that the Hilbert space of hadrons is a direct sum of mesonic states and baryonic states,
$|\hadron\rangle=|\meson\rangle\oplus|\baryon\rangle $, which results in the vanishing commutator
\begin{align}
[\hat{\calH},\hat{\calN}]=0.
\end{align}
The fact that mesons and baryons are mutually exclusive (leading to the factorization into mesonic and baryonic subvolumes) results in one of the two blocks being zero in both operators $\hat\calH$ and $\hat\calN$.
The meson states $|\meson\rangle$ count pseudoscalars, and we will denote them as pions $\pi$ (despite the fact they are flavorless for $\Nf=1$ and they cannot be distinguished from the $\eta$ or $\eta'$ mesons). The pion current is conserved, but only in the chiral limit. Monomers would generate a mass to the pion. Since Pauli saturation holds on the level of the quarks and pions have a fermionic substructure, we cannot have more than $\Nc$ pions per spatial site.
Due to the conservation of the pion current, if we start on each site with $\Nc$ pions, or with no pions at all, there cannot be any spatial dimer that transfers a meson to an adjacent site: either all sites are already saturated with mesons, or there is no meson to be transferred.
If we omitted the additive constant $\Nc/2$ from Eq.~(\ref{MesonState}),  particle-hole symmetry becomes evident. To see this, consider
the anti-commutator of the mesonic operators (restricted on mesonic states):
\begin{widetext}
\begin{align}
[\hat{J}^+,\hat{J}^-]&=\left(
\begin{array}{cccc}
-\hat{v}_\lsteel^2 & 0 & 0 & 0  \\
0 & \hat{v}_\lsteel^2-\hat{v}_\tsteel^2 & 0 & 0  \\
0 & 0 & \hat{v}_\tsteel^2-\hat{v}_\lsteel^2 & 0 \\
0 & 0 & 0 & \hat{v}_\lsteel^2  \\
\end{array}\right)=
\left(
\begin{array}{cccc}
-1 & 0 & 0 & 0  \\
0 & -1/3 & 0 & 0  \\
0 & 0 & 1/3 & 0 \\
0 & 0 & 0 & 1  \\
\end{array}\right).
\end{align}
\end{widetext}
The corresponding algebra has the structure of a spin, and it generalizes via Eq.~(\ref{Vertices}) to arbitrary $\Nc$:
\begin{widetext}
\begin{align}
 \hat{J}_1&= \frac{\sqrt{\Nc}}{2}\left(\hat{J}^+ + \hat{J}^-\right), \quad
 \hat{J}_2= \frac{\sqrt{\Nc}}{2i}\left(\hat{J}^+ - \hat{J}^-\right), \quad \hat{J}_3=i[\hat{J}_1,\hat{J}_2]=
 {\rm diag}\left(-\frac{\Nc}{2}, -\frac{\Nc}{2}+1, \ldots \frac{\Nc}{2}\right)
=\frac{\Nc}{2}[\hat{J}^+,\hat{J}^-]\nn
\hat{J}^2&=\frac{\Nc}{2}{\rm diag}(\hat{v}_0^2,\hat{v}_1^2+\hat{v}_0^2,\ldots \hat{v}_{\Nc-1}^2 +\hat{v}_{\Nc-2}^2, \hat{v}_{\Nc-1}^2)+\frac{1}{4}{\rm diag}(\Nc^2,(\Nc-2)^2,\ldots \Nc^2) = 
\frac{\Nc(\Nc+2)}{4}\mathbbm{1}
\end{align}
\end{widetext}
with $\hat{v}_k^2=(\Nc-k)(1+k)/\Nc$.
The ``spin''-representation is $d=\Nc+1$-dimensional, with $S=\Nc/2$.
For $\Nc=1$, $\hat{J}_\pm=\frac{1}{2}(\sigma_x\pm i\sigma_y)$  is expressed in terms of the Pauli matrices, and the continuous time partition function becomes that of the quantum XY model.
Although the algebra resembles that of a particle with spin, it has nothing to do with the spin of mesons or quarks. 
The alternating chains are simply expressing the fact that for staggered fermions, the lattice spacing is $2\at$ rather than $\at$. By  shifting the pion occupation numbers by its average value, we can identify the quantum state corresponding to this algebra:
\begin{widetext}
\begin{align}
 \meson  \mapsto\spin&=\meson-\frac{\Nc}{2}:& \hat{J}_3 \left|\frac{\Nc}{2},\spin \right\rangle&=\spin \left|\frac{\Nc}{2},\spin \right\rangle,& \hat{J}^2 \left|\frac{\Nc}{2},\spin\right\rangle&=\frac{\Nc\left(\Nc+2\right)}{4} \left|\frac{\Nc}{2},\spin \right\rangle,& [\hat{J}^2,\hat{J}_3]&=0.
\end{align}
\end{widetext}
This remarkable result is due to the fact that pion occupation numbers on the lattice are not just bounded from below but also from above.
We conclude this section by providing a physical interpretation of the dynamics on the hadronic states:
the pion dynamics encoded in the Hamiltonian is that of relativistic pion gas \cite{deForcrand2016}. In contrast, the fact that baryon becomes static is due to its non-relativistic nature. Its restmass is large but finite (see Sec.~\ref{baryonMassNucInt}).

\section{Continuous Time Worm Algorithm}
\label{Algorithm}

\subsection{Poisson Process}
\label{PoissonProcess}

Before we address the algorithm that samples the partition function Eqs.~(\ref{ParFuncCT}, \ref{ParFuncHam}), we want to emphasize an important property: spatial dimers are distributed uniformly in time.
The interval length (interpreted as the inter-arrival time between spatial dimers) are then exponentially distributed and the number of spatial dimers in a fixed time interval is Poisson distributed. Hence they can be  generated via a Poisson process:
\begin{align}
P(\Delta t)&=\exp(-\lambda \Delta t),& \Delta t& \in [0,1]
\label{PoissonProb}
\end{align}
with $\lambda$ the ``decay constant'' for spatial dimer emissions.
Due to the presence of baryons, $\lambda$ is space dependent: 
\begin{align}
\lambda&=d_M(\vec{x})/(4\bareT),& d_M(\vec{x})&=2d-\sum_{\langle \vec{x},\vec{y}\rangle} |B(\vec{y})|
\label{DecayProb}
\end{align}
where $d_M(\vec{x})$ is the number of mesonic sites adjacent to $\vec{x}$ where the Poisson process operates.
Note that in Eq.~(\ref{PoissonProb}) we have rescaled the compact time interval \linebreak $[0,1/\bareT] \mapsto [0,1]$ and thus have put the temperature into the decay constant $\lambda$.

The Poisson process of emitting pions from $(\vec{x},t)$ to an adjacent site $(\vec{y},t)$  with probability $\lambda$ gives rise to a decomposition of vertices into emission sites $(\vec{x},t)\in \calE$ and absorption sites  $(\vec{y},t)\in \calA$. Spatial dimers can be oriented consistently due to the underlying even/odd decomposition of lattice sites, but is also evident in the Hamiltonian representation, where $J^-$ is an emission and
$J^+$ is an absorption event. 
The emission sites $\calE$ are simply those that reduce the pion occupation number $\meson$ in Euclidean time by one unit, the absorption sites $\calA$ are those that increase $\meson$. Every spatial dimer corresponds to a pion hopping to an adjacent site and connects an $\calE$-site with an $\calA$-site. The number of $\calE$-sites equals the number of $\calA$-sites, and due to the periodic boundary conditions in time this even holds for every spatial site $\vec{x}$: 
\begin{align}
|\{t_i\,|\, (\vec{x},t_i) \in  \calE\}|&= 
|\{t_i\,|\, (\vec{x},t_i) \in  \calA\}|
\end{align}

The continuous time worm algorithm (CT-WA) needs to fulfill detailed balance, such that the emission process is counterbalanced by an absorption process to obtain the equilibrium distribution of spatial dimers according to temperature and chemical potential.

\subsection{Details of the Continuous Time Worm Algorithm}

Worm algorithm at discrete time (DT-WA) are well established also for strong coupling lattice QCD \cite{Adams2003,Forcrand2010,Fromm2010}.
Designing an algorithm that operates directly in the Euclidean continuous time limit has several advantages: 
(1) the ambiguities arising from the functional dependence of observables on the anisotropy $\xi(\gamma)$ - in particular non-monotonic behavior - will be circumvented and (2) we do not need to perform the continuum extrapolation $\Nt \rightarrow \infty$. This will allow us (3) to measure the phase boundaries unambiguously, as the baryonic part of the partition function simplifies such that the sign problem is completely absent, and
(4) at all temperatures of interest, the CT algorithm is considerably faster than its discrete version, in particular since the baryon update does not require a worm algorithm but can be replaced by a heat bath update.

\begin{figure}[h]
\includegraphics{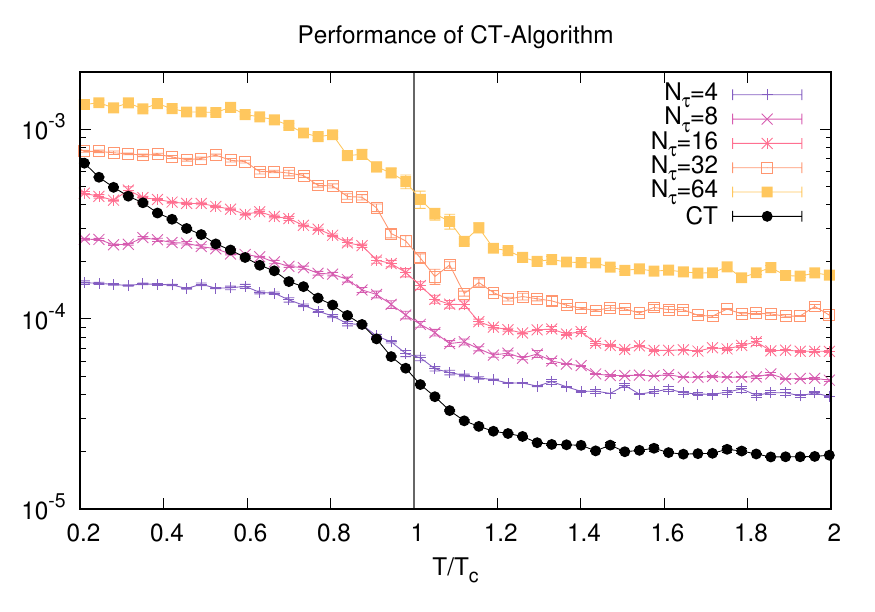}
\caption{
The performance of the continuous time algorithm CT-WA compared to the discrete time algorithm DT-WA for various $\Nt$. For a large range of temperatures, and in particular at the chiral transition, the CT-WA performs even better then DT-WA for $\Nt=4$. The lower the temperature, the larger $\Nt$ is required to obtain correct results for the various observables (see also Fig.~\ref{Suppression}).
}
\label{Performance} 
\end{figure}

\begin{figure*}[t]
\includegraphics[width=\textwidth]{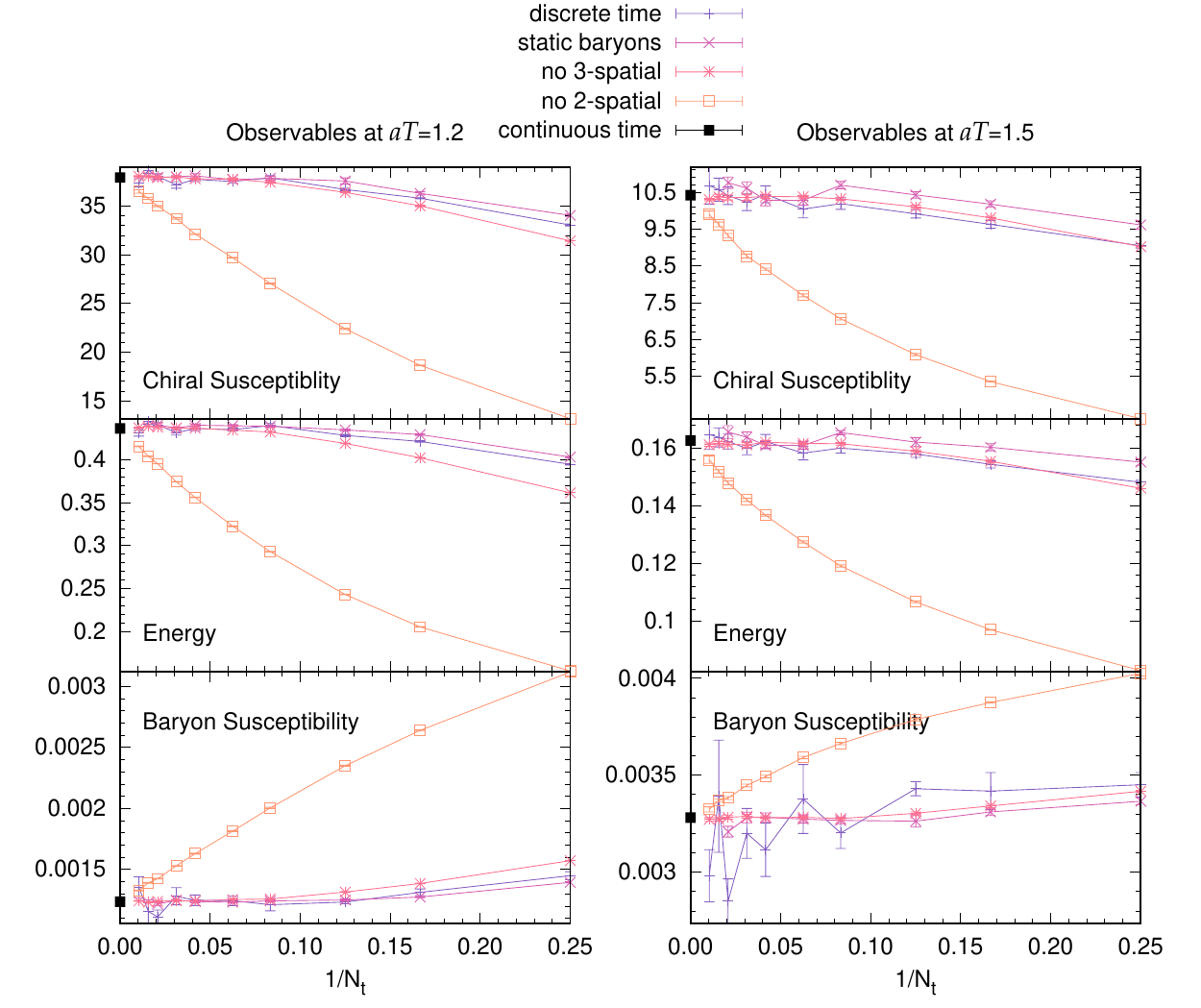}
\caption{$\Nt$-dependence of the chiral susceptibility (\emph{top}) and the energy (\emph{center}) and the baryon susceptibility (\emph{bottom}). We compare the full discrete simulations and various approximations according to the steps in deriving the continuous time limit (static baryon approximation, exclusion of spatial triple dimers, , exclusion of spatial double dimers).
We have fixed the bare temperature to $\bareT=1.2<\bareT_c$ and $\bareT=1.5>\bareT_c$ 
All observables extrapolate well into the continuum limit, with its Monte Carlo result at $1/\Nt=0$ having much smaller error.}
\label{Suppression}
\vspace{5mm}
\end{figure*}

In Fig.~\ref{Performance} we clearly see that the CT-worm algorithm outperforms the DT-worm algorithm at temperatures in the vicinity of the transition temperature and above. 
The lower the temperature, the more spatial dimers are sampled, which makes the average worm update longer. At first glance it seems that the CT-worm becomes more expensive, but one needs to keep in mind that lower temperatures require larger $\Nt$ to get valid estimates for observables. On a lattice with time extent $\Nt$, temperatures below $1/\Nt$ (which have $\gamma<1$) will have more spatial dimers than temporal dimers and suffer from saturation effects: The density of spatial dimers is limited to $\Nc \Nt/2$, whereas it is unlimited at continuous time. In Fig.~\ref{Suppression} we show the $\Nt$-dependence of various observables: they have a well-defined CT-limit. Also, this figure illustrates that the approximations which lead to $\calZ_{CT}$ in Eq.~(\ref{ParFuncHam}) are well justified. 
The extrapolation from discrete time to continuous time is difficult: large $\Nt$ require more statistics, and due to the sign problem, most observables get noisy due to sign reweighting. The first approximation is to make baryons static, which eliminates the sign problem. This step makes the extrapolation much more controlled, and even for $\Nt=4$, the static baryon approximation is not bad. 
Next we prohibit sites which have more than $3$ spatial dimers, which has only a mild effect at the temperatures considered here. If we also prohibit sites with more than 2 spatial dimers, the deviation at finite $\Nt$ is drastic, but also this approximation extrapolates to the same CT-limit for the observable. The point at $1/\Nt=0$ in Fig.~\ref{Suppression} is the outcome of the CT-WA, which has much smaller error bars and better performance with the same number of worm updates.

Continuous time (CT) algorithms as for Quantum Monte Carlo are now widely used in condensed matter (see e.g. \cite{Beard1996, Gull2011}), whereas CT methods in quantum field theories is rather new \cite{Unger2011,Unger2012,Huffman:2017swn,Huffman:2019efk}
The basic idea of a worm algorithm introduced in \cite{Prokofev2001} is to sample an enlarged configuration space with defects on the lattice known as worm tail $x_T$ and worm head $x_H$. Every worm algorithm consists of two kinds of updates: (1) move updates, which move the head $x_H$ and tail $x_T$ to a new site $x_0$, and (2) shift updates, which move the head $x_H$ through the lattice until the worm recombines with the tail. 
Worm algorithms are highly efficient: after recombination, the configuration has been globally updated, similar to cluster algorithms. Moreover, during the shift update, 2-point correlation functions can be measured.
In order to apply a worm algorithm, the partition function needs to be written in terms of bond variables. Those representations are typically available in spin models from the high-temperature expansion. In the case of lattice QCD, a dual representation based on the strong coupling expansion also admits the applicability of worm algorithms.

Our Continuous Time Worm Algorithm (CT-WA) can be derived from the Discrete Time Directed Path Worm Algorithm (DT-WA) that has been developed for $\U(\Nc)$ gauge group in the strong coupling limit \cite{Adams2003}, which does not include baryons.
This worm algorithm is based on an even-odd decomposition of weights: if the parity of the head $\epsilon(x_H)$ is the same as 
that of the tail $\epsilon(x_T)$, then the head has an active site location, if the parities differ the head is an a passive site.
The active sites correspond to the absorption sites $\calA$, and the passive sites correspond to the emission sites $\calE$ as discussed above. 

\begin{figure}[h!]
\begin{minipage}{0.5\textwidth}
\includegraphics[width=\textwidth]{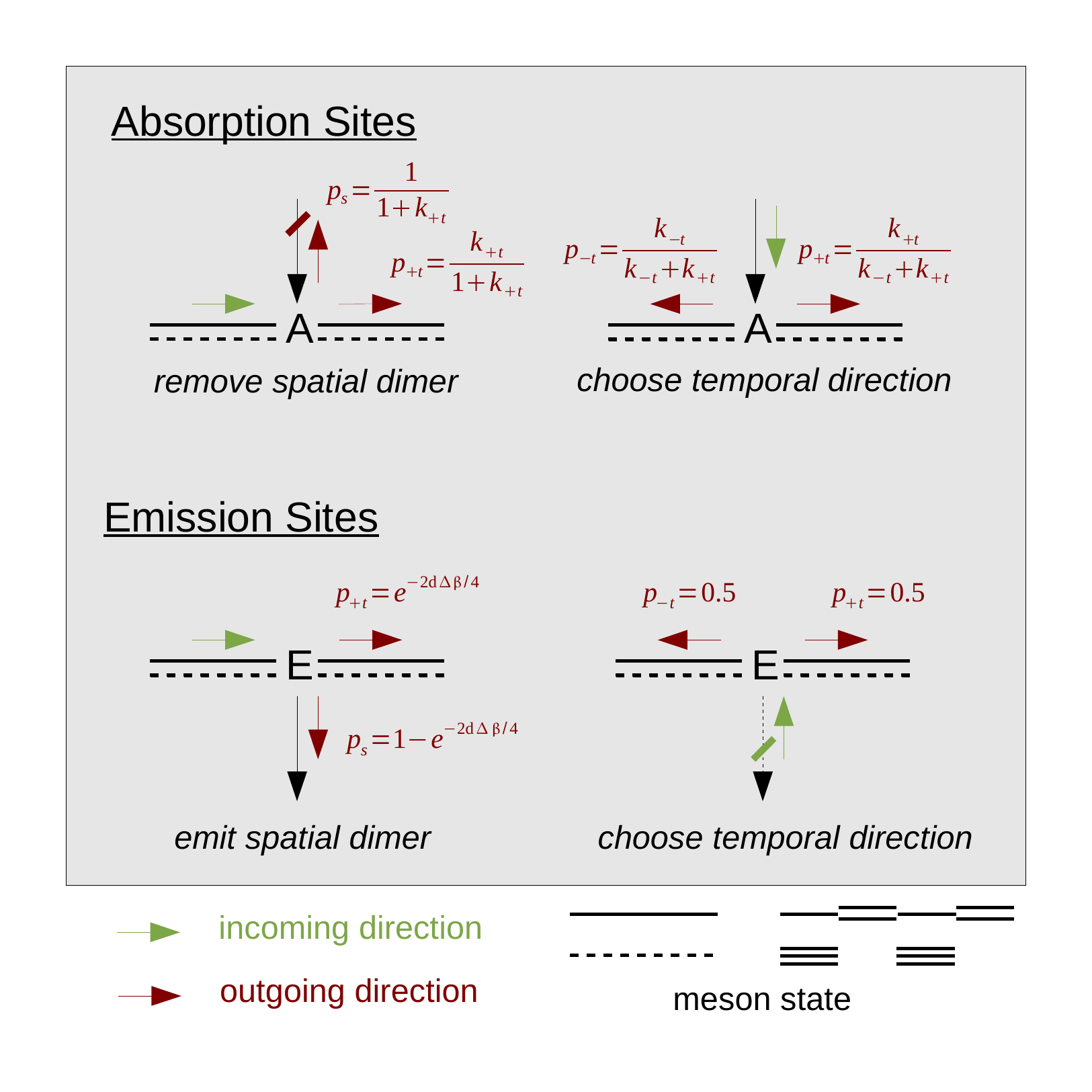}
\end{minipage}
\caption{Updating rules for the continuous time algorithm. 
\emph{Top:} an absorption site can be approached either from the temporal direction (\emph{left:} a spatial dimer may be removed) or from the spatial direction (\emph{right:} a dimer was emitted in the previous step). 
\emph{Bottom:} an emission site can be approached either from the temporal direction (\emph{left:} a spatial dimer may be emitted) or from the spatial direction (\emph{right:} a dimer was removed in the previous step). 
}
\label{absorptionemission}
\end{figure}

For $\SU(\Nc)$ gauge group, it is required to have two separate worms, one in the mesonic sector and one in the baryonic sector \cite{Fromm2010}. The mesonic worm for the $\SU(\Nc)$  group differs from the directed path Worm for $\U(\Nc)$ in one important aspect: In the directed path version backtracking is prohibited to evolve faster through configuration space (if the update shifts the worm head from $x$ to the adjacent site $y$, then in the next shift update the worm is not allowed to go back). With the simple baryon loop geometries in the CT-limit, we can supplement the continuous time version of the directed path worm algorithm by an additional heat bath update: after the mesonic worm has recombined, we propose for all sites $\vec{x}$ where no spatial dimers are attached (the so-called static sites) a new hadronic state with the probabilities
\begin{align}
 p(\meson)&=\frac{1}{\Nc+1+2\cosh(\mu_B/T)},& \meson=0,\ldots \Nc,\nn
 p(B^\pm)&=\frac{e^{\pm \mu_B/T}}{\Nc+1+2\cosh(\mu_B/T)}.
\end{align}
The consequence is that if the worm head propagates in positive/negative temporal direction, it will continue to do so until it will either emit or absorb a pion, i.~e. it will either add or delete a spatial dimer. It will not change the direction and diffuse: the CT-WA can be regarded as a Poisson process.
The updating rules are outlined in Fig.~\ref{absorptionemission}. The probabilities for the various cases (approaching/leaving an absorption site $\calA$ or emission site $\calE$) depend on the involved states $\meson$:  (1) if an $\calA$-site is approached from the temporal direction, the spatial dimer is removed with a heat bath probability determined by $J^-$, (2) if an $\calA$-site is approached from a spatial direction, the new temporal direction is also determined by $J^{-}$, (3) if an $\calE$-site is approached from temporal direction, the emission probability to insert a spatial dimer is $1-e^{-\lambda \Delta 
\tau}$ and the probability to continue in temporal direction is $e^{-\lambda \Delta \tau}$, (4) if it is approached from spatial direction, the new temporal direction is chosen equally likely.
At high temperatures, $\lambda=\lambda(\bareT)\ll 1$ according to Eq.~(\ref{DecayProb}) and the worm head will very likely continue in temporal direction by some time $\Delta t$ with probability $p_\tau\simeq 1-\lambda \Delta t$ and emit a spatial dimer with probability $p_\sigma\simeq \lambda \Delta \tau$.
The higher the temperature, the longer the worm propagates in temporal direction, possibly looping through the periodic boundary back to where it started.

In the discrete time algorithm, 
during worm evolution, whenever the worm head is on a site with opposite parity compared to the worm tail, $\epsilon(x_H)=-\epsilon(x_T)$, both worm head and tail can be interpreted as monomers (if $\epsilon(x_H)=\epsilon(x_T)$, the head is a sink rather a source for monomers). Even in the chiral limit, the monomer 2-point function can be accumulated in a histogram (due to translation invariance, only the relative lattice vector $z=x_1-x_2$ is needed):
\begin{align}
H_2(z) \mapsto  H_2(z)+\frac{\Lat}{d_M(x_x)+2\gamma^2}\delta_{z,x_T-x_H}
\label{CorrHistoIncDT}
\end{align}
with $d(x)$ defined in Eq.~\ref{DecayProb}.
An equivalent definition holds in the CT-limit:
\begin{align}
H_2(\vec{z},\tau) &\mapsto H_2(\vec{z},\tau) +\frac{\LatSpat}{2\bareT}\delta_{\vec{z},\vec{x}_T-\vec{x}_H}\delta(\tau-(\tau_T-\tau_H))\nn
G(\vec{z},\tau)&=\expval{
\cbc_{0}\,\cbc_{\vec{z},\tau}}
\simeq \frac{\Nc}{Z}H_2(\vec{z},\tau)
\label{CorrHistoIncCT}
\end{align}
with $Z$ the number of worm updates and $G(\vec{z},\tau)$ the connected chiral 2-point function approximated by accumulated and normalized histogram.
Details on how this and other mesonic 2-point correlation function are determined in practice are given in Sec.~\ref{CorrelationFunction}.

\onecolumngrid
\twocolumngrid
\clearpage

\subsection{Observables}
Almost all observables that can be measured via the DT-WA version can also be measured via CT-WA.
This is obviously the case for all observables that can be obtained as derivatives of $\log\calZ_{CT}$.
The discrete time observables in terms of the dual variables 
\begin{align}
N_M&=\sum\limits_x n_x,\quad N_{Dt}=\sum\limits_x k_{x,0},\quad N_{Bt}=\sum\limits_{x}|b_{x,0}|,\nn
 N_q&=2N_{Dt}+\Nc N_{Bt},\quad N_B=\sum_x \omega_x
\end{align}
are discussed in \cite{Bollweg2018}. The corresponding dimensionless thermodynamic observables in the CT-limit simplify because
\begin{align}
\lim_{\gamma \rightarrow \infty}\frac{\xi(\gamma)}{\gamma}\frac{d\gamma}{d\xi}&=\frac{\kappa\gamma^2}{\gamma}\frac{1}{  2\kappa\gamma}=\frac{1}{2},
\end{align}
which should be compared to the isotropic case based on Eq.~(\ref{AnsatzGamma02}):
\begin{align}
\left.\frac{\xi(\gamma)}{\gamma}\frac{d\gamma}{d\xi}\right|_{\gamma=1}
&=\frac{1}{2+4\kappa(\kappa-1)}\simeq 0.760(1).
\end{align}
Also, in the CT-limit we have no longer temporal dimers but only spatial dimers, and have to consider the chiral limit:
\begin{align}
 N_{q} &=\Nc\Nt\Ns^3-2N_{Ds},&
 N_M    &=0.
 \end{align}
We are now able to define the continuous time observables in terms of dual variables, which are always in dimensionless units with $a=\as$ and $V=\Ns^3 a^3$. Important observables are (1) the baryon density:
\begin{align}
a^3n_B&=\left.a^3\frac{T}{V}\frac{\partial \log \calZ}{\partial \mu_B}\right|_{V,T}=\frac{\expval{N_B}}{\Ns^3}=\expval{\omega}
\label{BaryonDensity}
\end{align}
which is given by the average winding number; (2) the energy density
\begin{align}
a^4\epsilon&=\bareMu a^3 n_B-\frac{a^4}{V}\left.\frac{\partial \log \calZ}{\partial T^{-1}}\right|_{V,\mu_B}
= C-\kappa\,\bareT \expval{n_{Ds}},
\label{energydensity}
\end{align}
where the irrelevant additive constant $C=\frac{1}{2}\Nc\Lat$ can be neglected compared to discrete time as we dropped the prefactor $\gamma^{\Nc \Lat}$ in Eq.~(\ref{PFLargeGamma}) which contained both the contribution from static mesons and static baryons;
(3) the pressure 
\begin{align}
a^4p&=\left.a^3 \kappa\,\bareT\frac{\partial \log \calZ}{\partial V}\right|_{T,\mu_B}=\frac{a^4\epsilon}{3}=\frac{1}{3}(C-\kappa\,\bareT\expval{n_{Ds}}),
\label{Pressure}
\end{align}
which in the strong coupling limit and chiral limit is just proportional to the energy density  such that the interaction measure $\epsilon -3p$ vanishes. At finite quark mass, the interaction measure is proportional to the chiral condensate, which here is zero in a finite volume as $\expval{\cbc}\propto \expval{n_M}$ (but see Sec.~\ref{ZeroT});
(4) the chiral susceptibility 
\begin{align}
a^6\chi_q&\equiv\chi_\sigma=\frac{\partial^2 \log \calZ}{\partial (2\hat{m_q})^2}
=\expval{n_M^2}\nn
&=a^4 \bareT \sum_{\vec{\vec{z}}}\int_{0}^{1/\bareT}d\tau\, G_2(\vec{z},\tau)
\end{align}
with only has the connected contribution non-zero in the chiral limit and 
$G_2(\vec{x}_1,t_1;\vec{x}_2,t_2)\equiv G_2(\vec{x}_1-\vec{x}_2,t_1-t_2)$ is the translation invariant monomer 2-point function that is measured during worm evolution, see Eq.~(\ref{CorrHistoIncCT});
(5) the entropy density
\begin{align}
a^3s&=\frac{a^3}{V T}\left(\frac{4\epsilon}{3}-\mu_B n_B\right).
\end{align}
The chiral condensate vanishes in the chiral limit in a finite volume. This is also evident from the absence of monomers in the dual representation. It is possible to obtain the chiral condensate from a 1/V expansion via chiral perturbation theory in a finite box, as explained in Sec. \ref{ZeroT}. 
Note that the pressure defined in Eq.~(\ref{Pressure}) is not equal to 
\begin{align}
a^4p'= a^3\frac{\kappa \bareT}{V}\log \calZ
\label{Pressure2}
\end{align}
because on the lattice the system is not homogeneous. The identity $p=p'$ only strictly holds in the continuum.

\subsection{Polymer Formulation and Wang Landau Method}
\label{WangLandau}

\newcommand{\DB}{{B}}

So far we have treated the mesonic and baryonic sector separately, 
and there is no need for the resummation known as the Karsch-M\"utter trick \cite{Karsch1989} for real chemical potential as there is no sign problem in the CT-limit.
However, a resummation of static mesons and baryons proves to be advantageous in several respects: (1) it allows to extend simulations to imaginary chemical potential 
beyond the value of $\at\mu_q=i \pi T/6$, where the baryon density becomes zero (discussed in Sec. \ref{ImagMu}), (2) we are able to adapt the Wang-Landau method 
\cite{WangLandau2001}
for determining the first order transition at low 
temperatures very accurately, and obtain also the canonical phase diagram from the density of states at high precision, see Sec. \ref{PhaseDiag}.

Apart from the usual (anti-) baryons denoted by $\calB$,  will discuss here two kinds of resummations of quantum states: the superposition of baryons and anti-baryons  ($\calP$-Polymers), and including static mesons ($\calQ$-Polymers):
\begin{align}
 |\calB\rangle_{\vec{x}}&=|B^+\rangle_{\vec{x}}-|B^-\rangle_{\vec{x}},&  B(C)&=\sum_{\vec{x}} b_{\vec{x}}(C),\nn
 |\calP\rangle_{\vec{x}}&=|B^+\rangle_{\vec{x}}+|B^-\rangle_{\vec{x}},&  P(C)&=\sum_{\vec{x}} p_{\vec{x}}(C),\nn
 |\calQ\rangle_{\vec{x}}&=|P\rangle_{\vec{x}}+\sum_{\meson=0}^{\Nc}|\meson\rangle_{\vec{x}} ,&Q(C)&=\sum_{\vec{x}}q_{\vec{x}}(C),
\label{PolymerDef}
\end{align}
where for a given configuration $C$, on each spatial site, the baryon and polymer numbers $B\leq P\leq Q$ are related via (in the following $V=\Ns^3$):
\begin{align}
b_{\vec{x}}&\in \{0,\pm 1\},& B&\in \{-V,\ldots V\}\nn
p_{\vec{x}}=|b_{\vec{x}}|&\in \{0, 1\},& P&\in \{0,\ldots V\}\nn
q_{\vec{x}}=p_{\vec{x}}+m_{\vec{x}}&\in \{0, 1\},& Q&\in \{0,\ldots V\}
\end{align}
with $m_{\vec{x}}=1$ iff the site is mesonic and static. The corresponding single site weights are:
\begin{align}
w_\calB(\mu_B/T)&=\exp\lr{\pm\frac{\mu_B}{T}},\nn
w_\calP(\mu_B/T)&=2 \cosh\lr{\frac{\mu_B}{T}},\nn
w_\calQ(\mu_B/T)& =\Nc+1+2 \cosh\lr{\frac{\mu_B}{T}},
\label{PolymerWeight}
\end{align}
These weights will be used for the following binomial/trinomial distributions:
\begin{widetext}
\begin{align}
 D^{\calQ\calP}_{\mu_B/T}(Q,P)&=\binom{Q}{P}\frac{(\Nc+1)^{Q-P} w_p(\mu_B/T)^P}{w_q(\mu_B/T)^Q},\qquad
 D^{\calP \calB}_{\mu_B/T}(P,B)=\binom{P}{(B+P)/2}\frac{e^{B\mu_B/T}}{w_p(\mu_B/T)^P},\nn
 D^{\calQ \calB}_{\mu_B/T}(Q,B)&=\sum_{P=|B|}^{Q}\binom{Q}{\frac{P+B}{2},\frac{P-B}{2},Q-P}
\frac{e^{B\mu_B/T}(\Nc+1)^{Q-P} }{w_q(\mu_B/T)^Q},
\label{BinomTrinom}
\end{align}
\end{widetext}
  with $B^\pm=\frac{P\pm B}{2}$ the number of (anti-) baryon sites and
$Q-P$ is the number of static mesons.
For some observables we need higher moments of the baryon number. We then only keep track of the histogram for $\calQ$-polymers, $H^\calQ_{V,\bareT,\bareMu}(Q)$ (normalized accordingly to be a probability distribution), and get the histogram in the baryon number $H^\calB_{V,\bareT,\bareMu}(B)$ from the above distributions:
\begin{align}
 H^{\calP}_{V,\bareT,\bareMu}(P)&=\sum_{Q=P}^{V}D^{\calQ\calP}_{\mu_B/T}(Q,P)\, H^{\calQ}_{V,\bareT,\bareMu}(Q),\nn
 H^{\calB}_{V,\bareT,\bareMu}(B)&=\sum_{P=B}^{V}   D^{\calP\calB}_{\mu_B/T}(P,B)\, H^{\calP}_{V,\bareT,\bareMu}(P)\nn
 &= \sum_{Q=P}^{V}  D^{\calQ \calB}_{\mu_B/T}(Q,B)\, H^{\calQ}_{V,\bareT,\bareMu}(Q).
\end{align}
For large spatial volumes $V$, the distributions in Eq.~(\ref{BinomTrinom}) involve large numbers. In practice we use the logarithmic versions of both histograms and binomial/trinomial distributions.
The polymer resummation will turn out to be crucial for the measurement of baryon fluctuations for the Taylor coefficients, see Sec.~\ref{TaylorExpansion}.\\

\begin{figure}[h!]
 \includegraphics[width=0.49\textwidth]{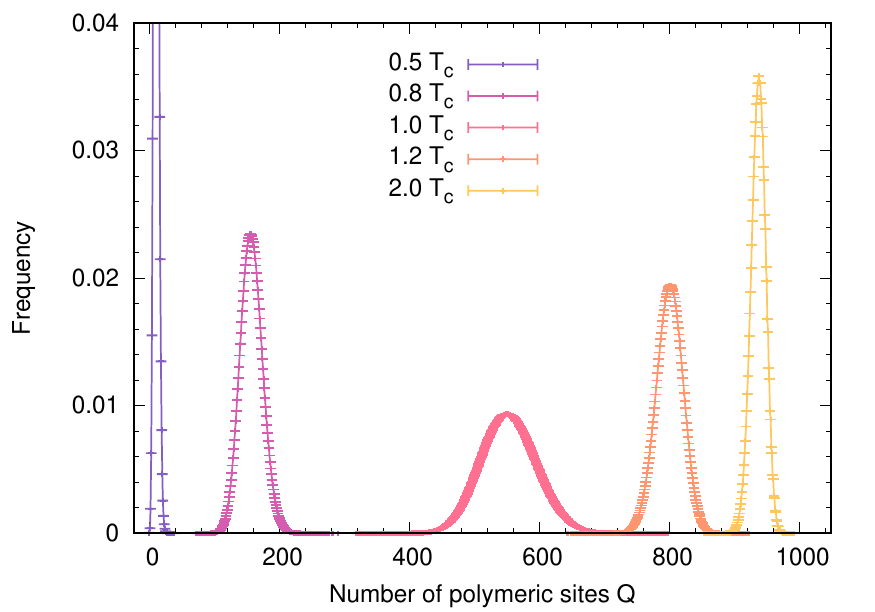}
 \caption{
 The $\calQ$-polymer histograms are shown for various temperatures, evaluated on a $10^3$-CT volume. At low temperature, almost all sites have spatial dimers attached, most configurations have low polymer number. At high temperature, almost all sites are static, most configurations have high polymer number $Q\leq \Ns^3$. In the vicinity of $T_c$, the distribution is broad. 
 }
 \label{HistogramPolymer}
\end{figure}

The expectation value of very high moments of baryonic observable such as higher moments of $B^+$, $B^-$ or of the baryon number $B$ given by some function $f$ can be computed from the above histogram, 
\begin{align}
 \expval{f(B^+,B^-)}&=H^\calP_{V,\bareT,\bareMu}(B^++B^-)\,f(B^+,B^{-}),\nn
 \expval{f(B)}&=H^{\calB}_{V,\bareT,\bareMu}(B)\,f(B),
 \label{BaryonFromHistogram}
\end{align}
which improves drastically over the usual measurement of higher moments.
In Fig.~\ref{HistogramPolymer} we show histograms $H^\calQ_{V,\bareT,\bareMu}$ for various temperatures and $\mu_B=0$.
The temperature dependence gives insight into the number of static vs.~dynamic sites: at high temperatures, almost all sites are static, and at low temperatures almost all sites are dynamic, e.g.~they interact via pion exchange with adjacent sites. The critical temperature is characterized by a broad distribution.

Another important application of histogram techniques is the Wang-Landau method, which computes the density of states $g(\bareT,B)$. It will allow us to obtain the canonical phase diagram, see Sec.~\ref{PhaseDiag}. 
We use that the grand-canonical partition sum is related to the canonical partition sum via the Laplace transformation
\begin{align}
\calZ_{GC}(\bareT,\bareMu)=\sum_{B=-V}^V \calZ_{C}(\bareT,B) e^{B \mu_B/T}
\end{align}
One method to determine the canonical partition sum $\calZ_{C}(\bareT,B)$ in the context of QCD is to obtain the $\calZ_{GC}$ for imaginary chemical potential and reweighting for the resulting Fourier coefficient \cite{deForcrand2006}.
In the dual representation, $\calZ_{C}(\bareT,B)$ can be determined directly by the Wang-Landau method, 
since it is in fact the density of states with respect to the canonical conjugate to $\bareMu$ and it is approximated by
$g(\bareT,B)$ up to the target precision. Then, 
observables in the GC-ensemble are immediately obtained:
\begin{align}
\expval{\mathcal{O}}_{GC} &= \frac{\sum_\DB \mathcal{O} \calZ_C(\bareT,B) e^{B\mu_B/T }}{ \sum_\DB \calZ_C(\bareT,\DB) e^{B\mu_B/T}}.
\end{align}
The accuracy even improves when the density of states using the polymer resummation $g(\bareT,P)$ is determined via Wang-Landau, and the canonical partition sum is recovered by the binomial transformation Eq.~(\ref{BinomTrinom}):
\begin{align}
 \calZ_{C}(\bareT,B)&=\sum_{P=0}^{V}\sum_{\substack{B^+=0\\B=2B^+-P}}^V\binom{P}{B}\,g(\bareT,P)
\end{align}

The Wang Landau method applied to $g(\bareT,P)$ consists of the following steps:
\begin{enumerate}
\item[(1)] A CT-worm update is run (which makes $g(\bareT,P)$ temperature dependent). 
\item[(2)] We loop through all spatial sites $\vec{x}$ and check whether the site is static (has no spatial dimers attached).
\begin{enumerate}
\item[(2a)] If so, we propose a new configuration with uniform probability distribution to generate one of the $\Nc+2$ states ($\Nc+1$ mesonic states and one P-state) having equal weight, possibly resulting in a change $P\mapsto P+\Delta P$ with 
\begin{align}
|\meson\rangle &\mapsto  |P\rangle : \Delta P=1,&
|P\rangle &\mapsto  |\meson\rangle : \Delta P=-1,\nn
|P\rangle &\mapsto  |P\rangle : \Delta P=0,&
|\meson\rangle &\mapsto  |\meson\rangle : \Delta P=0.
\end{align}
\item[(2b)] If not, the configuration is unchanged and $\Delta P=0$.
\end{enumerate}
\item[(3)] The new configuration is accepted with a metropolis acceptance step:
\begin{align}
 p_{acc}&=\min(1,(2\cosh(\mu_B/T))^{\Delta P})
\end{align}
\begin{enumerate}
\item[(3a)] If accepted, $P'=P+\Delta P$ is the new polymer number, 
\item[(3b)] If rejected $P'=P$.
\end{enumerate}
\item[(4)]In any case, even if the site was non-static and $P'=P$ (option (2b))
the histogram and density of states are updated:
\begin{align}
H(P')&\mapsto H(P')+1,&  \log(g(P'))&\mapsto \log(g(P'))+\log(f)
\end{align}
with $f$ the modification factor.
\end{enumerate}
We loop through (1-4) until the histogram $H(P)$ is flat 
enough:
\begin{align}
\sum_{P=0}^V |H(P)-\bar{H}|&<\delta&\Rightarrow && f&\mapsto \sqrt{f},& H(P)&=0
\label{FlatnessCondition}
\end{align}
with $\bar{H}$ the histogram average and $\delta$ defining the flatness condition. This step, which refines $g(P)$, is repeated until the final precision is reached, $\log(f)\leq\log(f_{\rm final})$. Then $g(P)$ approximates the true density of states with that precision.
In Sec.~(\ref{PhaseDiag}) we will show the density of states and the canonical phase diagram for various temperatures.

We perform simulations at a set of fixed temperatures and weight the obtained density of states to the critical $a\mu_c$, which is characterized by equal probability
of the low and high density phase. In practice, we determine $a\mu_c$ at which both peaks in the first order region have the same height (see Fig. (\ref{BPN})).

\subsection{Crosschecks}
\label{Crosschecks}

\begin{figure}[h!]
\includegraphics[width=0.49\textwidth]{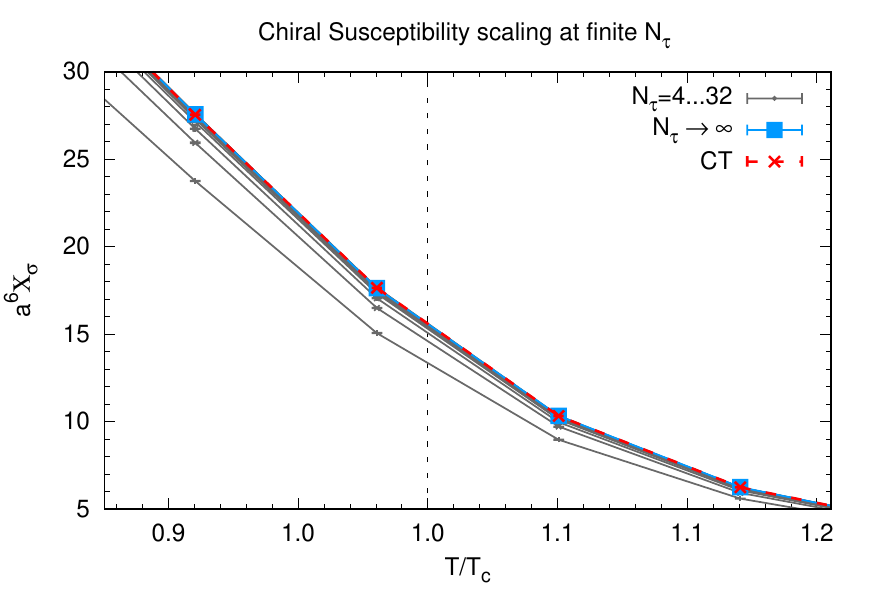}
\caption{Extrapolation of the chiral susceptibility from the finite lattices $\Nt=4,\ldots, 32$ towards $\Nt\rightarrow\infty$ and comparison with the continuous time result, showing excellent agreement.}
\label{CTExtrapSusc}
\end{figure}

To check the correctness of our CT-WA implementation, we have made extensive crosschecks. 
A comparison of the CT-algorithm on volumes with an analytic result extrapolated from $2\times \Nt$ lattice for gauge group U(1) is discussed in the appendix Sec.~\ref{AnalyticU1}.
Since there does not seem a simple analytic expressions for $\Nc>1$, we are left with comparing continuous time simulations with the extrapolation of discrete time simulations. We already discussed the suppression mechanism that lead to the continuous time results for various observables in 
Fig.~\ref{Suppression}. In Fig.~\ref{CTExtrapSusc} we show a comparison of the discrete time extrapolation and the continuous time simulations for the chiral susceptibility as a function of the temperature, which agree within errors for all temperatures. 

\begin{figure}[h!]
\includegraphics[width=0.49\textwidth]{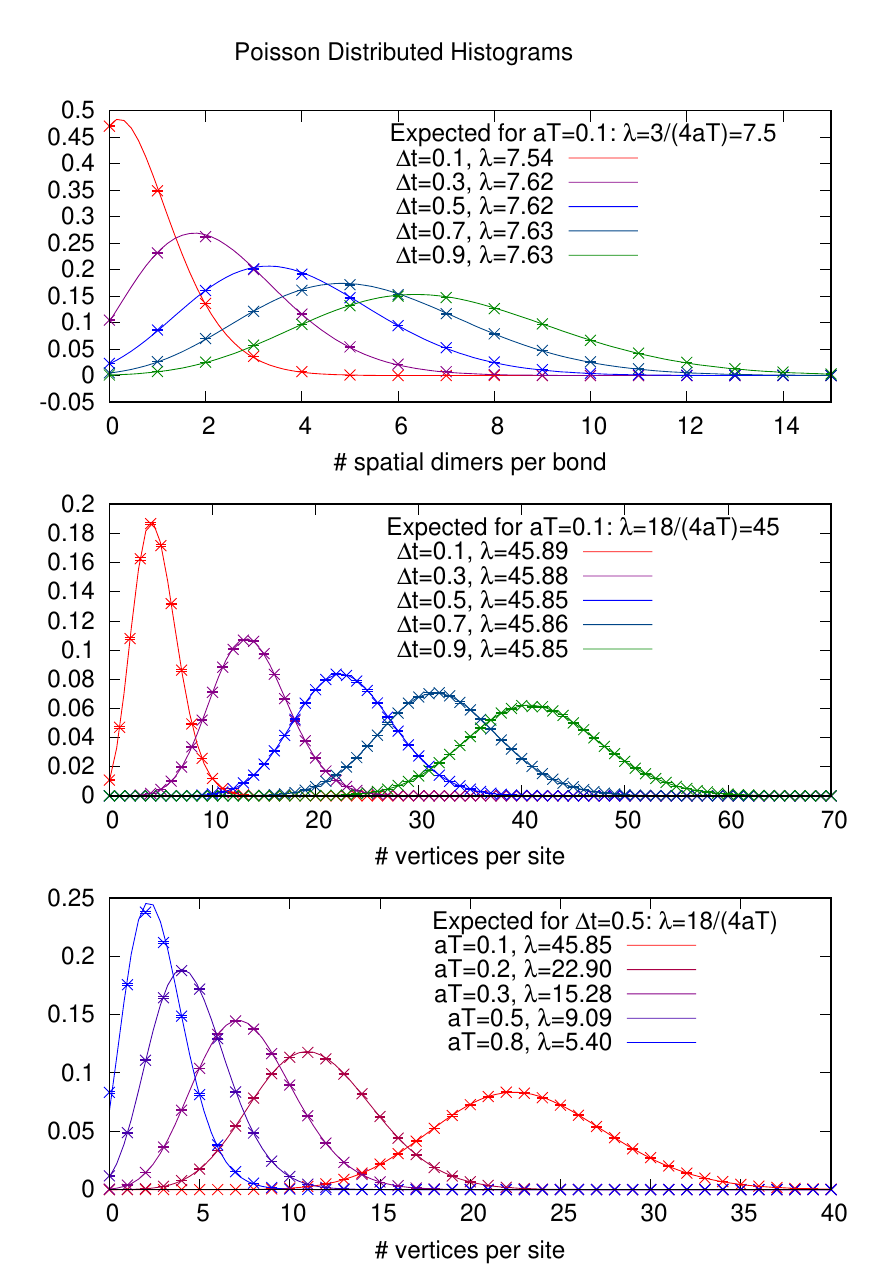}
\caption{
Distributions of the number of spatial dimers per bond (\emph{top}) and number of vertices per site (\emph{center}) for various interval lengths
$\Delta \tau$, and for various temperatures (\emph{bottom}). 
The quantities are Poisson distributed, with $\lambda$ fitted according to Eq.~(\ref{PoissonDist}) to the data,
reproducing the expected value. Small deviations for $\Delta \tau \leq 1$ occur as the Poisson process is on a circle rather a infnite line, and due to the presence of static baryons (which are highly suppressed at low temperatures).
}
\label{Poisson}
\end{figure}

Another aspect is to verify that the distribution of spatial dimers is indeed Poissonian, due to the fact that the weight of a configuration does not depend on the interval lengths between subsequent spatial dimers. This is illustrated in Fig.~\ref{Poisson}. The Poisson distribution 
\begin{align}
 P(N(\Delta \tau)=n)=\frac{(\lambda \tau)^n}{n!}e^{-\lambda \Delta \tau}
 \label{PoissonDist} 
\end{align}
has been fitted to histograms from Monte Carlo via CT-WA. The comparison with the expected values of $\lambda$ 
(with $\lambda=\frac{3}{4T}$ for the distribution of spatial dimers per bond
and $\lambda=\frac{6d}{4T}$ for the distribution of vertices per sites, with $d=3$)  is very good for small intervals $\Delta \tau<1$. The deviations to the expected 
$\lambda$ for large intervals $\Delta \tau\simeq 1$ is due to the periodic boundary conditions, where the Poisson distributions start to overlap.

\section{Zero Temperature}
\label{ZeroT}

\subsection{Determination of $\kappa$ and Pion Decay Constant}
\label{MeasureKappa}

The first task that is also relevant to define the temperature and chemical potential non-perturbatively (Eqs.~(\ref{CTLimit}, \ref{CTLimitMu})) is to determine the anisotropy correction factor $\kappa$, see Eq.~(\ref{DefAnisoCorrFactor}).
The procedure of anisotropy calibration is discussed for anisotropic lattices at strong coupling in discrete time in \cite{Chandrasekharan2003,Chandrasekharan2004,deForcrand2017} in detail.
The coefficient $\kappa$ is the strong coupling analogue of the Karsch coefficients at weak coupling that have been analyzed in \cite{Burgers1987,Karsch1989}
and numerically studied at fixed physical scale in \cite{Levkova2006}. Anisotropic lattices are also relevant when determining mesonic correlators, e.~g.~in the FASTSUM collaboration \cite{Aarts2018}.

\begin{figure}[h!]
\includegraphics[width=0.49\textwidth]{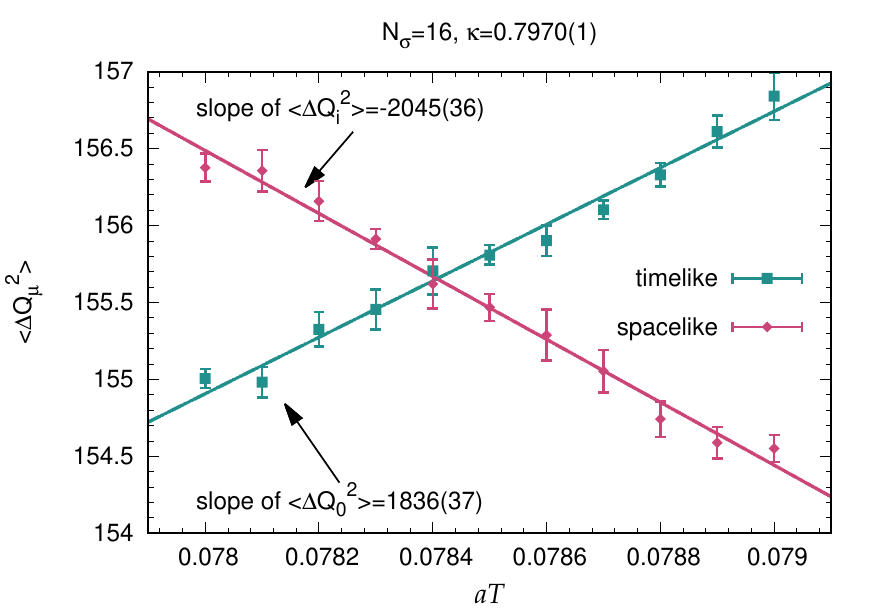}
\caption{Anisotropy calibration in the CT-limit, measured on a lattice 
$16^3\times \CT$, with $\bareT_0=0.07841(1)=\frac{1}{16\kappa}$,
resulting in $\kappa=0.7970(1)$.}
\label{AnisoCalibration}
\end{figure}

\begin{figure}[h!]
\includegraphics[width=0.49\textwidth]{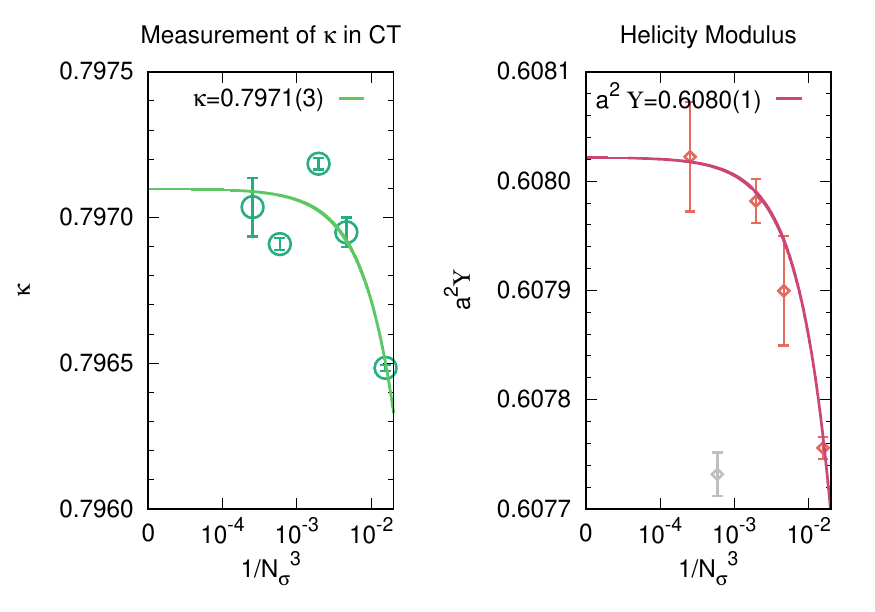}
\caption{
\emph{Left:} Thermodynamic extrapolation of anisotropy correction factor $\kappa$ needed to rescale the temperature and chemical potential. The result is compared to former result from the extrapolation of discrete time lattices (Ansatz 3).
\emph{Right:} Thermodynamic extrapolation of the helicity modulus $a^2\Upsilon$, from which we extract pion decay constant at zero temperature.
}
\label{KappaDetermination} 
\end{figure}

Our strategy to obtain $\kappa$ is based on the variance of the pion current.
In the chiral limit, the pion current for discrete time  
\begin{align}
j_\mu(x)&=\epsilon(x)\lr{k_\mu(x)-\frac{\Nc}{2}|b_\mu(x)|-\frac{\Nc}{2d}}
\end{align}
is a conserved current:  
\begin{align}
\sum_{\hmu}\lr{j_\mu(x)-j_\mu(x-\hmu)}=0
\end{align}
Likewise, also 
the corresponding currents in the CT-limit  
\begin{align}
\meson_0(\vec{x},\tau)\equiv j_0(\vec{x},\tau)&=\meson(\vec{x},\tau)-\frac{\Nc}{2}\\
\meson_i(\vec{x},\tau)\equiv j_i(\vec{x},\tau)&=\epsilon(\vec{x},\tau) k_i(\vec{x},\tau),
\end{align}
see Eq.~(\ref{MesonState})
- where we have dropped the baryonic contributions and the constant, as they do not contribute at continuous time - are conserved and now directly linked to the meson occupation numbers:
\begin{align}
\meson(\vec{x},\tau_1) + \int_{\tau_1}^{\tau_2}d\tau \sum_{i=1}^3\lr{\meson_i(\vec{x},\tau)-\meson_i(\vec{x}-\hat{i},\tau)}=\meson(\vec{x},\tau_2)
\end{align}
for all $\tau_2>\tau_1$, and the temporal/spatial charges are
\begin{align}
Q_0&=\sum_{\vec{x}} \meson_0(\vec{x},\tau)\equiv\mathfrak{M}_0,\nn
Q_i&=\sum_{{\vec{x}\perp \vec{e}_i}} \int_0^{1/\bareT} d\tau\, \meson_i(\vec{x},\tau)\equiv\mathfrak{M}_i.
\end{align}
which have the expectation values
\begin{align}
\expval{\mathfrak{M}_0}&=\expval{\mathfrak{M}}-\frac{\LatSpat\Nc}{2}=0, & \expval{\mathfrak{M}_i}&=0.
\end{align}
The variances are however temperature dependent.
If the spatial and temporal variances are equal,
\begin{align}
 \expval{(\Delta Q_0)^2}= \expval{\mathfrak{M}_0^2}\stackrel{!}{=}\expval{\mathfrak{M}_i^2}=\expval{(\Delta Q_i)^2},
\end{align}
that corresponds to equal physical extent in space and time:
\begin{align}
L&=\frac{1}{T}&\Rightarrow && \Ns&=\frac{1}{aT}=\frac{1}{\kappa \bareT}. 
\end{align}
This allows us to measure $\kappa$: given the lattice extent $\Ns$, we scan the bare temperature $\bareT$ to determine its value $\bareT_0$ that corresponds to a physically isotropic lattice:
\begin{align}
\kappa_{\Ns}&=\frac{1}{\Ns \bareT_0},& \kappa&=\lim_{\Ns\rightarrow \infty} \kappa_{\Ns} 
\end{align}
This calibration is shown in Fig.~\ref{AnisoCalibration}, the results for $\kappa$ for various volumes are shown in Table \ref{AnisotropyTable} and its extrapolation in Fig.~\ref{KappaDetermination} (\emph{left}). The finite size effects are very small.
Note that in contrast to the previous study \cite{deForcrand2017}, there is no reason to distinguish $\kappa$ for gauge group $\U(3)$ and $\SU(3)$: the thermodynamic extrapolation $\Ns\rightarrow \infty$ coincides with the zero temperature extrapolation, and since the calibration is performed at $\bareMu=0$, static baryons are virtually absent (see also Fig.~\ref{HistogramPolymer}). This is not the case at finite $\xi$ (finite $\at$). 
As discussed in Sec. \ref{FiniteTCTLimit}, the determination of $\kappa$ in \cite{deForcrand2017} suffers from systematic uncertainties as the extrapolation in $\xi$ is based on rather small $\xi\leq 8$. Our final continuous time result $\kappa=0.797(1)$ is consistent with the extrapolations, favoring Ansatz 3.

\begin{table}[h!]
\begin{tabular}{rrr}
\hline
\,$V$ & $\kappa$\quad\phantom{.} & $a^2\Upsilon$\quad\phantom{.}\\
\hline
4 	& 	0.7965(1)	&	0.6078(1)\\ 
6 	& 	0.7970(1)	&	0.6079(1)\\
8 	& 	0.7972(1)	&	0.6080(1)\\   
12 & 0.7969(1)	&	0.6077(1)\\
16 & 0.7970(1)	&	0.6080(1)\\ 
\hline
$\infty$ & 0.7971(3) & 0.6080(1)\\
\hline 
\end{tabular}
\caption{The values of $\kappa$ and the helicity modulus for various $\Ns$ and the extrapolation to the thermodynamic limit, as shown in Fig.~\ref{KappaDetermination}.}
\label{AnisotropyTable}
\end{table}
In Fig.~\ref{KappaDetermination} (\emph{right}) we show the thermodynamic extrapolation of the helicity modulus, which yields the square of the pion decay constant:
\begin{align}
a^2 F_\pi^2&=\lim_{\Ns \rightarrow \infty} a^2 \Upsilon, &  a^2 \Upsilon&=\frac{1}{\Ns^2}\left.\expval{\mathfrak{M}_0^2}\right|_{\bareT_0},
\end{align}
resulting in $aF_\pi=0.7797(1)$. This compares well with the extrapolation of discrete time \cite{deForcrand2017} which yields \linebreak $aF_\pi=0.78171(4)$, taking into account that the extrapolation of $a^2 F_\pi^2$ has similar uncertainties as $\kappa$, which are overcome by the continuous time simulations.

The method of anisotropy calibration has also been extended by us to finite quark mass \cite{Bollweg2018} and recently also to finite $\beta$.
These results are a clear indication that it is possible to define the continuous time limit unambiguously for finite $m_q$ and finite $\beta$ in the strong coupling regime, with $\kappa=\kappa(m_q,\beta)$. 

\subsection{Chiral Condensate and Chiral Susceptibility}
\label{ChiralCondensateFiniteSize}

\begin{figure}[h!]
\includegraphics[width=0.49\textwidth]{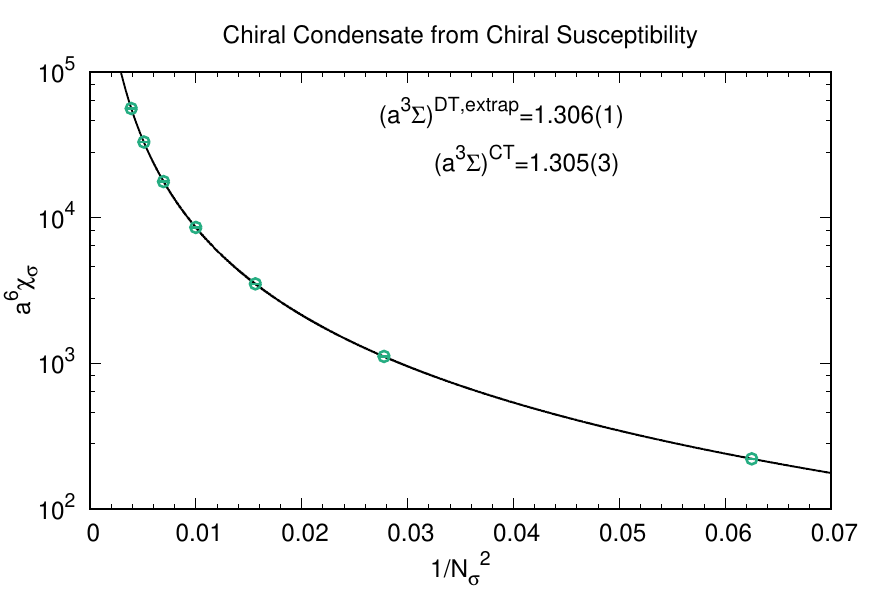}
\caption{The infinite volume chiral condensate obtained in the chiral limit via chiral perturbation theory in a finite box, corresponding to the $\epsilon$-regime. In the range $1/\Ns^2$ considered here, the fit of 
$a^3\Sigma$ is dominated by the leading order and next to leading order term $\mathcal{O}\lr{(L F_\pi)^{-2}}$.
}
\label{ChiPTCondensate}
\end{figure}

Despite the fact that in the chiral limit, the chiral condensate is zero in a finite volume - in the dual representation this is due to the absence of monomers - it is nevertheless possible to extract the chiral condensate from the chiral susceptibility $\chi_\sigma$ (which is non-zero in a finite volume). The corresponding chiral perturbation theory in a finite box - the so-called $\epsilon$-regime - is an expansion in the inverse volume \cite{Hasenfratz:1989pk}, and for the O(2) model in $d=4$:
\begin{align}
 a^6 \chi_\sigma &\simeq \frac{1}{2}a^6 \Sigma^2 \Ns^4\lr{ 1+ \frac{\beta_1}{a^2 F_\pi^2 \Ns^2}+\frac{\alpha}{2a^2 F_\pi^4 \Ns^4}}
 \label{AnsatzCondensate}
 \\
\alpha&=\beta_1^2+\beta_2+\frac{1}{8\pi^2}\log 
\frac{a\Lambda_
\Sigma^2\Ns}{\Lambda_M},
 \end{align}
where $\beta_1=0.140461$ and $\beta_2=-0.020305$ are shape coefficients of a finite 4-dim.~box. Note that the value $\Sigma$ that can be extracted from this equation corresponds to the chiral condensate in the thermodynamic limit. In Fig.~\ref{ChiPTCondensate} we show the fit according to Ansatz Eq.~(\ref{AnsatzCondensate}) to obtain the chiral condensate from the  Monte Carlo data of the chiral susceptibility for various volumes, all in the CT limit. 
Apart from $\Sigma$, we also treat $\alpha$ as a fit parameter as we do not know the values of the renormalization group invariant scales $\Lambda_\Sigma$ and $\Lambda_M$, but it turns out that $\alpha$ is consistent with zero. 
The value of $aF_\pi$ determined in the previous section is used. 
The thermodynamic extrapolation $\Ns\rightarrow \infty$ coincides with the zero temperature extrapolation as the bare temperature is set to $\bareT=
(\kappa \Ns)^{-1}$ to always obtain a physically isotropic lattice.
Our result from continuous time simulations 
yields $a^3\Sigma=1.305(3)$ and agrees well with the extrapolation of the Monte Carlo data at discrete time as discussed in \cite{deForcrand2017}. 

\subsection{Energy and Baryon Mass}
\label{baryonMassNucInt}

\begin{figure}[h!]
\includegraphics[width=0.49\textwidth]{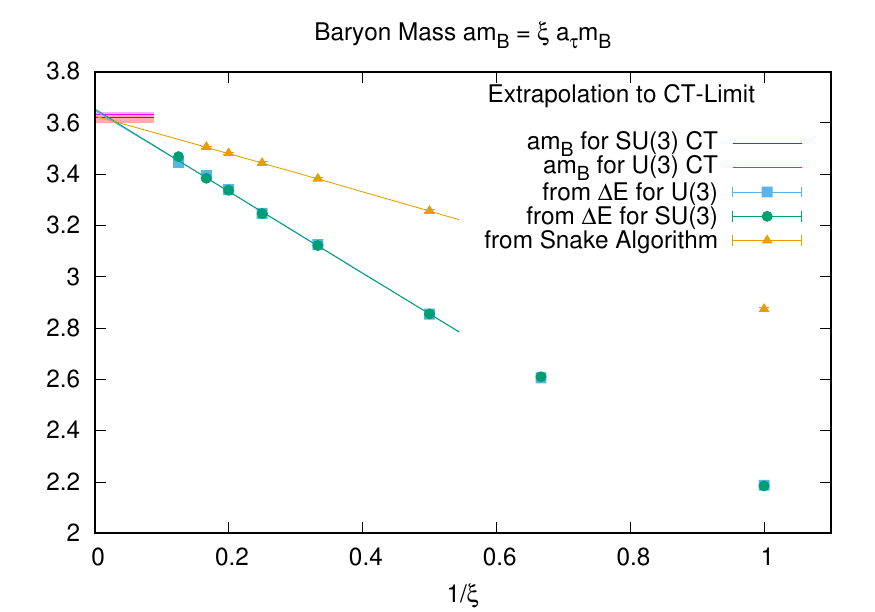}
\caption{The baryon mass as obtained from the valence baryon mass, and from the free energy in the zero temperature limit.}
\label{baryonmass}
\end{figure}

The baryon mass $m_B$ is an important quantity to understand the nature of nuclear interaction, and its value in lattice units $am_B$ is also a good choice to scale other quantities to dimensionless ratios, such as $T/m_B$, $\mu_B/m_B$. 
At zero temperature, where the free energy $F=E-TS$ coincides with the internal energy $E$, the static baryon mass in the strong coupling limit is given by the probability of a baryon to propagate in temporal direction. This can be immediately expressed by the probability of having a static baryon in the ensemble: 
\begin{align}
p_B&=e^{-\Delta F/T}, & \Delta F&=-T\log\frac{Z_B}{Z},& m_B&=\lim_{T\rightarrow 0} \Delta F
\end{align}
The extrapolation of the static baryon mass towards continuous time has been discussed in \cite{deForcrand2017} with the result 
$am_B=\xi a_\tau m_B =3.556(6)=\kappa\, am^{\rm MF}_{B}$, $am^{\rm MF}_{B}=4.553(7)$, which is about 20\% larger than the 
isotropic value \linebreak $am_B=2.877(2)$. Since $p_B\ll 1$, the mass is 
evaluated via the so-called snake algorithm at discrete time: 
\begin{align}
\frac{Z_B}{Z}&=\frac{Z_{\Nt}}{Z_{\Nt-2}}\frac{Z_{\Nt-2}}{Z_{\Nt-4}}\ldots \frac{Z_{2}}{Z_{0}},\; Z_{0}\equiv Z,\; Z_{\Nt}\equiv Z_B\nn
 a\Delta F&=\frac{\xi}{\Nt}\log \frac{Z_{B}}{Z}=\frac{\xi}{\Nt}\sum_{k=0}^{\Nt-2}\log \frac{Z_{k+2}}{Z_k}. 
\end{align}
The ratio $\frac{Z_{k+2}}{Z_k}$ is the probability to extend a static baryon segment of length $k$ by two segments, and the sum results in a static baryon of length $\Nt$. 
The method unfortunately does not extent straight forwardly to continuous time: the ratios $\frac{Z_{k+2}}{Z_k}$ cannot be measured, since at the end of a static baryon segment there is a finite probability that two spatial dimers are attached at the same location, in contrast to other observables discussed above (Fig.~\ref{Suppression}). 
However, we are able to determine the baryon mass from the energy difference based on Eq.~(\ref{energydensity}):
\begin{align}
a\Delta E&=aE_B-aE_0=\LatSpat a^4 (\epsilon_B-\epsilon_0)\nn
&= \LatSpat \kappa\, \bareT \lr{\expval{n_{Ds}}_0-\expval{n_{Ds}}_B}.
\end{align}
The energy density at zero temperature in the CT-limit, if one does not take the irrelevant constant $C$ in Eq.~(\ref{energydensity}) into account (rendering it negative), can be measured at very high accuracy:
\begin{align}
a^4\epsilon^{\U(3)}_0=-1.82471(2),
\quad
a^4\epsilon^{\SU(3)}_0=-1.82475(8),
\label{ZeroEnergyDensity}
\end{align}
where the value for gauge group $\U(3)$  (which does not have baryons) coincides with the value for gauge group $\SU(3)$ (where baryons become suppressed with decreasing temperature). 
The fact that \linebreak $a^4\epsilon_0=-\lim_{T\rightarrow 0} aT \expval{n_{Ds}}$ is finite implies that the number of spatial dimers diverges as 
$\propto 1/T$.
Note that a previous determination of $\epsilon_0$ at discrete time \cite{deForcrand2016} includes the diverging constant:  $a^4\epsilon_0= 0.66(2) \xi$.
We measured the energy density without ($\epsilon_0$) and with a static baryon ($\epsilon_B$), both on discrete and continuous time lattices.
The discrete time measurements of $a\Delta E$ are extrapolated via a polynomial Ansatz in $1/\xi$, as shown in Fig.~\ref{baryonmass}. The fit results are summarized in Tab.~\ref{Table_BaryonMass}, and are compared with the continuous time results. Indeed, we find very good agreement of all extrapolated estimates of the baryon mass with its continuous time result within errors. It should be pointed out that at $\gamma=1$, where $\expval{k_0}=\frac{\Nc}{2d}=\frac{3}{8}$, the static baryon mass from $\Delta F$ (via the snake algorithm) differs substantially from the baryon mass obtained from $\Delta E$. But towards the CT limit, both definitions agree. 
The extrapolation of the discrete time data (obtained from $\Delta E$ or $\Delta F$) is in $1/\xi$ rather than $1/\xi^2$: it is more suitable as the extrapolation appears to be almost linear in $1/\xi$, but clearly there are additional uncertainties related to the derivative $d\xi / d\gamma$ that are bypassed by simulations directly in the CT limit.

\begin{table}[h!]
 \begin{tabular}{lll}
 \hline
\vspace{-3mm}\\
 Method & $am_B^{\rm extrap}$ & $am_B^{{\rm CT}}$ \\
 \hline
 $\Delta E$ for $\U(3)$  & 3.644(20)  &	3.640(7)\\ 
 $\Delta E$ for $\SU(3)$ & 3.649(20)  &	3.628(22)\\
 $\Delta F$ with snake alg.  & 3.627(6) &	- \\
 \hline
 \end{tabular}
 \caption{The baryon mass from extrapolation or direct measurement, as shown in Fig.~\ref{baryonmass}. $\Delta E$ has been evaluated at various temperatures and extrapolated to zero temperature.
 The result for the snake algorithm valid for SU(3) differs slightly from the value $am_B=3.556(6)$ given in \cite{deForcrand2017} due to the improved extrapolation used here.
 }
 \label{Table_BaryonMass}
 \end{table}
We distinguish between $\U(3)$ and $\SU(3)$ results for the baryon mass: in $\U(3)$ gauge theory, there is only the valence baryon and no 
$\mu_B$-dependence of the partition function, whereas $\SU(3)$ gauge theory has intrinsic baryon fluctuations. 
At zero temperature, those baryon fluctuations are largely suppressed. Even though $\U(3)$ gauge theory has no baryons, there is no obstacle in measuring the baryon mass in $\U(3)$ via the response of a valence baryon to the pion bath, resulting in less statistical noise. Our best estimate of the baryon mass is thus the $\U(3)$ result in the CT-limit, as it does not suffer from any ambiguities due to extrapolation:
\begin{align}
am_B=3.640(7).
\end{align}
 This baryon mass receives contributions from a pion cloud surrounding the static point-like baryon. 
%

\section{The SC-LQCD Phase Diagram}
\label{PhaseDiag}

\subsection{Chiral Transition}
\label{ChiralTransition}
\begin{figure}[h!]
\includegraphics[width=0.49\textwidth]{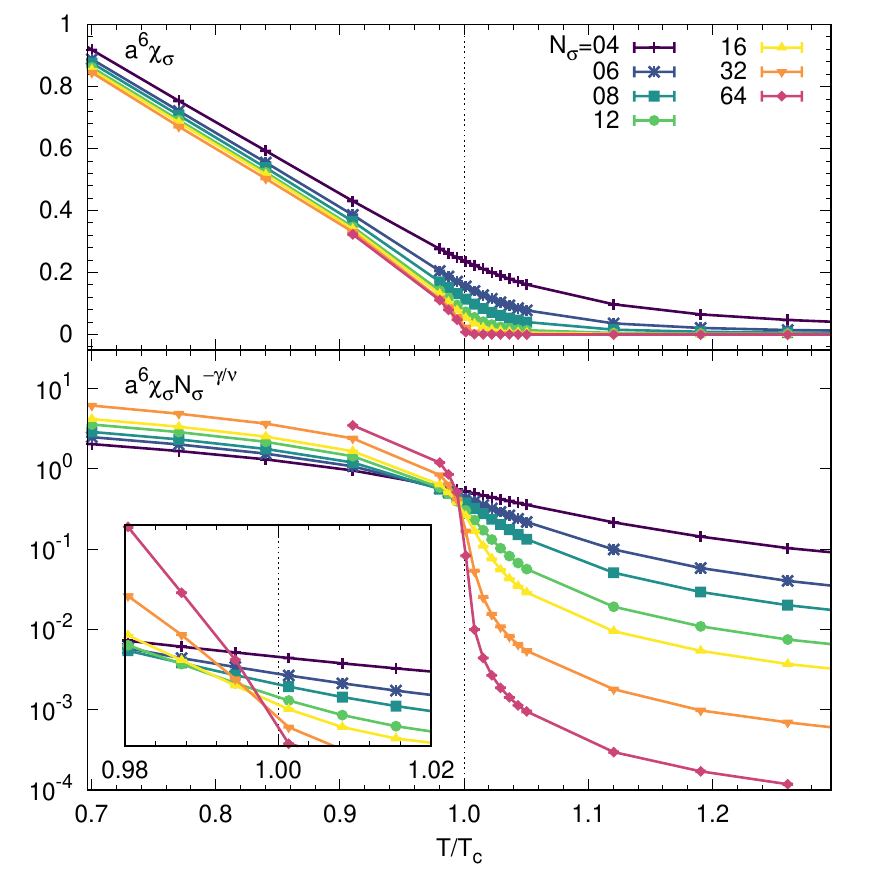} 
\caption{Finite size scaling for the chiral susceptibility to determine the chiral transition temperature in the chiral limit. 
Intersections signal the chiral transition temperature $\bareT_c$. 
Note the chiral susceptibility is rescaled using O(2) critical exponents, and it does not develop a peak in the chiral limit.}
\label{ChiralSuscCT}
\end{figure}

\begin{figure}[h!]
\includegraphics[width=0.49\textwidth]{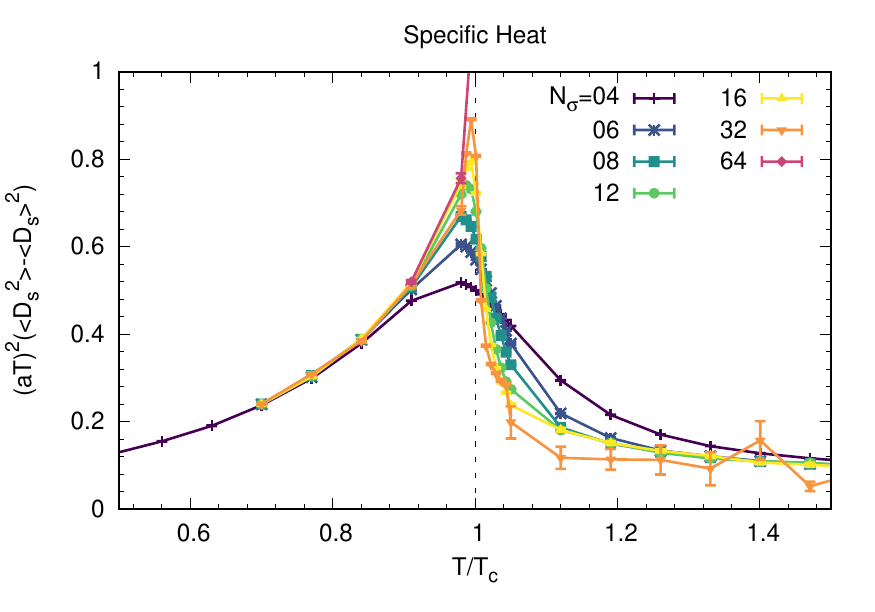}
\caption{
The specific heat, which is proportional to the susceptibility of spatial dimers (see Eq. \ref{energydensity}).
The typical $\lambda$-shape is apparent in the transition region
}
\label{SpecificHeat}
\end{figure}

In Sec.~\ref{ChiralCondensateFiniteSize} we have determined the chiral condensate in the chiral limit at zero temperature. In principle this can be extended to finite temperature, and the chiral transition could be determined by the vanishing of the chiral condensate.
It suffices in practice to determine the chiral transition from the chiral susceptibility, which is obtained from the worm algorithm to high precision.
Also, this method readily extends to finite density: the chiral transition can be easily obtained from finite size scaling of the chiral susceptibility up to the chiral tricritical point $(a\mu_B^{\rm TCP},aT^{\rm TCP})$.
The finite size scaling of the susceptibility in the $\epsilon$-regime is illustrated in
Fig.~\ref{ChiralSuscCT} for volumes up to $64^3\times\CT$ at $\mu_B=0$. 
We expect critical behavior in the ${\rm O}(2)$ universality class in 3 dimensions, resulting the scaling law \cite{PhysRevB.63.214503}
\begin{align}
\lim_{L\rightarrow \infty} \chi(L,T_c) &\propto L^{\gamma/\nu},&
\gamma&=1.3177(5),\nn
&&\nu&=0.67155(27).
\label{FSS}
\end{align}
The result for the transition temperature is 
\begin{align}
\bareT_c&=1.4276(2),& aT_c=\kappa\,\bareT_c&=1.1379(4). 
\end{align}
We find that also the specific heat is sensitive to the chiral transition: Fig.~\ref{SpecificHeat} shows that a weak cusp develops in the vicinity of $T_c$.
Although the strong coupling limit is far away from the continuum for realistic quarks, we can nevertheless compare dimensionless ratios $T/m_B$ with continuum extrapolated ratios. With $m_B\simeq 938 {\rm GeV}$ and the pseudo-critical crossover temperature $T_{pc}\simeq 154 {\rm MeV}$ \cite{Bazavov:2011nk} we find that the ratio at strong coupling and in the chiral limit is more than twice as large:
\begin{align}
 \left.\frac{T_c}{m_B}\right|_{CT-SC}&=0.379(1),&  \left.\frac{T_{pc}}{m_B}\right|_{\rm cont.}&=0.164(9).
\end{align}
The comparison improves when a finite quark mass is considered at strong coupling, as the pseudo-critical transition temperature drops rapidly with the mass while the baryon mass is quite insensitive \cite{Kim2016}. 
We note that the continuous time transition temperature for $\U(3)$ gauge group and its comparison with the $\Nt\rightarrow \infty$ extrapolation have been discussed in \cite{Unger2011}, with $\bareT_c^{\U(3)}=1.8843(1)$.\\

\begin{figure}[h!]
\includegraphics[width=0.49\textwidth]{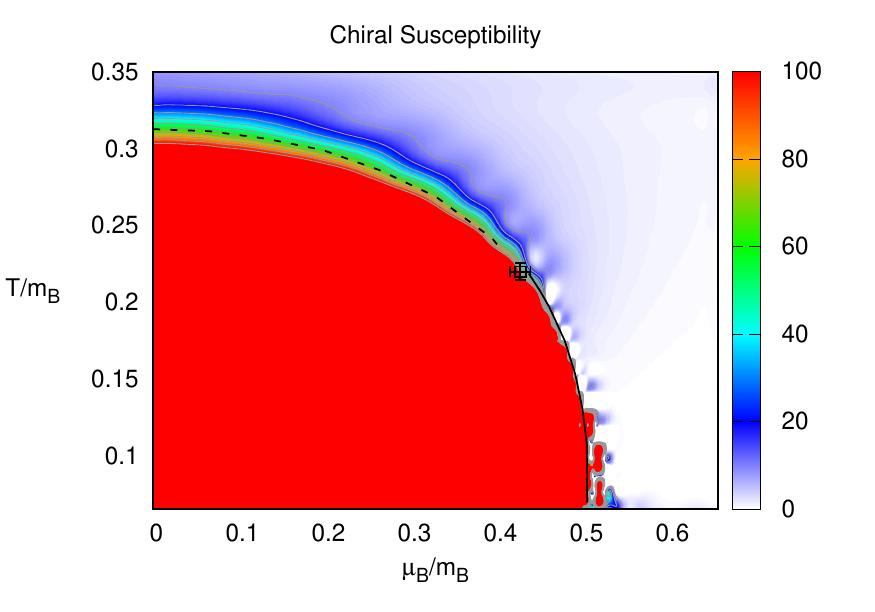}
\caption{
The chiral susceptibility in the full $\mu_B - T$ plane on a $8^3\times CT$ lattice, also indicating the first order (solid) and second order (dashed) lines and the location of the tricritical point, obtained from finite size scaling according to Eq.~\ref{FSS}.
The chiral susceptibility diverges in the chirally broken phase, but is much smaller in the chirally restored phase.
Along the first order transition which is strong already for $aT<0.7$ and hinders reliable results below $\bareT<0.3$, artificial wiggles appear due to hysteresis of the overlapping low and high density branches. 
}
\label{PhaseDiagChiralSusc}
\end{figure}

\begin{figure}[h!]
\includegraphics[width=0.49\textwidth]{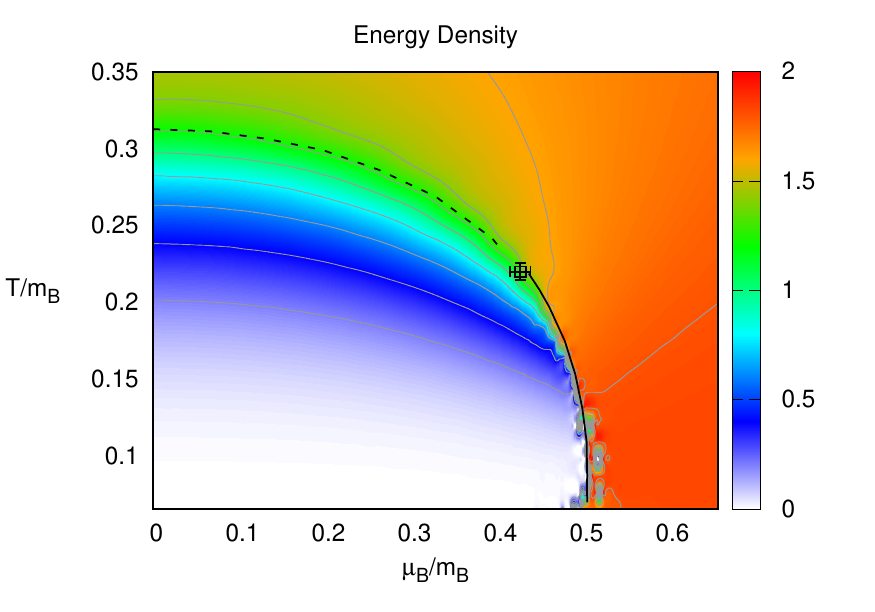}
\caption{
The energy density in the full $\mu_B - T$  plane on a $8^3\times \CT$ lattice. It is sensitive to the chiral transition. 
It also shows a strong first order behavior and at low temperatures becomes insensitive to the chemical potential below $\mu_B^{1st}$.
}
\label{PhaseDiagEnergy}
\end{figure}

The determination of $aT_c$ at finite chemical potential is straight forward up to the tricritical point. 
Fig.~\ref{PhaseDiagChiralSusc} illustrates the chiral susceptibility $\chi_\sigma$ in the full $\mu_B - T$ plane.
The second order chiral phase transition turns into first order for $\mu_B>\mu_B^{\rm tric}$, and the chiral susceptibility - which is 
$\propto \expval{(\bar{\psi}\psi)^2}$ in the chiral limit - behaves as an order parameter and develops a gap.
There is no back-bending of the first order transition, in contrast to discrete time (due to saturation of spatial dimers, $N_{Ds}\leq \Nc\Omega/2$), which has been discussed in \cite{deForcrand2017}.
Similarly, the energy density $\epsilon(T)-\epsilon_0$ can be measured in the full $\mu_B - T$ plane, as shown in Fig.~\ref{PhaseDiagEnergy}. For small chemical potential and for temperatures below $T_c$ it behaves according to the Stefan-Boltzmann law \cite{PhysRevLett.46.1497}:
\begin{align}
\epsilon(T)-\epsilon_0&=\sigma T^4,& \sigma&=\frac{\pi^2}{30},
\end{align}
which corresponds to an ideal pion gas and has already been discussed at zero chemical potential for discrete time \cite{deForcrand2016}.
At zero temperature, the energy density jumps at the first order transition to the finite value $-\epsilon_0$ given in Eq.~(\ref{ZeroEnergyDensity}), which is the maximal value corresponding to the absence of spatial dimers.


\subsection{The Nuclear Transition}

\begin{figure}[h!]
\includegraphics[width=0.49\textwidth]{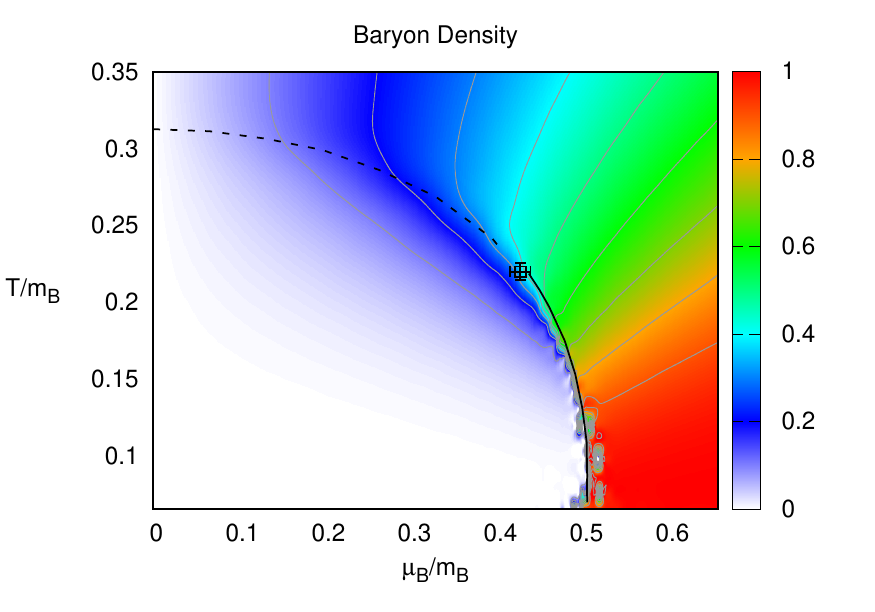}
\caption{
The baryon density in the full $\mu_B - T$  plane on a $8^3\times \CT$ lattice. 
It also shows a strong first order behavior and at low temperatures becomes insensitive to the chemical potential below $\mu_B^{1st}$, which is known as Silver-Blaze property. The first order line terminates in a critical end-point which coincides with the chiral tricritical point.
The baryon density is not sensitive to the second order chiral transition.
}
\label{PhaseDiagBaryonDensity}
\end{figure}

\begin{figure}[h!]
\includegraphics[width=0.49\textwidth]{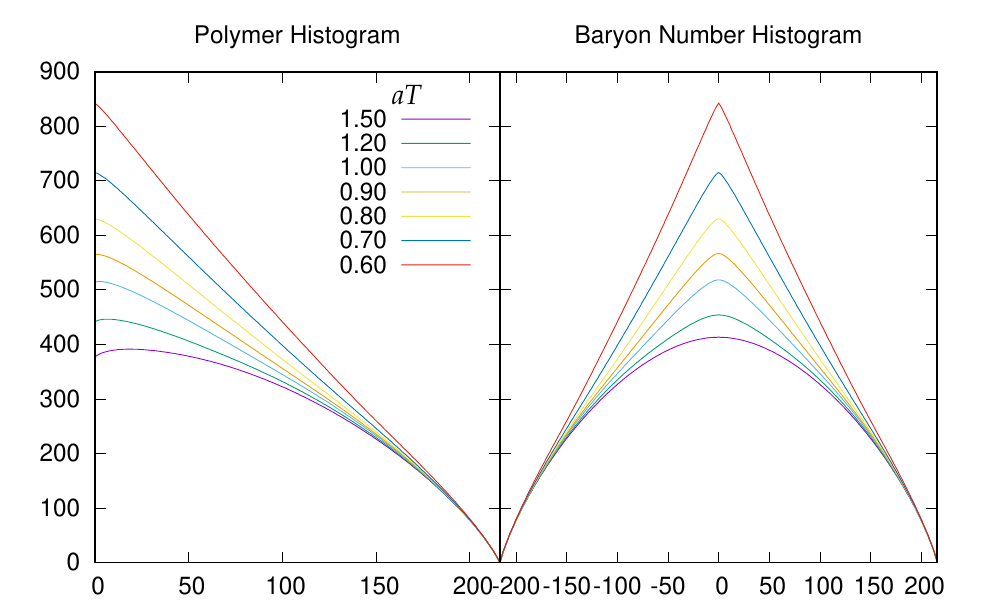}
\caption{
The logarithmic density of states for $\calP$-polymers $\ln g(P)$ (\emph{left}) and baryon number
$\ln g(B)$ (\emph{right}) on a $6^3\times \CT$ lattice, obtained via the Wang-Landau method for various temperatures $\bareT$
For all temperatures, the accuracy was set to $f_{\rm final}=10^{-8}$ and the flatness condition is $\delta=0.1$, see Eq.~(\ref{FlatnessCondition}).
}
\label{DensityOfStates}
\end{figure}

\begin{figure}[h!]
\includegraphics[width=0.49\textwidth]{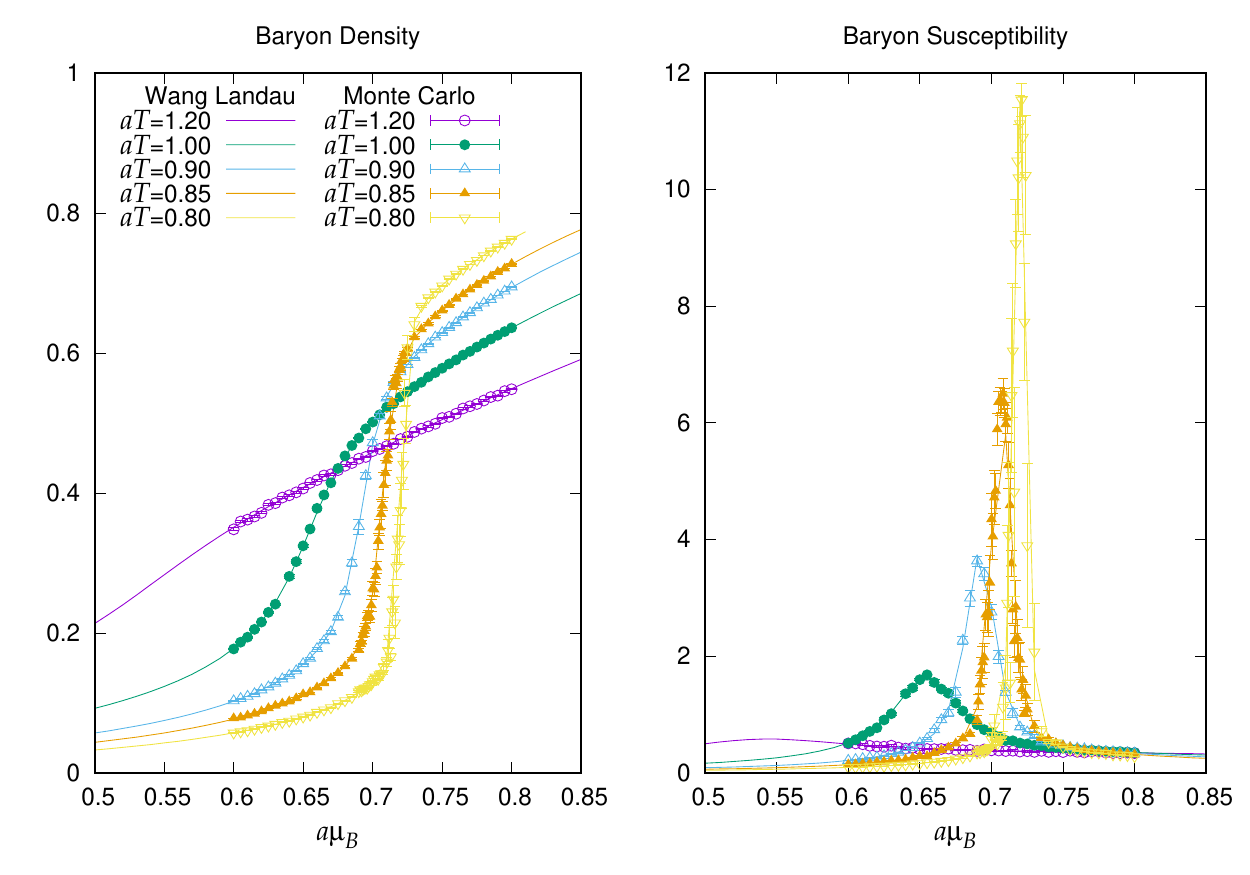} 
\caption{Comparison of the Wang Landau method with Monte Carlo data from CT-WA 
for baryonic observables. \emph{Left:} the baryon density $a^3\expval{n_B}$. \emph{Right:} the baryon susceptibilty $a^6\LatSpat(\expval{n_B^2}-\expval{n_B}^2)$. All data are shown as a function of $\bareMu$ for various temperatures $\bareT$ and on a $6^3\times \CT$ lattice. 
The Monte Carlo data are in perfect agreement with the more precise data form the density of states. The error bands are obtained by 10 independent Wang Landau simulations and are too small to be visible.
}     
\label{baryonDensityWL}
\end{figure}

\begin{figure}
\includegraphics[width=0.49\textwidth]{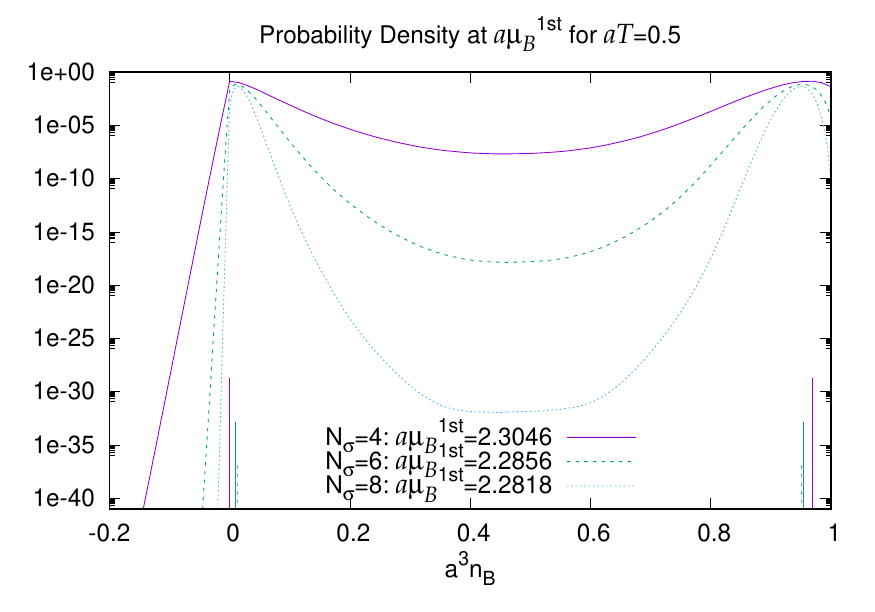}
\caption{The probability density, obtained from reweighting the density of states to $a\mu_B^{1st}(\Ns)$ such that the two maxima are of the same height, for various volumes and at a fixed temperature $\bareT=0.5$. 
The first maximum denotes the baryon density $a^3n_{B}^{(1)}$ where the mixed phase in the canonical phase diagram begins, 
The second maximum denotes the baryon density $a^3n_{B}^{(2)}$ where the mixed phase ends. The peak densities are indicated as vertical lines, the thermodynamic extrapolation resulting in the values given in Tab.~\ref{TableExtrapWL}.
}
\label{BPN}
\end{figure}

Strong coupling lattice QCD exhibits not only spontaneous chiral symmetry breaking and its restoration along a second and first order boundary, but also  a nuclear liquid gas transition signaled by the baryon density. In order to determine the first order transition line in the phase diagram, we measure the baryon density and its susceptibility, both by direct simulations at finite chemical potential, and by the Wang-Landau method explained in Sec.~\ref{WangLandau}.
The  baryon density in the $\mu_B - T$ plane is shown in Fig.~\ref{PhaseDiagBaryonDensity}.
The volumes considered are $4^3\times \CT$, $6^3\times\CT$ and $8^3\times\CT$ at low temperatures and additionally $12^3\times\CT$, $16^3\times\CT$ 
in the vicinity of the chiral tricritical point.
Simulations at low temperatures across the first-order transition are challenging: for $\mu_B<\mu_B^{1st}$, the phase is described as an ideal pion gas, for $\mu_B>\mu_B^{1st}$ the phase is that of a baryon crystal (liquid), resulting in a large latent heat. 
In a Monte Carlo simulation, tunneling between the phases is exponentially suppressed by the volume and hysteresis between the low and high density phase shows up.
This difficulty is overcome by the Wang-Landau method: in Fig.~\ref{DensityOfStates} we show the logarithmic density of states for $\calP$-polymer and baryon number, and in Fig.~\ref{baryonDensityWL} the density of states are applied to recover the baryon density via Eq.~(\ref{BaryonFromHistogram}).
We find that the full first order nuclear transition coincides with the chiral first order transition. 
The determination of $\mu_B^{1st}$ and the boundaries of the mixed phase is illustrated in Fig.~\ref{BPN} for various volumes. The result of the thermodynamic extrapolation according to 
\begin{align}
\bareMu^{1st}(\Ns) &= \bareMu^{1st}+ c \Ns^{-3},\nn
 a^3n_B^{(i)}(\Ns) &= a^3n_{B,c}^{(i)} + \tilde{c} \Ns^{-1} & (i=1,2),
\label{extrapolationWL}
 \end{align}
based on the volumina with $\Ns=4,6,8$ (which is sufficient due to the strong first order behavior) is given in Tab.~\ref{TableExtrapWL}.
Even though we cannot get lower than $\bareT=0.3$ ($T/m_B=0.066$), we can attempt a zero temperature extrapolation which yields
\begin{align}
a\mu_B^{1st}&=\kappa\,\bareMu^{1st}=1.86(2),& \bareMu^{1st}&=2.34(3),
\end{align}
which is not very different from the discrete time determination $a\mu_B^{1st}= 1.78(1)$ valid for isotropic lattices, $\gamma=1$  \cite{Forcrand2010}.
Nuclear matter at strong coupling is in fact a quark saturated phase: the baryon density at zero temperature jumps from $\expval{n_B}=0$ to 
the maximal value $\expval{n_B}=1$, where every lattice site is occupied by a static baryon. 
It is no coincidence that chiral symmetry is restored in the nuclear phase: mesons cannot occupy baryonic sites, leaving no room for spontaneous chiral symmetry breaking. Away from the strong coupling limit, where baryons are no longer pointlike and become spread over several lattice spacings, the nuclear phase may have a non-vanishing chiral condensate. 

We want to conclude this section by quantifying the interaction strength between baryons. In the CT-limit we find
\begin{align}
 \frac{m_B-\mu_B^{1st}}{m_B}\simeq 0.489(6),
\end{align}
which should be compared to the discrete time ($\gamma=1$) ratio \cite{Forcrand2010}
\begin{align}
 \frac{m_B-\mu_B^{1st}}{m_B}\simeq 0.381(3).
\end{align}
Hence the nuclear interactions are enhanced in the CT-limit.

\begin{table}
\begin{tabular}{rrrr}
\hline
$\bareT$ & $\bareMu^{1st}$ & $a^3 n_B^{(1)}$ &  $a^3 n_B^{(2)}$ \\
\hline
0.4  & 2.301(7) & 0.0037(9) & 0.967(2) \\
0.5  & 2.2784(4)  & 0.0275(3) & 0.931(1) \\
0.6  & 2.2538(1) & 0.0059(1) & 0.8632(6) \\
0.7  & 2.2102(1) & 0.0979(1) & 0.741(3) \\
0.75 & 2.1800(3) & 0.149(1) & 0.675(1) \\
0.8  & 2.1444(2) & 0.192(1) & 0.6062(7) \\
0.85 & 2.1037(4) & 0.2685(9) & 0.535(1) \\
0.9  & 2.0587(3) & 0.3535(1) & 0.4796(3) \\
0.92 & 2.0395(2) & 0.399(1) & 0.455(1) \\
0.95 & 2.009(2) & 0.415(1) & 0.454(4) \\
\hline
\end{tabular}
\caption{Result of the thermodynamic extrapolation of $\bareMu^{1st}$, $a^3 n_B^{(1)}$ and  $a^3 n_B^{(2)}$
according to Eq.~(\ref{extrapolationWL}) for various bare temperatures $\bareT$.
}
\label{TableExtrapWL}
\end{table}


\subsection{The SC-LQCD Phase Diagram}
\label{PhaseDiagram}

\begin{figure}
\includegraphics[width=0.49\textwidth]{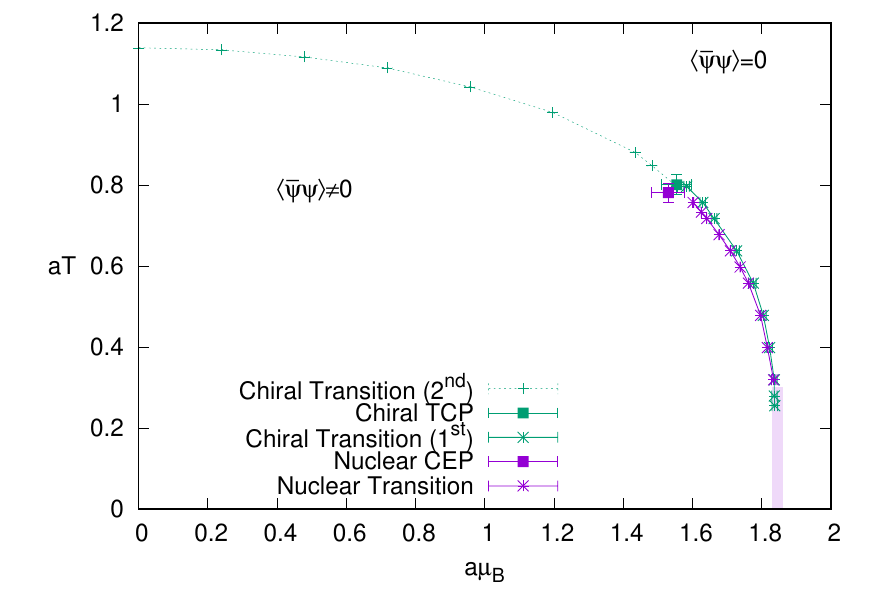} 
\includegraphics[width=0.49\textwidth]{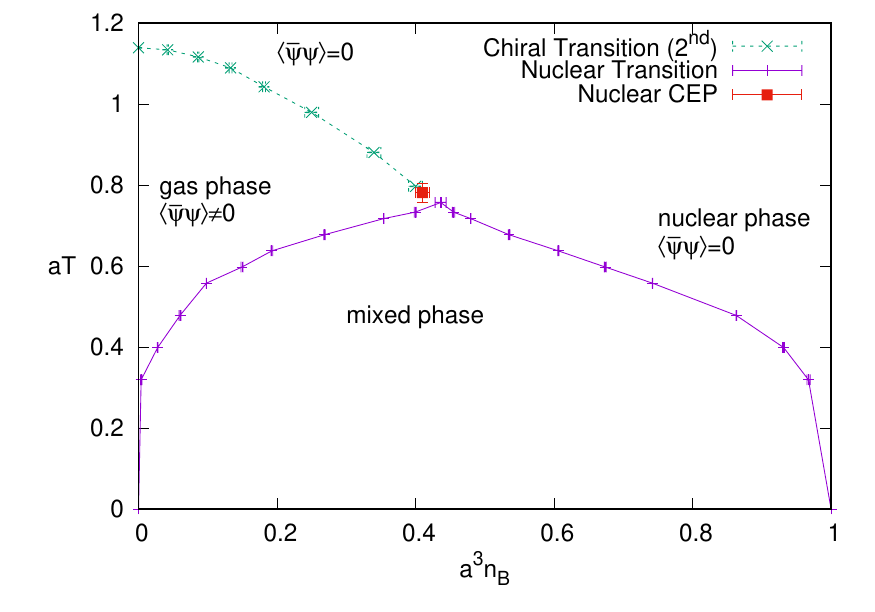} 
\caption{
The SC-QCD phase diagrams in the continuous time and the chiral limit. Results on the chiral transition are obtained via the Worm algorithm CT-WA and the 1st order nuclear transition is obtained via the Wang-Landau method. \emph{Top}: The grand canonical phase diagram in the $a\mu_B-aT$ plane. The chiral and nuclear first order transitions are on top within errors.
\emph{Bottom}: The grand canonical phase diagram in the $a^3n_B-aT$ plane. Note that due to Pauli saturation, at zero temperature the mixed phase at zero temperature extends 
to the full range in $a^3n_B$
}
\label{PhaseDiagPlot}
\end{figure}

We now want to summarize the previous results on the chiral and nuclear transitions and establish the phase boundaries both of the grand-canonical and canonical phase diagram, shown in Fig.~\ref{PhaseDiagPlot}.
In the grand-canonical phase diagram, one can clearly see that the chiral first order phase boundary and the nuclear transition (obtained from the Wang Landau method, see Tab.~\ref{TableExtrapWL}) are on top.
In the grand-canonical phase diagram, a mixed phase of both nuclear gas and liquid persists. The low density boundary $a^3n_B^{(1)}$ tends to zero, whereas the high density boundary $a^3n_B^{(2)}$ tends to 1. A meaningful density of nuclear matter cannot be assigned at strong coupling. 

There are various strategies to locate the chiral tricritical point, which is characterized as the end point of a triple first order line where the existence of three phases cease to coexist (the nuclear phase and two chirally broken phases for positive and negative quark mass). According to the Gibbs' phase rule, the upper critical dimension is 3, such that the tricritical exponents are analytic:
\begin{align}
 \gamma&=1,& \nu&=\frac{1}{2}.
\end{align}
To distinguish tricritical second order behavior from O(2) critical behavior, Eq.~(\ref{FSS}), large volumes are required.
There is a better strategy, based on the fact that the tricritical point coincides with the nuclear critical endpoint (which can be made plausible via a percolation analysis, see Sec.~\ref{Percolation}).
This is clearly only expected in the strong coupling limit, but also holds for small values of $\beta$ at finite $\Nt$ \cite{deForcrand2014}. 
The nuclear end point is characterized by the vanishing of the mixed phase, resulting in $n_B^{(1)}=n_B^{(2)}$.
The corresponding density of states becomes flat as the double peak structure vanishes. 
Our best estimate for the tricritial point in the CT-limit is
\begin{align}
aT^{\rm TCP}&=0.78(2),&a\mu_B^{\rm TCP}&=1.53(5),\nn
&&a^3n_B^{TCP}&=0.43(2)
\end{align}
If one does not take into account the rescaling with $\kappa$, then $\bareT^{\rm TCP}=0.98(3)$ and $\bareMu^{\rm TCP}=1.92(6)$ compares quite well with its determination on a disrete lattice: $\bareT^{\rm TCP}_{\Nt=4}=0.94(7)$, $\bareMu^{\rm TCP}_{\Nt=4}=1.92(9)$ \cite{Fromm2010}, indicating that the $\Nt$ corrections are small 
up to the critical point and become only large at lower temperatures \cite{deForcrand2017}.
We also note that the mean field tricritial point deviates substantially: $\bareT^{\rm TCP}_{\rm MF}=0.866$, $\bareMu^{\rm TCP}_{\rm MF}=1.731$ \cite{Nishida2003}.
As soon as a small finite mass is introduced, the chiral tricritical point turns into a chiral critical end point of Z(2) universality class. Close to the chiral limit, we estimate 
\begin{align}
 \mu_B^{\rm CEP}/T^{\rm CEP}&\simeq \mu_B^{\rm TCP}/T^{\rm TCP}=1.96(7),
\end{align}
which may in principle be within reach with conventional hybrid Monte Carlo, based on the fermion determinant such as Taylor expansion \cite{PhysRevD.95.054504}. But with increasing quark mass also the ratio $\mu_B^{\rm CEP}/T^{\rm CEP}$ increases rapidly ($a\mu_B^{\rm CEP}$ increases whereas $aT^{\rm CEP}$ decreases), as has been studied for discrete time in \cite{Kim2016}. The critical endpoint is quickly out of reach for methods of circumventing the sign problem via HMC methods. In the appendix Sec. \ref{FiniteMass} we elaborate further on the prospects of finite quark masses in the continuous time limit.

Our new results eliminate systematic uncertainties in previous findings in Monte Carlo for fixed $\Nt$ \cite{Forcrand2010}.\\

\subsection{Extension to Imaginary Chemical Potential}
\label{ImagMu}

Lattice QCD at imaginary chemical potential is usually considered because in contrast to non-zero real chemical potential, the fermion determinant is sign problem-free and it allows to analytically continue to real chemical potential \cite{deForcrand2003}.
It is also interesting in its own right due to the Roberge-Weiss periodicity \cite{Roberge1986} and the Roberge-Weiss transition \cite{Cuteri2015}.

\begin{figure}
\includegraphics[width=0.49\textwidth]{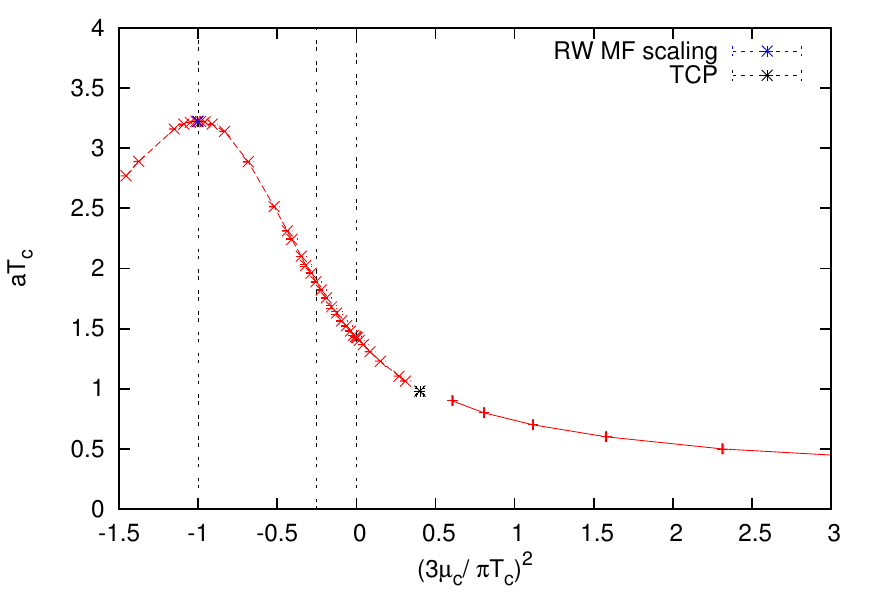}
\caption{Extension of the phase diagram to imaginary chemical potential. The chiral transition has the Roberge-Weiss periodicity. 
The transition at the Roberge-Weiss point $(\mu_B/(\pi T))=-1$ has a mean-field scaling behavior at strong coupling, but no signature of a first order transition. The second vertical line at $(\mu_B/(\pi T))=-1/4$ is characterized the absence of baryon, resulting in a $\U(3)$ gauge theory.
}
\label{I3}
\end{figure}

In the dual representation of SC-LQCD at discrete time, it is not straightforward to simulate at imaginary chemical potential. 
However, at continuous time where baryons are static, we can use \linebreak $\cosh(i\mu^{\rm im}_B/T)=\cos(\mu_B^{\rm im}/T)$, and with the $\calP$- and $\calQ$ polymer resummation (see Sec.~\ref{WangLandau}):
\begin{align}
\cos(\mu_B^{\rm im}/T) &\geq 0 \quad \text{for}&& \mu_B^{\rm im}/T\leq\frac{\pi}{2}\nn
\Nc+1+2\cos(\mu_B^{\rm im}/T)&\geq 0\quad  \text{for all} && \mu_B^{\rm im}/T
\end{align}
The second equation enables us to measure the chiral transition for arbitrary imaginary chemical potential.
Our result is shown in Fig.~\ref{I3}. 
At the Roberge-Weiss point $\mu_B^{\rm im}/T=\pi$ we do not find a cusp, in contrast what would be expected at weak coupling.
We also cannot observe a first order transition in the chiral observables, which is expected as the partition function becomes analytic in the high temperature limit. By integrating out the gauge links, the center sectors are no longer distinct.
Gauge observables such as the Polyakov loop should be able to signal a first order transition between the center sectors at high temperatures, which requires to include gauge correction. 
We also want to note that the point at $\mu_B^{\rm im}/T=\pi/2$ is special as it corresponds to the $\U(3)$ transition temperature $\bareT=1.8843(1)$ (as discussed in \cite{Unger2011}) as $\calP$-polymers have weight $w_p=0$ according to Eq.~(\ref{PolymerDef}).

\section{Taylor Expansion and Radius of Convergence}
\subsection{Taylor Expansion}
\label{TaylorExpansion}

\begin{figure}
 \includegraphics[width=0.49\textwidth]{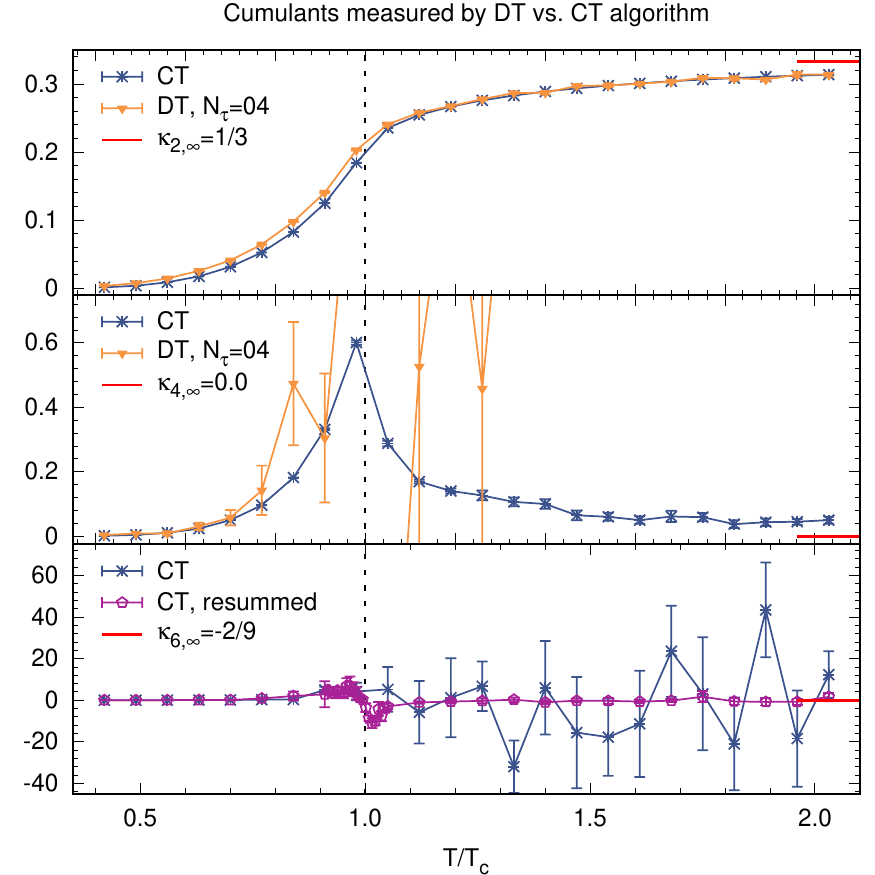}
 \caption{Measurement of the cumulants $\kappa_2$ (\emph{top}), $\kappa_4$ (\emph{center}) and $\kappa_6$ (\emph{bottom}) as a function of the temperature, comparing discrete time (DT) and continuous time (CT) results. Clearly, the continuous time cumulants are less noisy. Also indicated is the analytic value in the high temperature limit.
 }
 \label{CumulantDTCT}
\end{figure}

\begin{figure}
 \includegraphics[width=0.49\textwidth]{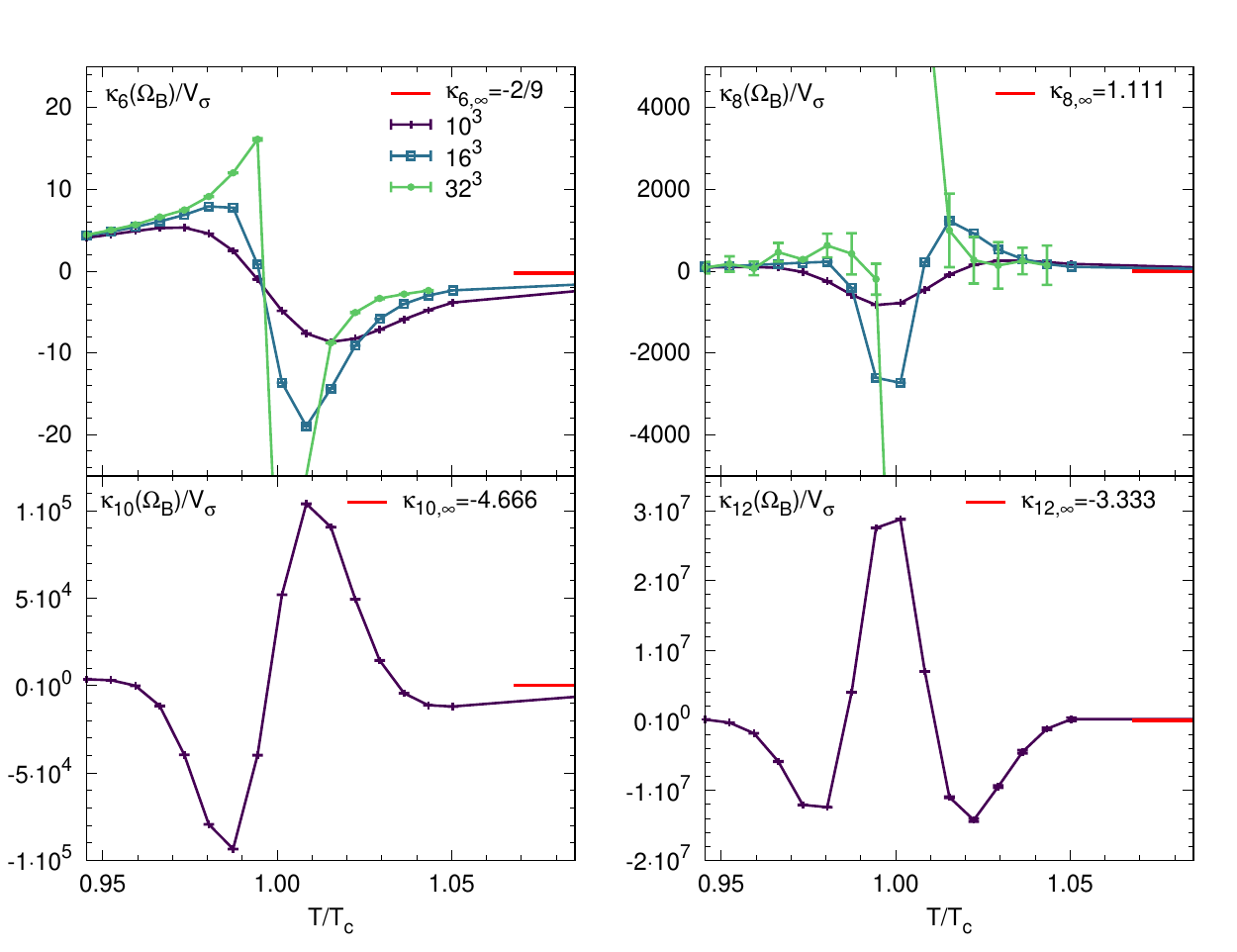}
 \caption{Measurement of the cumulants
 $\kappa_6$ (\emph{top left}),
 $\kappa_8$ (\emph{top right}),
 $\kappa_{10}$ (\emph{bottom left}) and
 $\kappa_{12}$ (\emph{bottom right}) as a function of the temperature in the vicinity of $T_c$ for volume $10^3$ or greater. Due to resummations and histogram methods explained in Sec.~\ref{WangLandau}, we are able to get the corresponding Taylor coefficients $c_{2n}$ completely under control. The number of extrema and inflection points increases with the order. 
 Also indicated is the high-temperature limit.
 }
 \label{Taylor_Histogram}
\end{figure}

The dual representation of SC-LQCD is a great laboratory to benchmark other methods to circumvent the sign problem. One of the prominent methods in the context of lattice QCD is the Taylor expansion \cite{Allton2005}, which might allow to estimate the location of a possible chiral critical endpoint based on estimates for the radius of convergence of the Taylor series. The standard thermodynamic observable that is Taylor expanded for that purpose is the pressure. This requires high orders of the Taylor series, but the current state of the art is limited to 6.~order (improved action) \cite{PhysRevD.95.054504} and 8.~order (unimproved action) \cite{PhysRevD.95.054512}.
It turns out that due to the continuous time limit and by taking into account both the polymer resummations and histogram method presented in Sec.~\ref{WangLandau}, we are able to determine higher orders of Taylor coefficients, both for the pressure and the baryon susceptibility.
The Taylor expansion of the pressure Eq.~(\ref{Pressure2}) at fixed temperature and about $\mu_B=0$, where only even orders contribute, is given by
\begin{align}
 p&=\frac{T}{V}\log \Zcal=p(T,\mu_B=0)+\sum_{n=1}^{\infty}c_{2n}\lr{\frac{\mu_B}{T}}^{2n}\nn
 c_{2n}&=\frac{T}{V}\frac{1}{(2n)!}\frac{\partial^{2n} \log \calZ}{\partial(\mu_B/T)^{2n}}=\frac{T}{V}\frac{1}{(2n)!}\kappa_{2n}(\omega)\LatSpat^{2n}
 \label{TaylorExpandPressure}
 \end{align}
where the cumulants $\kappa_n$ are defined in terms of the moments of the winding number $\omega$
via a cumulant-generating function $K(t)$:
\begin{align}
 M(t=\mu_B/T)&=\expval{e^{tx}}=\sum_{r=0}^\infty \mu_r \frac{t^r}{r!},\nn
 \mu_m&=\left.\frac{d^m M}{dt^m}\right|_{t=0}=\expval{\omega^m},\nn
 K(t)&=\log(M(t))=\sum_{r=0}^\infty \kappa_r \frac{t^r}{r!}.
\end{align}
We can measure all Taylor coefficients from the baryon density fluctuations, as $a^3n_B=\expval{\omega}$ according to Eq.~(\ref{BaryonDensity}).
We also obtain immediately from the Taylor coefficients of the pressure $c_{2n}$ those of the baryon susceptibility:
\begin{align}
\chi_B=\frac{\partial^2}{\partial(\mu_B/T)^2} p=\sum_{n=2}^{\infty}n(n-1)c_{2n}\lr{\frac{\mu_B}{T}}^{2n-2}. 
\end{align}
A comparison of discrete and continuous time evaluations of the first cumulants as shown in Fig.~\ref{CumulantDTCT} demonstrates the cumulants are less noisy in the CT-limit. But it further requires the polymer resummations and histogram method to determine the higher order cumulants up to 
$\kappa_{12}$, shown in Fig.~\ref{Taylor_Histogram}. From a thermodynamic extrapolation of the inflection points, we obtain an estimate for $T_c$ consistent with its determination in Sec.~\ref{ChiralTransition}. 

A comment on the definition of the pressure used in this section is in order: we have previously discussed that Eq.~(\ref{Pressure2}) is only valid in homogeneous systems, as is expected for the continuum limit of lattice QCD. In the strong coupling limit this is not the case. We can however only measure the pressure defined by a volume derivative according to Eq.~(\ref{Pressure}) in terms of dual variables, and it is of course possible to Taylor expand the spatial dimer density $\expval{n_{Ds}}$ as well. But this definition is proportional to the energy density and shows a gap along the first order transition. In contrast, \ref{TaylorExpandPressure} is well behaved as it proportional to the thermodynamic potential $F=-T\log \Zcal$, which is continuous along any transition.

\subsection{Estimates for the Radius of Convergence}
\begin{figure}
 \includegraphics[width=0.49\textwidth]{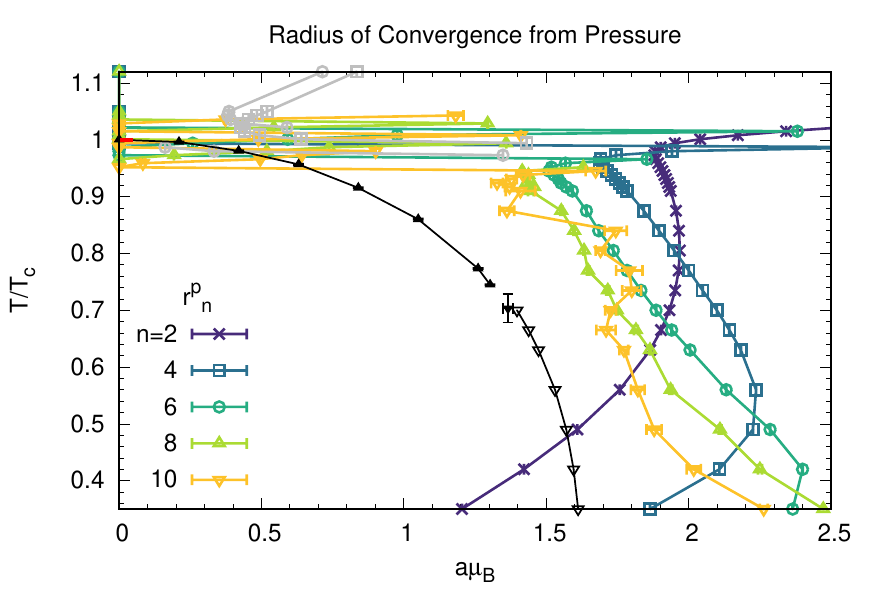}
 \caption{Radius of convergence, estimated from the pressure. The data in grey above $T_c$ correspond to imaginary chemical potential $i\bareMu$.}
 \label{RadiusOfConvergence1}

 \end{figure}

\begin{figure}
 \includegraphics[width=0.49\textwidth]{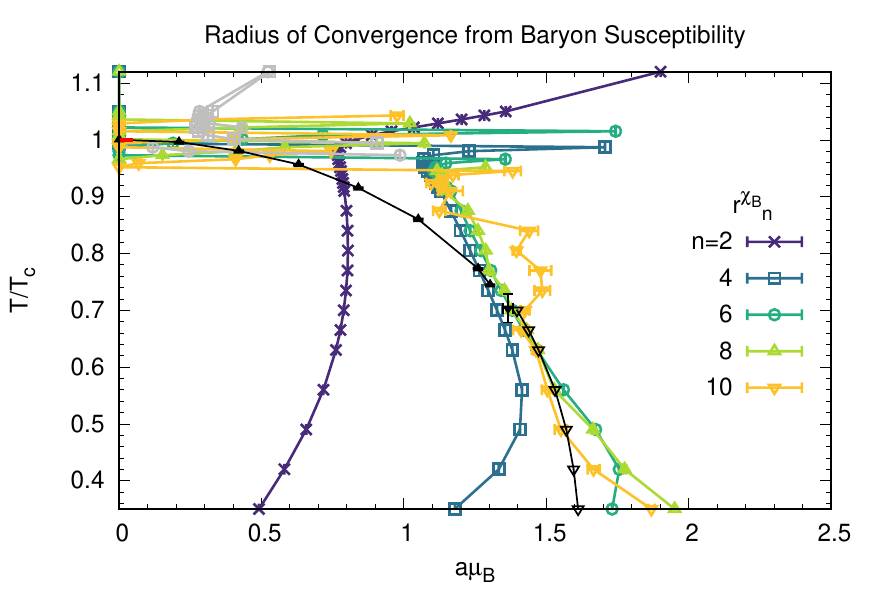}
\caption{Radius of convergence, estimated for the baryon susceptibility. It converges faster to the first order phase boundary compared to that of the pressure.}
 \label{RadiusOfConvergence2}
\end{figure}

We are now in a position to estimate the radius of convergence \cite{Karsch:2010hm} from these Taylor coefficients:
\begin{align}
 r^\alpha&=\lim_{n\rightarrow\infty} r^{\alpha}_n,& r^{\alpha}_n&=\sqrt{\left|\alpha\frac{\kappa_n}{\kappa_{n+2}}\right|},\nn
 \alpha_p&=\sqrt{(n+2)(n+1)},& \alpha_{\chi_B}&=\sqrt{(n-1)(n)}.
\end{align}
The corresponding results for the various $n$ are given in Figs.~\ref{RadiusOfConvergence1}, \ref{RadiusOfConvergence2}, where the radii for are plotted within the phase diagram. Above $aT_c$, the radius becomes imaginary (indicated in gray colors).
Note that we are still in the chiral limit where the whole phase boundary is either second or first order. Hence we expect that the radius of convergence drops to zero at $aT_c$. Below $aT_c$, the first singularity is given by the phase boundary, and we find indeed that the higher orders converge to the phase boundary. This is in particular observed for $r^{\chi_B}_n$, where the first order line is well approximated for $n=10$.



\section{Temporal Correlators and Meson Pole Masses}
\subsection{Staggered Euclidean Time Correlators}
\label{CorrelationFunction}

\onecolumngrid
\begin{table*}[t]
\centering
\begin{tabular}{|c | c c | c c | c c  |}
\hline
$g_{\vec{x},\tau}^S$ & \multicolumn{2}{c|}{$\Gamma^S\otimes\Gamma^T$} & \multicolumn{2}{c|}{$J^{PC}$} & \multicolumn{2}{c|}{Physical states} \\
& NO: $1^\tau$& O: $(-1)^\tau$& NO & O & NO & O \\
\hline
$1$ 		  				&	$1\otimes 1$																	& $\gamma_0\gamma_5\otimes(\gamma_0\gamma_5)^*$ 	& $0^{++}$ 	& $0^{-+}$ 	& $\sigma_S$ 	& $\pi_A$		\\
$(-1)^{x_i}$   		&	$\gamma_i\gamma_5\otimes(\gamma_i\gamma_5)^* $ 	& $\gamma_i\gamma_0\otimes(\gamma_i\gamma_0)^*$	& $1^{++}$ 	& $1^{--}$ 	& $a_A$  		& $\rho_T$	\\
$(-1)^{x_j+x_k}$	&	$\gamma_j\gamma_k\otimes(\gamma_j\gamma_k)^*$	& $\gamma_i\otimes\gamma_i^*$  									& $1^{+-}$ 	& $1^{--}$ 	& $b_T$  		& $\rho_V$\\
$(-1)^{x_i+x_j+x_k}$		&	$\gamma_0\otimes\gamma_0^*$ 									& $\gamma_5\otimes(\gamma_5)^*$ 								& $0^{+-}$ 	& $0^{-+}$ 	& $-_V$  			& \; $\pi_{PS}$ 	\\
\hline
\end{tabular}
\caption{
Table of mesonic operators for staggered fermions that are diagonal in the spin-taste basis $\Gamma^S\otimes \Gamma^T$ and the corresponding physical states. The oscillating and non-oscillating states within the same $g_x^S$ are distinguished.  
}
\label{SpinTasteOperators}
\end{table*}
\twocolumngrid

We have explained in Sec.~\ref{Algorithm} that the monomer 2-point correlation function is sampled during worm evolution. 
We are mainly interested in temporal correlation functions, from which we can extract the ground state energy corresponding to the meson pole mass.
In this section we will explain how to extract them and discuss their dependence on temperature and baryon chemical potential.

The basic definition of the temporal correlators at zero momentum $\vec{p}=0$ for 
staggered fermions $\bar{\chi}$, $\chi$, based on the local single-time-slice operators \cite{DeGrand:2006zz} is:
\begin{align}
C_S(\tau)&=\sum_{\vec{x}}C_S(\vec{x},\tau),\nn
C_S(\vec{x},\tau)&=\langle \bar{\chi}_{\vec{0},0}\chi_{\vec{0},0}\bar{\chi}_{\vec{x},\tau}\chi_{\vec{x},\tau}\rangle g^S_{\vec{x},\tau},
\label{signGD}
\end{align}
where the spin $S$ of the meson is given by the kernel operators $\Gamma^S$ in terms of phase factors $g_{\vec{x},\tau}^S \in \{\pm 1\}$. We will only consider operators that are diagonal in spin-taste space: $\Gamma^S\otimes \Gamma^T$ with $\Gamma^T={\Gamma^S}^*$.
We will not consider any flavor structure as $\Nf=1$ (but see App.~\ref{ParFuncNf2} for $\Nf=2$).
In every mesonic correlator specified by $\Gamma^S$, there is a non-oscillating part and oscillating part with additional phase factor $(-1)^\tau$, which is due to the even-odd decomposition for staggered fermions. This parity partner has opposite spin, parity and taste content.
Thus the non-oscillating and oscillating part correspond to different physical states, see Tab.~\ref{SpinTasteOperators}. 
Of particular interest is the pion $\pi_{PS}$ which is the Goldstone boson for the residual chiral symmetry, Eq.~(\ref{ChiralTrafo}). 
Throughout the worm evolution, monomer two-point correlation functions are accumulated whenever head and tail are at opposite parities: 
\begin{equation}
C_S(\vec{x}_H-\vec{x}_T,\tau_H-\tau_T)=C_S(\vec{x},\tau)=N_c\frac{O(C_S(\vec{x},\tau))}{Z}.
\end{equation}
with $Z$ the number of worm updates.
Such worm estimators are incremented as 
\begin{align*}
 O(C_S(\vec{x},\tau))\;\rightarrow \; O(C_S(\vec{x},\tau))+ f g^S_{\vec{x},\tau}\,\delta_{x_T,x_1}\delta_{x_H,x_2},
\end{align*}
\phantom{.}\\[-11mm]
\begin{align}
 f\equiv f(\gamma),&& \tau &\in[0,1, \dots \Nt]&& (\text{discrete time}),  \nonumber\\
 f\equiv f(T),     && \tau &\in[0, \bareT^{-1}] && (\text{continuous time}),
\label{increments}
\end{align}
with $f(\gamma)$ given in Eq.~(\ref{CorrHistoIncDT}) and $f(T)$ given in Eq.~(\ref{CorrHistoIncCT}).
Summing over the correlators yields immediately the corresponding discrete/continuous time susceptibilities:
\begin{align}
a^6\chi_S^{\rm DT}&=\frac{1}{\Ns^3 \Nt}\sum_{\vec{x},\tau} C_S(\vec{x},\tau),\\
a^6\chi_S^{\rm CT}&=\frac{\bareT}{\Ns^3}\sum_{\vec{x}}\int_{0}^{1/\bareT}d\tau \;C_S(\vec{x},\tau).
\end{align}
\phantom{.}

\begin{figure}
\includegraphics[width=0.49\textwidth]{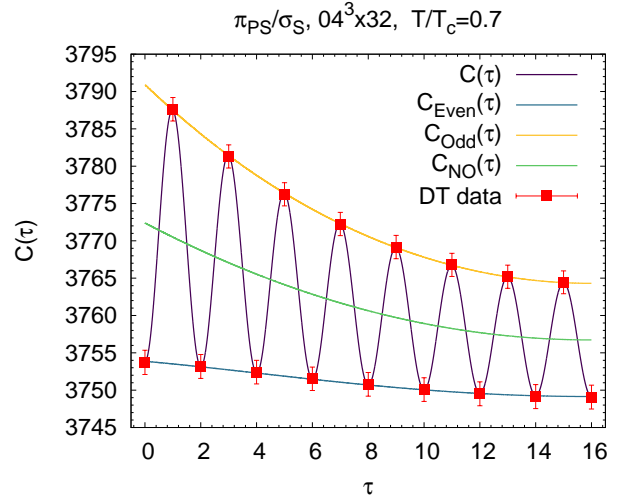}
\caption{
Discrete time pion correlator for $\Nt=32$ at \hspace{5mm}
$T/T_c=0.7$, showing the oscillating behavior, and the decomposition into even and odd contributions, according to Eq.~(\ref{CorrEvenOdd}). The fit $C(\tau)$ is reconstructed from the fits $C_{\rm Even}(\tau)$ and $C_{\rm Odd}(\tau)$.
}
\label{dtCorrFit}
\end{figure}

\begin{figure}
\includegraphics[width=0.49\textwidth]{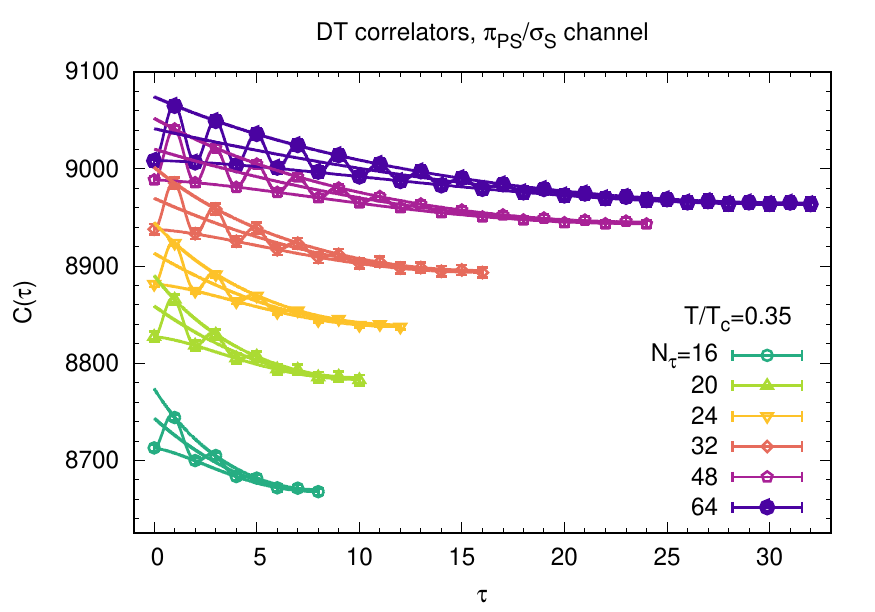}
\caption{Discrete time pion correlators for various $\Nt$, showing that the oscillatory behavior persists for larger $\Nt$, with a well behaved continuous time limit.}
\label{dtCorrFitVar}
\end{figure}

\begin{figure}
\includegraphics[width=0.49\textwidth]{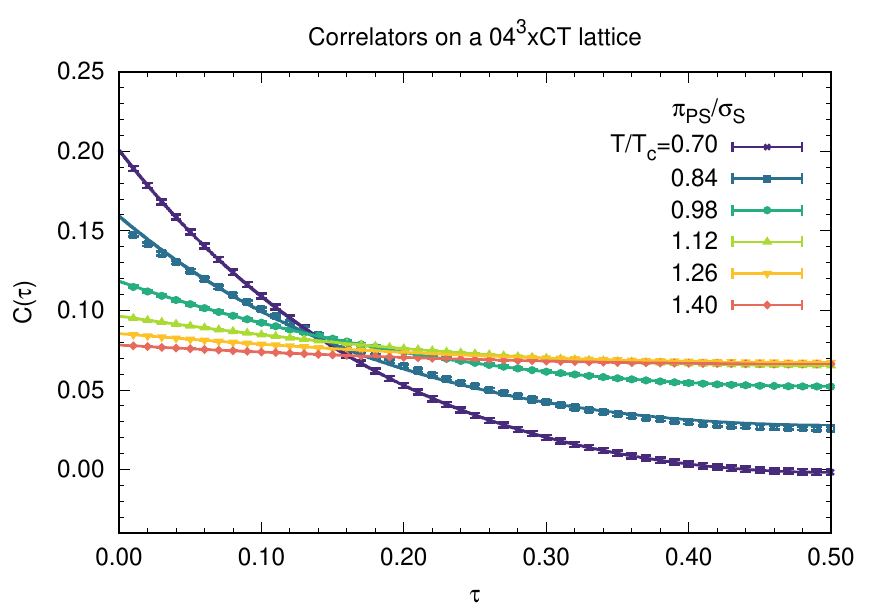}
\caption{Continuous time pion correlator fitted according to Eq.~(\ref{PionCorrelator}), for various temperatures, on the full range in Euclidean time $\tau$, for $N=100$ bins. This is sufficient to extract the pole masses $M_\pi(\bareT)$ to high precision.}
\label{CTCorrPi}
\end{figure}

The non-oscillating and oscillating parts of the correlators for discrete time 
\begin{align}
C(\tau)&=C_{\rm NO}(\tau)+(-1)^{\tau}C_{\text{O}}(\tau),\nonumber\\
C_{\rm NO}(\tau)&=A_{\rm NO}\cosh(\at M_{\rm NO}(\tau-\Nt/2)),\nonumber\\
C_{\rm O}(\tau)&=A_{\rm O}\cosh(\at M_{\rm O}(\tau-\Nt/2),
\end{align}
are shown in Fig.~\ref{dtCorrFit}. It is advantageous to consider the linear combinations 
\begin{align}
C_{\rm Odd}(\tau)&=C_{\rm NO}+C_{\rm O},\nn
C_{\rm Even}(\tau)&=C_{\rm NO}-C_{\rm O},
\label{CorrEvenOdd}
\end{align}
and fit the even/odd correlators instead: (1) the fit is more stable (2) it generalizes to the continuous time limit, where we can distinguish even and odd $\tau$ via emission and absorption events, see Sec.~\ref{ContTimeParFunc}.
 We can reconstruct the physical states by inverting Eq.~(\ref{CorrEvenOdd}).
The discrete time correlators for the pion are shown in Fig.~\ref{dtCorrFitVar}. We observe that the correlators for increasing $\Nt$ become more continuous and their range extends to $\Nt/2$. In Fig.~\ref{CTCorrPi}, the continuous time correlators for the pion $\pi_{\rm PS}$ is reconstructed from 
\begin{align}
C_\pi(\tau)&=\frac{1}{2}\lr{C_{\rm Odd}(\tau)-C_{\rm Even}(\tau)}\nn
&=A_\pi \cosh(M_\pi/T (\tau -1/2))
\label{PionCorrelator}
\end{align}
with $\tau\in[0,1/2]$ and spatial kernel $g^{\pi}_{\vec{x}}=(-1)^{x+y+z}$, 
and likewise for other mesons.
This requires book-keeping on which events contribute to $C_{\rm Odd}$ or $C_{\rm Even}$ depends 
on whether the worm head is located at an absorption event $x_H\in \calA$ or an emission event $x_H\in \calE$.\\

Even in the CT-limit, it is necessary to discretize the temporal correlators, due to memory limitations and finite statistics. The histograms will depend on the bin size 
\begin{align}
\Delta\tau&=\frac{1}{\bareT N},
\end{align}
with $N$ the number of bins. The finer $\Delta\tau$, the less events are placed in each bin, which makes the determination of the correlator more difficult. On the other hand, the coarser $\Delta\tau$, the less data are available to reconstruct the correlator. In principle one could measure the continuous time correlators without introducing a binning \cite{Berg2008}, but in practice this seems not necessary as our measurements for $N=100, 200, 400$ lead to almost identical results.


\subsection{Temperature and Density Dependence of Meson Pole Masses}
\begin{figure}
\includegraphics[width=0.49\textwidth]{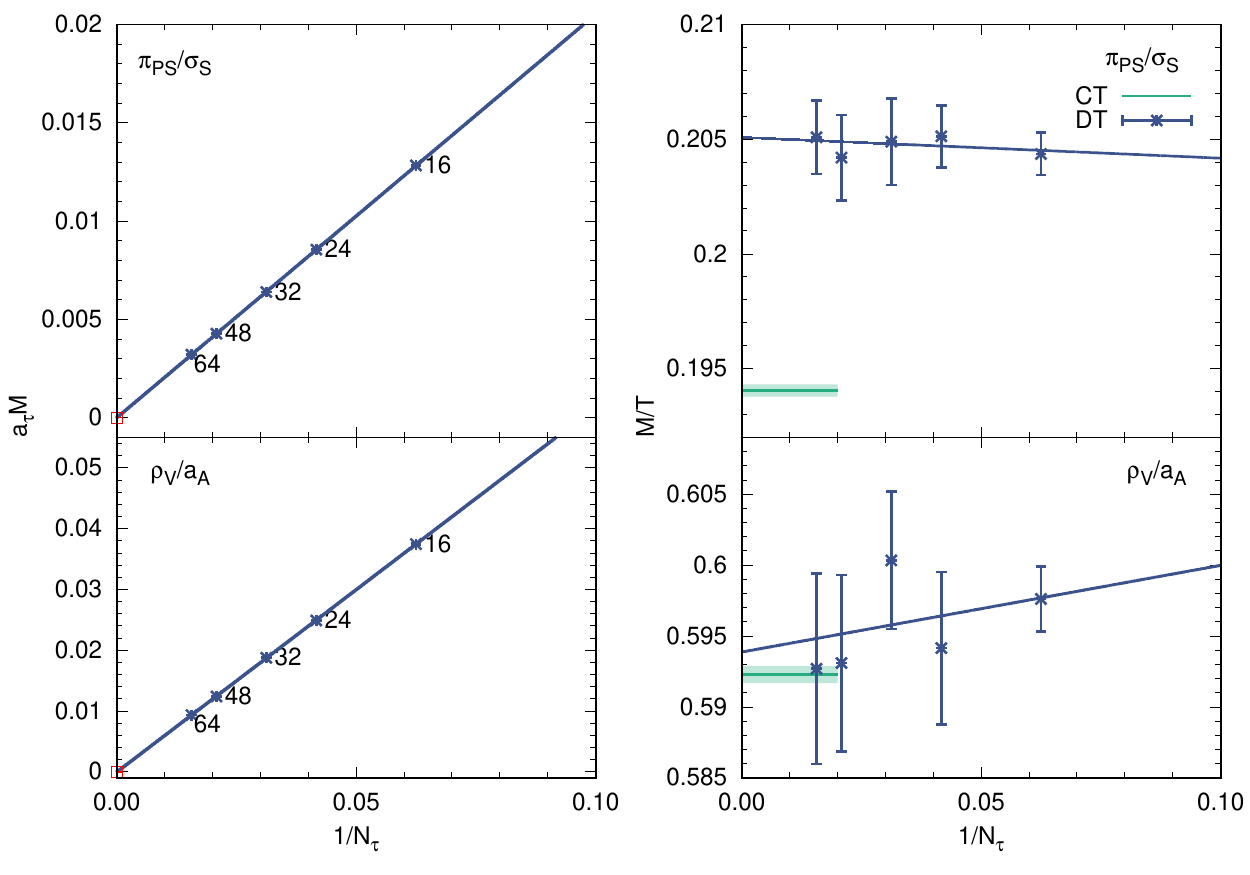}
\caption{Pole masses for $\bareT=1.8$. \emph{Left:} Extrapolation of the pion pole masses extracted from discrete time correlators to the continuous time limit. 
\emph{Right:} Comparison of the discrete time pole masses to the continuous time pole mass (green band) in units $M/T$, rescaled via Eq.~(\ref{ConvertMassesDTtoCT}).
The discrepancy for the pion mass \emph{(top)} of about 3\% may stem from uncertainties of distinguishing the ground states from the excited states at rather small $\Nt$. The continuous time pole masses have much smaller statistical errors compared to discrete time.
}
\label{dtExtrap}
\end{figure}

Since temporal correlators are measured at zero spatial momentum, the extracted meson masses are pole masses: $E_0(\vec{p}=0)=M$.
We extract the ground state mass $M$ as dimensionless quantity $M/T$ by multi-state fits (including excited states) and by varying the fit range $[\tau_{\rm min}/\bareT,1/(2\bareT)]$. To obtain good balance between the required number of states and the error estimation,  we apply the Aikaike Information Criterion \cite{10.1093/biomet/76.2.297}. 
We adjust $\tau_{\rm min}$ to be most sensitive to mass plateau. 
To compare discrete time (where we extract $\at M$ to continuous time, we convert via
\begin{align}
M/T&=\Nt \,\at M, &  aM&= \kappa \, \bareT\, M/T,
\label{ConvertMassesDTtoCT}
\end{align}
as shown in Fig.~\ref{dtExtrap}.
Making use of the same fitting scheme, the error bars for the extracted pole masses from CT-correlators are much smaller than the corresponding DT-correlators. Moreover, the uncertainties when extrapolating DT-correlators of about 3\% are circumvented.\\

\begin{figure}[h!]
\includegraphics[width=0.49\textwidth]{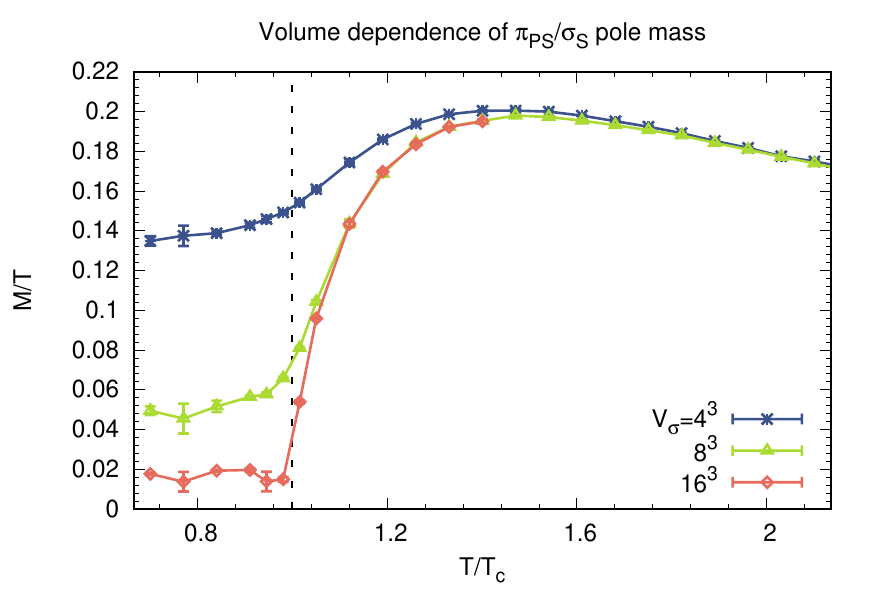}
\caption{Pole masses $M/T$ as a function of temperature, for different volumes but the same channel $\pi_{PS}/\sigma_S$, measured in continuous time. 
Due to finite volume effects, the pion mass is not strictly zero in the chiral limit ($\epsilon$-regime). 
}
\label{massesCTContinuumPi}
\end{figure}

\begin{figure}[h!]
\includegraphics[width=0.49\textwidth]{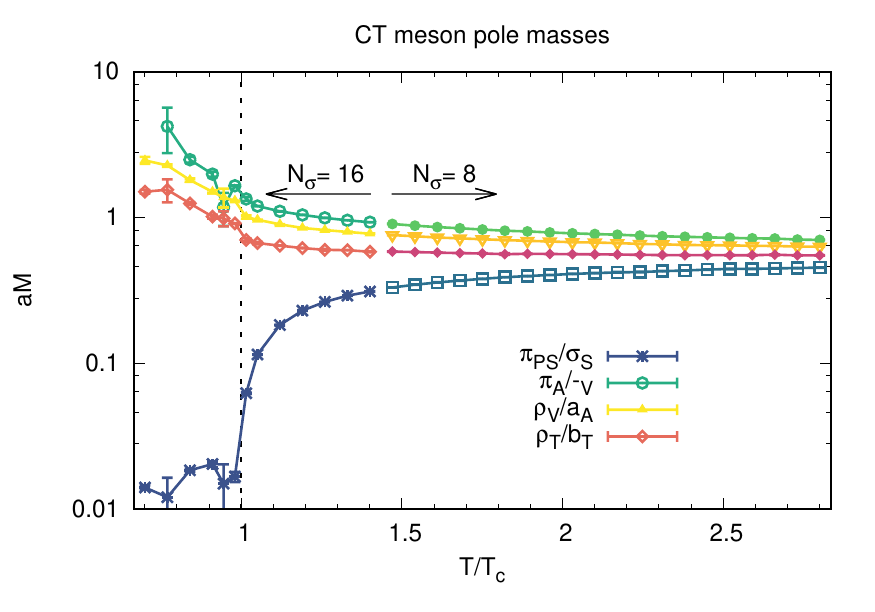}
\caption{Pole masses $aM$ for various quantum numbers $J^{PC}$ as a function of the temperature, measured in the CT-limit. The mass degeneracies are given in (\ref{massDegen}).
We observe an imprint of the chiral transition on the pole masses (based on $\Ns=16$), and the convergence to the same value $aM=0.411(1)$ in the large temperature limit (based on $\Ns=8$).
}
\label{massesCTContinuum}
\end{figure}

We have measured the temperature-dependence of the pole masses and find that in particular the pion becomes heavy at the chiral transition, see Fig.~(\ref{massesCTContinuum}). 
For $N_f=1$ we find a mass degeneracy for the pairs of states:
\begin{align}
\sigma_S &\leftrightarrow \pi_{PS},&
\pi_A &\leftrightarrow -_V,\nn
b_T &\leftrightarrow \rho_T,&
a_A &\leftrightarrow \rho_V,
\label{massDegen}
\end{align}
which corresponds to a multiplication by the parity $\epsilon(x)$, compare Tab.~\ref{SpinTasteOperators}.
This is due to the strong coupling and the chiral limit (i.e.~we are in the $\epsilon$-regime): e.g.~the pion $\pi_{PS}$ is mass degenerated with the sigma meson $\sigma_S$. This degeneracy is lifted as soon as $am_q>0$, see \ref{FiniteMass}.
The pion becomes indeed massless below $T_c$ in the thermodynamic limit, as seen in Fig.~\ref{massesCTContinuumPi}. But the pion and all other mesons do not acquire a thermal mass, as shown in Fig.~(\ref{massesCTContinuum}). Rather, they all tend to the same high temperature value  $aM=0.411(1)$. We suspect that this is an artifact  of the strong coupling limit: even at high temperatures, in the chirally restored phase, the quarks are still confined into mesons. Hence, they do not experience the anti-periodic boundary conditions \cite{Boyd:1994np} and will not receive contributions from the lowest Matsubara frequencies $\pi T$ above $T_c$.

\begin{figure}[h!]
\includegraphics[width=0.49\textwidth]{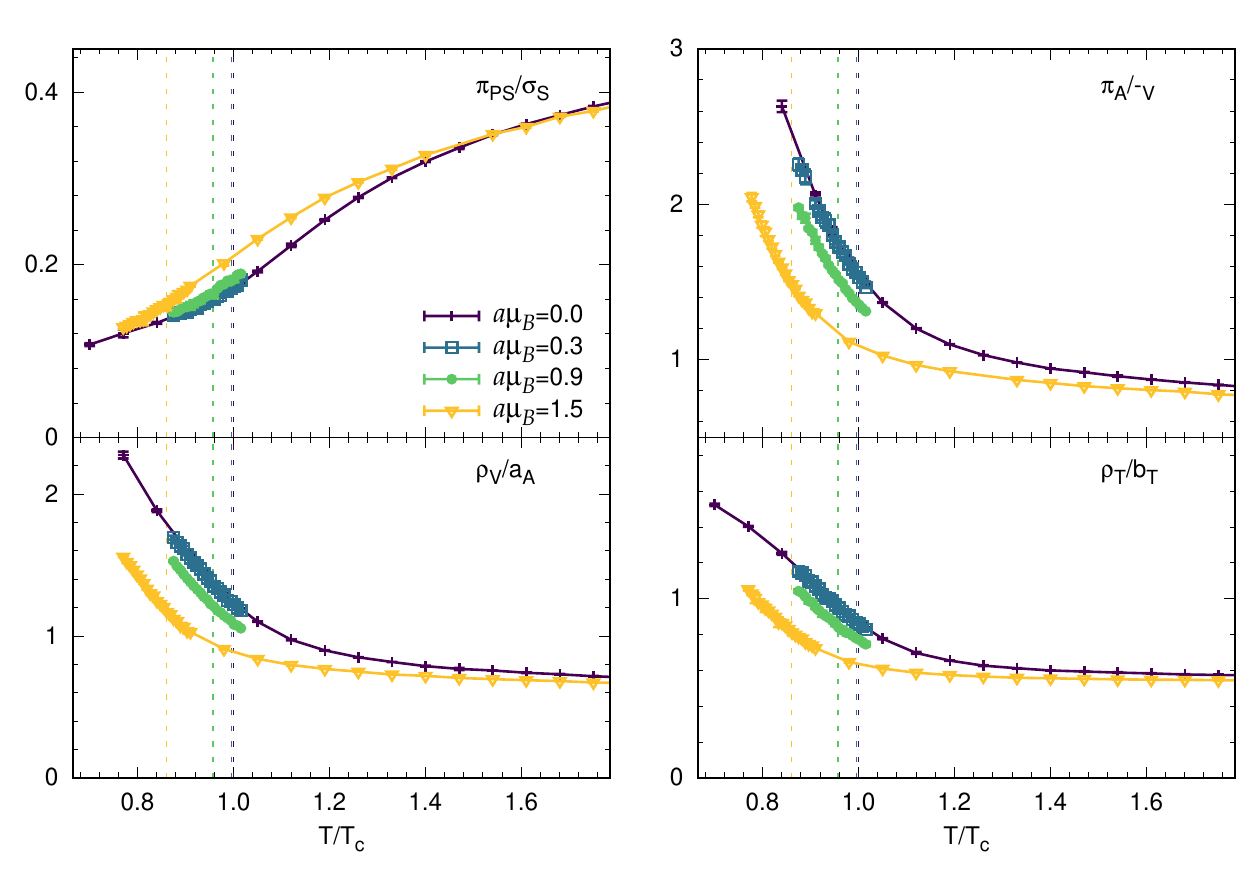}
\caption{Pole masses $aM$ for various chemical potentials $\bareMu$, as a function of the temperature. The vertical dashed lines indicate the transition temperatures for the chemical potentials considered. At high temperatures, the dependence on the chemical potential becomes weaker.
}
\label{massesCTFiniteDensity}
\end{figure}

The extension to finite chemical potential $\bareMu$ is straight forward, the results on the temperature dependence of the pole masses for various chemical potentials below $\bareMu^{\rm TCP}$ is shown in Fig.~\ref{massesCTFiniteDensity}. The pole masses change most at the transition temperature for the respective chemical potential. Their hign-temperature limits become independent of the chemical potential.  


\section{Conclusion}

We have demonstrated the power of continuous time simulations of lattice QCD in the strong coupling limit, which make extrapolations for $\Nt\rightarrow \infty$ obsolete. 
All ambiguities arising from such an extrapolation are removed. The Hamiltonian formulation gives further insight into the world-line formulation of strong coupling lattice QCD.
We discussed in detail the continuous time worm algorithm in terms of a Poisson process, the dual observables, and resummation and histogram techniques to determine the phase diagram both in the $\mu_B - T$ plane and $n_B - T$ plane via the Wang-Landau method. 
The phase boundary can be compared with estimates from the radius of convergence from Taylor coefficients which we can determine via baryon fluctuations at zero density up to $c_{12}$. We have also investigated temporal correlation functions, which we can measure with high resolution and higher statistics compared to discrete time, and from which we could determine the temperature dependence of the meson pole masses, both at zero and non-zero density. Whether the continuous time correlation functions can also be extended on the Schwinger-Keldysh contour to extract transport coefficients is under investigation. Real time simulations in the dual formulation of SC-LQCD are not completely sign-problem free, but  much less severe compared to the standard formulation based on the fermion determinant.

Some first steps to extend our Hamiltonian formulation to more flavors and finite quark mass are presented in the appendix. We plan to include the gauge corrections from the Wilson gauge action in continuous time in a similar way as we have already successfully implemented in discrete time \cite{Unger2017,Gagliardi:2019cpa,Kim:2020tcd}. As the continuous time limit is well defined also at finite lattice gauge coupling $\beta$, we may improve on the phase diagram by reducing the spatial lattice spacing directly in the continuous time limit via quantum Monte Carlo simulations.

\begin{acknowledgments}
W.~Unger is grateful to Philippe de Forcrand for providing the initial idea to consider the continuous time limit, and is thankful for the many discussions on the continuous time formulation.
We would like to thank our colleagues Olaf Kaczmarek and S\"oren Schlichting for discussions of some aspects related to Euclidean correlators, Christian Schmidt for discussions on Taylor expansion, Owe Philipsen for discussions on the canonical phase diagram and Jangho Kim for his contributions to our code for discrete time.
For the extraction of pole masses from temporal correlators, we are thankful to Hauke Sandmeyer for providing numerical tools.
We acknowledge contributions of the students Aaron von Kamen (on the Wang Landau method) and Ferdinand J\"unnemann (on percolation) to this project.

Numerical simulations were performed on the OCuLUS cluster at PC2
(Universit\"at Paderborn). This work is supported 
by the Deutsche Forschungsgemeinschaft (DFG) through
the Emmy Noether Program under grant No. UN 370/1
and through the CRC-TR 211 'Strong-interaction matter under extreme conditions'– project number 315477589 – TRR 211.
\end{acknowledgments}

\section{Appendix}

\newcommand{\kbs}{k_{b_\sigma}}
\newcommand{\kbt}{k_{b_\tau}}

\subsection{Derivation of the Continuous Euclidean Time Limit}
\label{DerivationCT}

In this section we want to explain how to derive Eq.~(\ref{ParFuncCT}) from Eq.~(\ref{ParFuncDTG})
We start from the discrete partition function for gauge group U($\Nc$), neglecting the baryonic part for a moment.
We have to investigate what sequence of vertices is admissible on each site and at the same time conserves the pion current.

We will use the vertices in the meson occupation numbers and introduce the shorthand notation 
\begin{align}
\hat{v}(k|l|m)\equiv \hat{v}(k|l)\hat{v}(l|m).
\end{align}
We classify admissible sequences via the length of the interval: whether it is even or odd.  This is determined by the sequence of emission sites $\calE$ or absorption sites $\calA$. The discussion applies to $\Nc=3$ but generalizes straightforwardly to odd $\Nc$. For even $\Nc$, the meson state $\meson=\Nc/2$ needs a special treatment which will not be address here. We distinguish via even-odd parity:
\begin{enumerate}
 \item \emph{Odd intervals} are those where an $\calA$-site is followed by an $\calE$-site, or an $\calE$-site is followed by an $\calA$-site:
\begin{align}
&\hat{v}(0|1|0),& &\hat{v}(1|2|1),& &\hat{v}(2|3|2),\nonumber\\
&\hat{v}(1|0|1),& &\hat{v}(2|1|2),& &\hat{v}(3|2|3).
\end{align}
which is exactly the case when we have two subsequent $\lsteel$-vertices
or two subsequent $\tsteel$-vertices.
\item \emph{Even intervals} are those where an $\calA$-site is followed by an $\calA$-site site, or an $\calE$-site is followed by an $\calE$-site:
\begin{align}
&\hat{v}(0|1|2),& &\hat{v}(1|2|3),\nonumber\\
&\hat{v}(3|2|1),& &\hat{v}(2|1|0).
\end{align}
wich is exactly the case when a $\lsteel$-vertex is followed by a $\tsteel$-vertex or vice versa.
\end{enumerate}

Since $\Nt$ is even, the number of odd intervals must be an even number. Also, on each site, the number of $\calE$-sites equals the number of $\calA$-sites.
Any CT-configuration $\calG$ with $\Nc$ odd is completely determined by specifying the location and kind of the vertices and whether an interval is even or odd: an interval between two vertices of the same type is always of odd length, between two different vertices it is of even length.

If we now consider a spatial bond given by the nearest neighbor pair $b=\langle\vec{x},\vec{y}\rangle$ such that there is at least one spatial dimer on $b$, then the sequence of $\calE$-sites and $\calA$-sites is exactly opposite if we ignore vertices which do not belong to dimers on $b$ (see Fig.~\ref{DimerChains}).
This implies that it is completely determined by the type of vertex whether we have an even or odd interval. 
The first dimer on $b$ can be put on any of the $\Nt$ temporal locations, but the second dimer can only be put on $\Nt/2$ locations, and all subsequent spatial dimers $(\Nt-k)/2$ temporal locations. Given that the maximal number of spatial dimers is given by the order in $\mathcal{O}(\gamma^{-\ka_b})$, in the limit $\Nt\rightarrow \infty$ the probability of two spatial dimers on $b$ to be at the same location (effectively forming a double-dimer) is zero, and we can disregard the finite $\Nt$ corrections $\Nt-k$. 
In this limit, we have $\Nt(\Nt/2)^{k_b}$ possible temporal locations. 
We have however not yet considered symmetry factors as in the above argument, the spatial dimers added to the bond are not time ordered. Time ordering is however a global aspect that cannot be considered in isolation of a single bond. 
If we force the whole set of spatial dimers with $\ka=\sum_b{\ka_b}$ to be time ordered, we have to divide by $\ka!$ as only one of the permutations are time-ordered sequence.
Another way to see how the symmetry factor arises from time ordering for $\Nt\rightarrow \infty$ with $t\mapsto \tau/\Nt$, $\tau\in [0,1[$
is to replace the sums by integrals:
\begin{align}
               \int_0^1 d\tau_1 \int_{\tau_1}^1 d\tau_2 \ldots \int_{\tau_{k-1}}^1 d\tau_k\,& w(\tau_1,\tau_2,\ldots \tau_k)\nn
 =\frac{1}{\ka!}\int_0^1 d\tau_1 \int_0^1 d\tau_2     \ldots \int_0^1 d\tau_k      \,& w(\tau_1,\tau_2,\ldots \tau_k)
 \end{align}
where $t_i$ is the temporal location of the $i$-th spatial bond. This holds because the weight $w(t_1,t_2,\ldots t_k)$ does not depend on the locations but just on the number of vertices that appear. 
We conclude that the total weight of a $\U(3)$ configuration is: 
\begin{align}
 \sum_{\ka\in 2\mathbbm{N}}\frac{1}{\ka!}\lr{\frac{\Nt/2}{\gamma^2}}^{\ka}\sum_{\calG\in \Gamma_\ka}v_\lsteel^{N_\lsteel(\calG)}v_\tsteel^{N_\tsteel(\calG)}
\label{DerivationCTFormula}
 \end{align}
where $\Gamma_\ka$ is the set of topologically inequivalent configurations (which differ in the distribution of $\lsteel$- and $\tsteel$-vertices
over sites.
In summary, a CT-configuration $\calG$ is completely determined by specifying whether the intervals between vertices are even or odd, up to translation by $\at$
which corresponds to time reversal, 
\begin{align}
&\mathcal{T}:&\meson\mapsto\Nc-\meson,&& \calG\in\Gamma_\ka \mapsto \calG^{\mathcal{T}}\in\Gamma_\ka
\end{align}
 with equal weight: $w(\calG)=w(\calG^{\mathcal{T}})$.
 With $\Nt/\gamma^2=1/\bareT$ we arrive at the mesonic part of the continuous time partition function Eq.~(\ref{ParFuncCT}).

\newcommand{\Dhb}{\Delta \hat{\beta}}

It remains to discuss the baryonic part of the partition function. Since baryons form self-avoiding loops,
it suffices to note that spatial baryon hoppings are suppressed by $\gamma^{-\Nc}$.
Hence, for $\Nc \geq 3$ spatial hoppings are essentially absent as $\gamma\rightarrow \infty$ and baryons become static. This does not happen for $\Nc=1$ (electrons) and  $\Nc=2$ (di-quarks). In physical terms, only for $N_c\geq3$ the baryon is heavy and non-relativistic. Hence a baryon-antibaryon pair cannot be created from the vacuum.\\

\subsection{Analytic Result for U(1)}
\label{AnalyticU1}
We have derived an analytic expression for strong coupling U(1) on $2\times \Nt$ lattices for arbitrary values of $\Nt$ and $\gamma^2$, 
enabling us to obtain the continuous time result. A generalization to $\SU(\Nc)$ is not straight forward.
The continuous time assumption that spatial dimers with spatial multiplicity $k_i>1$ are suppressed is (trivially) exact in U(1). 

The partition function in the chiral limit for U(1) can be derived from considering all spatial dimers, making use of the fact that the interval length in units of $\at$ between subsequent spatial dimers must be odd:
\begin{widetext}
\begin{align}
\calZ_0(\gamma,\Nt) &= \gamma^{2\Nt}\left(4 + \sum_{\ka\in2\mathbb{N}^+}^{\Nt}2^\ka\alpha_\ka(\Nt)\gamma^{-2\ka}\right),
&\alpha_\ka(\Nt)&=\frac{2}{\ka!}\prod_{k=0}^{\ka/2-1}\lr{\lr{\frac{\Nt}{2}}-k^2} = \frac{\Nt}{\ka}\binom{(\Nt+\ka)/2-1}{\ka-1}
\end{align}
\end{widetext}
Note that $\ka$ always has to be even as spatial hoppings have to come in pairs to be consistent with the boundary conditions in time. The factor $2^n$ is due to the fact that each spatial dimers can hop either in forward or backward direction due to the periodic boundary conditions in space. 
Note that we have not approximated $\alpha_n$ by $\frac{2}{n!}(\frac{\Nt}{2})^{n/2}$, as we did in the steps leading to Eq.~(\ref{DerivationCTFormula}). 

\begin{figure}[h!]
 \includegraphics[width=0.49\textwidth]{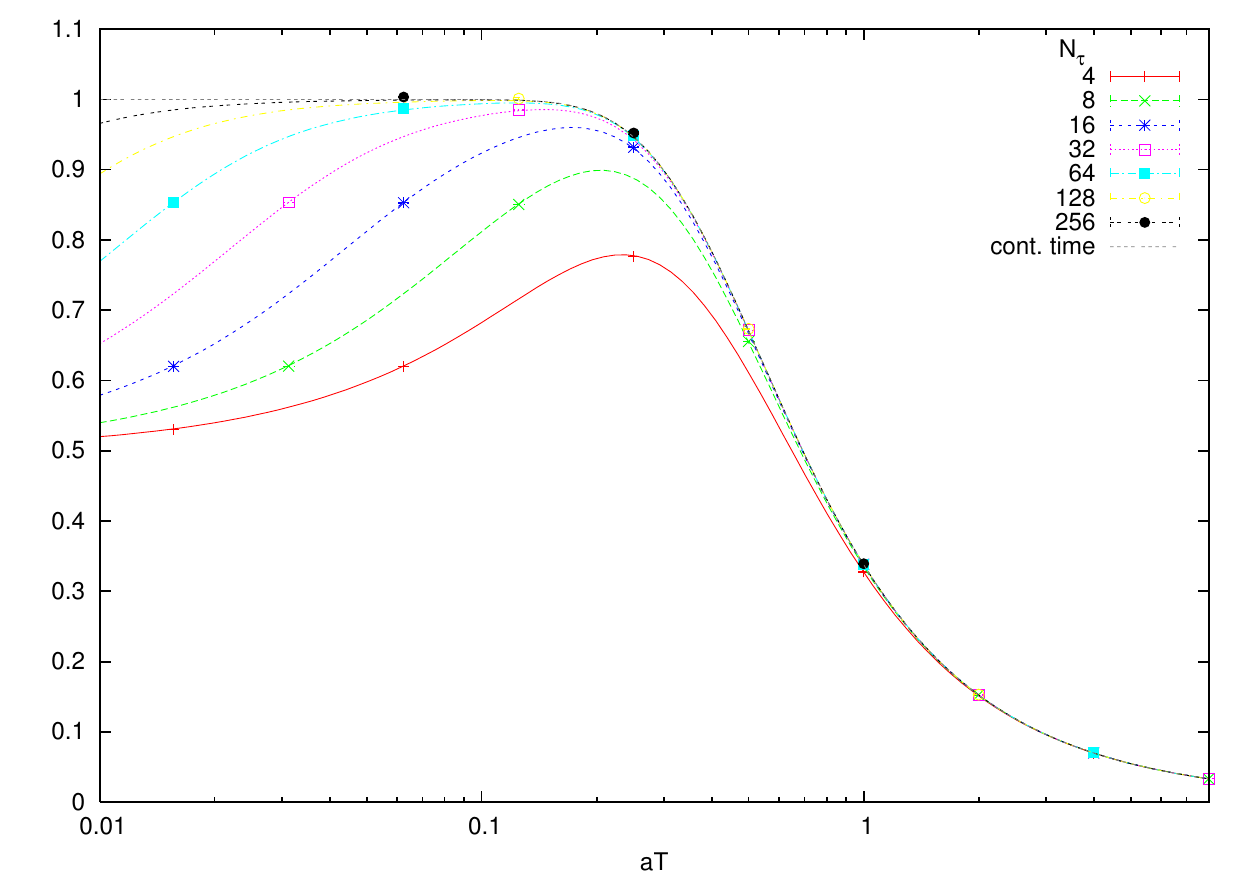}\\
\caption{
The chiral susceptibility for U(1) gauge group at strong coupling as a function of $\bareT$. Monte Carlo data from the worm algorithm (both discrete and continuous time) are compared to the analytic results  Eqs.~(\ref{DTResult},\ref{ContResult}). Note that the discretization errors are maximal in the low temperature region.}
\label{U1Fig}
\end{figure}

We also want to consider the contribution to the partition sum with a total number of 2 monomers. 
For lattices with spatial extend $\Ns=2$, the situation is considerably simple because it is not possible 
to separate the monomers by spatial dimers. If we decompose configurations into a piece with the monomers located but no spatial hoppings, 
and a piece with no monomers, where the first piece has length $D$ and the second piece has length $\Nt-D$, 
we can factorize the possible configurations by considering those on the $2\times D$ lattice and the $2\times (\Nt-D)$ lattice 
where it is required not to make use of periodic boundary conditions. This restricts the possible configurations further 
(no temporal dimers connecting the first and the last site allowed). $D$ may be odd or even, depending on whether the two monomers are
on the same spatial site ($D$ even) or on different spatial sites ($D$ odd).
The corresponding result in the 2-monomer sector is:
\begin{widetext}
\begin{align}
\calZ_2(\gamma,\Nt) &= \gamma^{2\Nt-2}\left(
\lr{\frac{\Nt}{2}}^2 \lr{4+2\Nt \gamma^{-2}}
+ 2\Nt \sum_{\ka\in \mathbbm{N}^+}^{\Nt-2}
2^\ka \sum_{D=1}^{\Nt-\ka} 
\tilde{\alpha}_\ka(\Nt-D)\beta(D)\gamma^{-2\ka-2 (D\hspace{-2mm}\mod2))}\right),\nn
\tilde{\alpha}_\ka(C)&=
\begin{cases}
\binom{(C+\ka-4)/2}{\ka-2} &\text{for $C$ even}  \\
\binom{(C+\ka-3)/2}{\ka-1} & \text{for $C$ odd} \\
\end{cases},\qquad
\beta(D)=\frac{1}{4}
\begin{cases}
\frac{1}{2}D\lr{D+2}&\text{for $D$ even}  \\
\lr{D+1}^2& \text{for $D$ odd} \\
\end{cases}
\end{align}
\end{widetext}
Both $\calZ_0$ and $\calZ_2$ are divergent series in $\gamma$, but their ratio is not, and gives the chiral susceptibility in the chiral limit:
\begin{widetext}
\begin{align}
\chi(\bareT,\Nt)&=\frac{1}{\Nt}\frac{\calZ_2(\sqrt{\bareT\Nt},\Nt)}{\calZ_0(\sqrt{\bareT\Nt},\Nt)}=
\frac{1}{2}\tanh\lr{\frac{\Nt}{2} \acsch(\Nt \bareT)}\lr{\frac{1}{\sqrt{1+(\Nt \bareT)^{-2}}}+\tanh\lr{\frac{\Nt}{2}\acsch(\Nt \bareT)}}
\label{DTResult}
\end{align}
\end{widetext}
where we have used the definition of $\bareT$, Eq.~(\ref{CTLimit}).
This result is explicitly temperature dependent, with $\Nt$ quantifying the cut-off dependence. 
In the limit $\Nt\rightarrow \infty$, $\acsch(x)\simeq 1/x$, and the chiral susceptibility has a well-defined continuous time limit:
\begin{align}
\chi(\bareT)=\frac{1}{2} \tanh\lr{\frac{1}{2\bareT}} \lr{1+\tanh\lr{\frac{1}{2\bareT}}}.
\label{ContResult}
\end{align}
In Fig.~\ref{U1Fig} the agreement of Monte Carlo data with the exact result is shown, both at finite $\Nt$ and continuous time. 

\subsection{Mean field and Percolation Analysis}
\label{Percolation}
The mean-field analysis for SC-LQCD based on a $1/d$-expansion has been studied for many decades  \cite{Kawamoto1981,KlubergStern1983,Faldt1985,Bilic1991a,Bilic1992a,Nishida2003,Kawamoto2005,Miura2016}. 
Also the continuous time partition function derived here can be used as a starting point for a mean-field analysis.
Our mean-field analysis assumes that a single site only couples to a mean-field bath of spatial dimers, where the location of $\calE$-sites and $\calA$-sites on its $2d$ nearest neighbors does not matter. 
This is well justified at high temperatures, where bonds with spatial dimers are isolated, but may also hold approximately at lower temperatures.
Our partition sum has only one dynamical site, and all other sites have a fixed number of vertices determined by a self-consistency relation. Neglecting the baryon sector, the resulting partition function in $d$ spatial dimension is:
\begin{align}
\calZ_{\rm MF}(\bareT) 
&=\exp\frac{\LatSpat d(1+v_\tsteel)}{2\bareT}
\end{align}
This implies for the energy density:
\begin{align}
a^4 \epsilon_{MF} &= \frac{\bareT}{\LatSpat}\frac{\partial }{\partial \bareT^{-1}} \log \calZ_{\rm MF}(\beta) =  \frac{d (1+v_\tsteel)}{2},
\end{align}
resulting in $a^4 \epsilon_{MF}=\frac{3}{2}+\sqrt{3}$ which should be compared to the discrete time value $a^4 \epsilon=\frac{3}{4}$ at $\gamma=1$.\\

\begin{figure}
\includegraphics[width=0.5\textwidth]{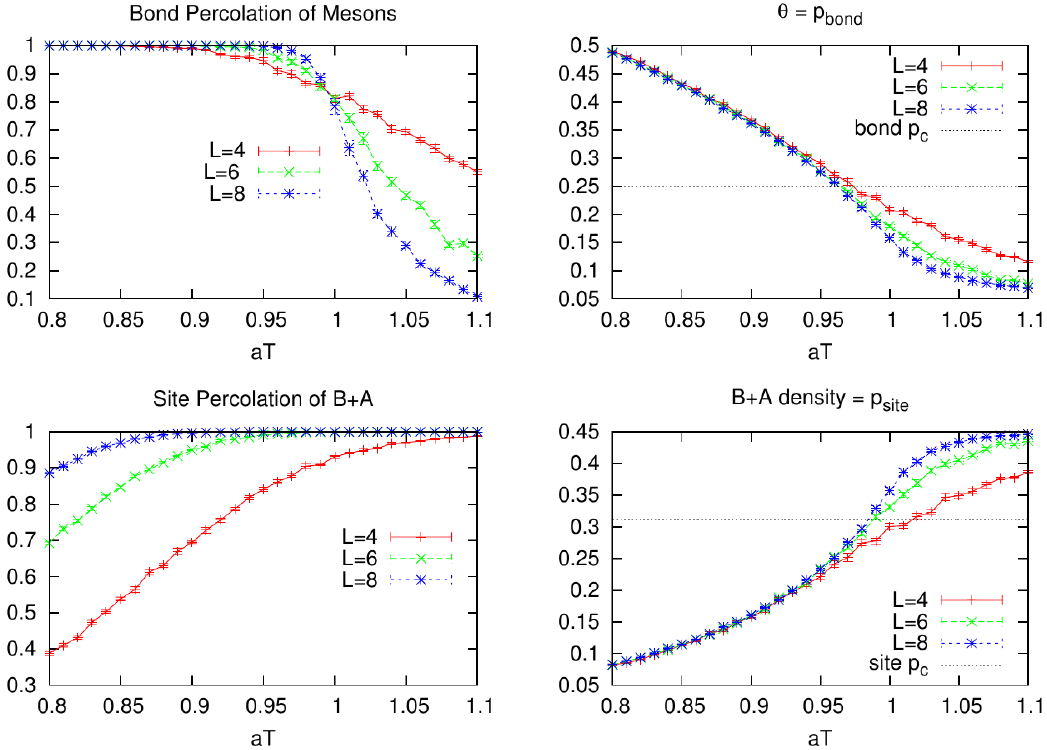}
\caption{Percolation analysis at $\bareMu=0.65$ (close to the tricritical value). \emph{Left:} Volume dependence of percolation for mesons \emph{(top)} and $\calQ$-polymers \emph{(bottom)}. 
\emph{Right:} Comparison with percolation theory, where the critical percolation thresholds are known (for $d=3$, on bonds: $p_{c}=0.2488$, on sites: $p_c=0.3116$), indicated as dashed lines.
If the observable $\theta$ is taken to be $p_{\rm bond}$ and the density of $\calQ$-polymers as $p_{\rm site}$, the critical percolation value is obtained close to
$\bareT^{TCP}=1.005(25)$.
}
\label{PercolationAnalysis}
\end{figure}
A qualitative understanding of the phase diagram can also be obtained via a percolation analysis on the spatial volume. We consider mixed percolation, both on bonds and sites \cite{Essam_1980}: 
\begin{enumerate}
\item[(1)] In the chirally broken phase, the pion correlation length diverges, thus the phase is characterized by bond percolation, where a bond is activated whenever there is at least one spatial dimer at that bond (for some time location). It can be related to the average bond occupation probability $\theta\simeq p_{\rm bond}$ with
\begin{align}
\theta &= \frac{1}{\LatSpatM}\sum_{b\in \LatSpatM }\expval{\theta(n_b)},& 
\theta(n_b)&=
\begin{cases}
 0 & n_b=0\\
 1 & n_b>0\\
\end{cases}
\end{align}
\item[(2)] In the nuclear phase, where every site is activated if it is occupied by a baryon or anti-baryon ($\calP$-Polymer), it can be related to the average site occupation probability, $\expval{n_\calP}\simeq p_{\rm site}$.
\end{enumerate}
Our criterion for percolation is that in the statistical average, the probability that a cluster spans around the periodic lattice in at least one spatial direction is close to 1. The percolation threshold is characterized by a step function in the thermodynamic limit.
Clearly, at low temperatures, the vacuum phase is characterized by bond percolation and the nuclear phase is characterized by site percolation. At higher temperatures, we may have a phase where bond and site percolation coexist. It turns out that this mixed phase exists along the first order transition. The percolation threshold for each bond and site percolation is reached close to the tricritical point, see~Fig.~\ref{PercolationAnalysis}. 
It is not surprising that the identification of the bond occupation probability with $\theta$ works extremely well, as the pions form a free relativistic gas \cite{deForcrand2016}. 
In contrast, the identification of $p_{\rm site}$ with the baryons density is not as good, as they do not form a free gas but are subject to strong nuclear interactions.\\ 

\subsection{Continuous Time Limit for $\Nf=2$}
\label{ParFuncNf2}
\newcommand{\I}{\mathcal{I}}
\renewcommand{\J}{\mathcal{J}}
\newcommand{\Qmd}{\QMat^\dagger}
\newcommand{\M}{\mathcal{M}}
\newcommand{\Md}{\mathcal{M}^\dagger}
\newcommand{\Ud}{U^\dagger}
\newcommand{\MMd}{\mathcal{M}\mathcal{M}^\dagger}
\newcommand{\Mxy}{M_{xy}}
\newcommand{\mmd}{\QMat\Qmd}
\newcommand{\floor}[1]{\lfloor #1 \rfloor}
\newcommand{\ceil}[1]{\lceil #1 \rceil}
\setlength{\tabcolsep}{2pt}
The two-flavor formulation admits more than one baryon per site and the Grassmann constraint allows for pion exchange between them, modifying nuclear interactions substantially. It also compares better to the strong coupling limit with Wilson fermions in a world-line formulation, as discussed in the context of Polyakov effective theory \cite{Fromm2012} which integrates out the spatial, but not the temporal gauge links.
SC-LQCD with $\Nf=2$ in the dual formulation has been discussed on discrete lattices for $\U(1)$ gauge group in \cite{Cecile2007}.
For $\U(3)$ gauge group, the link integrals have been addressed in \cite{Fromm2010}.
Here, we report on first steps towards a Hamiltonian formulation. The suppression of spatial bonds $\gamma^{-k}$, $k>2$ also applies here.
Let us first consider the static lines. We want to establish the basis of quantum states that generalize the $\Nf=1$ states $|\meson\rangle$ and  $|\baryon\rangle$.
To arrive at this basis, we consider the $\SU(3)$ one-link integrals \cite{Eriksson1981}:
\begin{widetext}
\begin{align}
\J(\QMat,\Qmd) &=\int_{\SU(3)} dU e^{\tr[ U \Qmd+U^\dagger \QMat]}= 2\sum_{n_0,n_1,n_2,n_3=0}^\infty \frac{1}{(n_0+n_1+2n_2+3n_3+2)!(n_0+n_2+2n_3+1)!}\prod_{k=0}^3\frac{x_i^{n_k}}{n_k!}\nn
x_0&=\det[\QMat]+\det[\Qmd],\qquad  x_1=\tr[\mmd], \qquad x_2=\frac{1}{2}\left(\tr[\mmd]-\tr[(\mmd)^2]\right),\nn
x_3&=\frac{1}{6}\left(\tr[\mmd]^3-3\tr[\mmd]\tr[(\mmd)^2]+2\tr[(\mmd)^3]\right)=\det[\mmd]
\end{align}
\end{widetext}
with $\QMat$ the quark matrix. The color trace $\tr$ can be converted to a sum over colors and a flavor trace $\Tr$: 
\begin{align}
\tr[(\mmd)^k]&=(-1)^{k+1}\Tr[(M_x M_y)^k],&
 M_z&=\begin{pmatrix}
      \bar{u}u_z & \bar{u}d_z\\
      \bar{d}u_z & \bar{d}d_z
     \end{pmatrix}
\end{align}
The sum over $n_i$ $(i=0,\ldots 3)$ terminates due to the Grassmann integration. The corresponding invariants $x_i$ can be evaluated for 
$\Nf=2$:
\newcommand{\kU}{k_U}
\newcommand{\kD}{k_D}
\newcommand{\kP}{k_{\pi^+}}
\newcommand{\kM}{k_{\pi^-}}
\newcommand{\kPPUD}{k^{(2)}_{\pi^+\pi^-,UD}}
\newcommand{\kUDPP}{k^{(2)}_{UD,\pi^+\pi^-}}
\newcommand{\mZ}{\mathfrak{m}_0}
\newcommand{\mPZ}{\mathfrak{m}_{\pi^0}^2}
\newcommand{\mSZ}{\mathfrak{m}_{\bar{\pi}^0}^2}
\newcommand{\mU}{\pi_U}
\newcommand{\mD}{\pi_D}
\newcommand{\mP}{\pi_+}
\newcommand{\mM}{\pi_-}
\newcommand{\mPPUD}{\mathfrak{m}_{\pi^+\pi^-,UD}}
\newcommand{\mUDPP}{\mathfrak{m}_{UD,\pi^+\pi^-}}
\renewcommand{\mZ}{\mathfrak{m}_0}
\renewcommand{\mPZ}{\pi_0^2}
\renewcommand{\mSZ}{\bar\pi_0^2}
\newcommand{\xS}{x_S}
\begin{widetext}
\begin{align}
 x_0&=B_{uuu}+B_{uud}+B_{udd}+B_{ddd}+\bar{B}_{uuu}+\bar{B}_{uud}+\bar{B}_{udd}+\bar{B}_{ddd}\nn
 x_1&=\Tr[M_x M_y]=\kU+\kD+\kP+\kM\nn
 x_2&=\frac{1}{2}\lr{\Tr[M_x M_y]^2+\Tr[(M_x M_y)^2]}
 = x_1^2+\xS\nn
 x_3&=\frac{1}{6}\lr{\Tr[M_x M_y]^3+3\Tr[M_x M_y] \Tr[(M_x M_y)^2] +2\Tr[(M_x M_y)^3]}
 =x_1^3+ \frac{3}{2}x_1 \xS \nn
\xS&= \kPPUD+\kUDPP-\kU\kD-  \kP\kM. 
\end{align}
with the fluxes and dimers defined as:
\begin{align}
B_{uud}&=\bar{u}\bar{u}\bar{d}_x uud_y,&
B_{udd}&=\bar{u}\bar{d}\bar{d}_x udd_y,&
B_{uuu}&=\bar{u}\bar{u}\bar{u}_x uuu_y,&
B_{ddd}&=\bar{d}\bar{d}\bar{d}_x ddd_y,\nn
\bar{B}_{uud}&=uud_x\bar{u}\bar{u}\bar{d}_y,&
\bar{B}_{udd}&=udd_x\bar{u}\bar{d}\bar{d}_y,&
\bar{B}_{uuu}&=uuu_x\bar{u}\bar{u}\bar{u}_y,&
\bar{B}_{ddd}&=ddd_x\bar{d}\bar{d}\bar{d}_y,\nn 
 \kU&=\bar{u}u(x)\bar{u}u(y),&  \kD&=\bar{d}d(x)\bar{d}d(y),& \kP&=\bar{d}u(x)\bar{u}d(y),&\kM&=\bar{u}d(x)\bar{d}u(y),\nn
\kPPUD&=\bar{u}d(x) \bar{d}u(x)\bar{u}u(y)\bar{d}d(y),&&& \kUDPP&=\bar{u}u(x)\bar{d}d(x)\bar{u}d(y) \bar{d}u(y).
\label{LinkStatesNf2}
\end{align}
\end{widetext}
Note that the baryonic fluxes are spinless and spin arises only when measuring baryonic correlators with the corresponding staggered kernels. 
We still have to integrate out the Grassmann variables to obtain the quantum states in the occupation number basis, and the corresponding Hamiltonian, where we consider the chiral limit only.
The Grassmann constraint then dictates that all quarks $u$,$d$ and anti-quarks $\bar{u}$,$\bar{d}$ are within mesons or baryons. 
The Grassmann integral in the chiral limit is
\begin{widetext}
\begin{align}
 I_G=\int \prod_{\alpha}[{\rm d}u_\alpha {\rm d}\bar{u}_\alpha {\rm d}d_\alpha {\rm d}\bar{d}_\alpha] (\bar{u}u)^{\kU} (\bar{d}d)^{\kD} (\bar{u}d)^{\kM} (\bar{d}u)^{\kP}=
(-1)^\frac{\kP+\kM}{2}(\Nc!)^2\left\{
\begin{array}{ll}
1 & (\kP+\kM)/2 \mod \Nc =0\\
\frac{1}{\Nc} & \text{otherwise}
\end{array}
\right.
\end{align}
\end{widetext}
\onecolumngrid
\begin{table*}
{\footnotesize
\begin{tabular}{|r|r|c|c|c|c|c|c|c|r|}
\hline
$B$ & $I$\; & $\meson=0$ & $\meson=1$ & $\meson=2$ & $\meson=3$ & $\meson=4$ & $\meson=5$ & $\meson=6$ & $\Sigma$ \\
\hline
 -2& 0 & $\bar{p}$\,$\bar{n}$ & & & & & & & 1 \\
\hline
-2 & $\Sigma$ & 1 & 0 & 0 & 0 & 0 & 0& 0 & 1\\
 \hline 
\hline
 -1&  $\frac{3}{2}$ & $\bar{B}_{uuu}$ &  $\bar{B}_{uuu}\mD$ & $\bar{B}_{uuu} \mD^2$ & $\bar{B}_{uuu} \mD^3$ &&&&  4 \\
 -1 & $+\frac{1}{2}$ & $\bar{B}_{uud}$ &  $\bar{B}_{uud}\, (\mU,\, \mD)$     & $\bar{B}_{uud}\,\mZ^2$, $\bar{B}_{uud}\,\mD^2$ & $\bar{B}_{uud} \mZ^2\mD$    &&&&  6 \\
 -1 & $-\frac{1}{2}$ & $\bar{B}_{udd}$       &  $\bar{B}_{udd}\,\mU$, $\bar{B}_{udd}\,\mD$     & $\bar{B}_{udd}\,\mZ^2$, $\bar{B}_{udd}\,\mU^2$ & $\bar{B}_{udd} \mZ^2\mU$    &&&&  6 \\
 -1&  $-\frac{3}{2}$ & $\bar{B}_{ddd}$ &  $\bar{B}_{ddd}\mU$ & $\bar{B}_{ddd} \mU^2$ & $\bar{B}_{ddd} \mU^3$ &&&&  4 \\
\hline
-1 & $\Sigma$ & 4 & 6 & 6 & 4 & 0 & 0& 1 & 20\\
\hline
\hline 
0 & -3& & & & $\mM^3$ &&&& 1\\ 
0 & -2& & & $\mM^2$ & $\mM^2\mU$,  $\mM^2\mD$ & $\mM^2 \mZ^2$ & & & 4\\
0 & -1& & $\mM$ & $\mM\mU$, $\mM\mD$ & $2\mM\mZ^2$, $\mM\mU^2$, $\mM\mD^2$ & $\mM\mZ^2\mU$, $\mM\mZ^2\mD$  &  $\mM \mZ^4$ & & 10 \\
0 & 0 & 1& $\mU$, $\mD$ & $\mPZ$, $\mSZ$, $\mU^2$, $\mD^2$ & $\mPZ\mU$,$\mSZ\mU$, $\mPZ\mD$, $\mSZ\mD$, $\mU^3$, $\mD^3$ & $\mPZ\mZ^2$, $\mSZ\mZ^2$, $\mZ^2 \mU^2$, $\mZ^2 \mD^2$ & $\mZ^4 \mU$, $\mZ^4 \mD$ & $\mZ^6$ & 20 \\
0 & -1& & $\mP$ & $\mP\mU$, $\mP\mD$ & $2\mP\mZ^2$, $\mP\mU^2$, $\mP\mD^2$ & $\mP\mZ^2\mU$, $\mP\mZ^2\mD$  &  $\mP \mZ^4$ & & 10 \\
0 & -2& & & $\mP^2$ & $\mP^2\mU$,  $\mP^2\mD$ & $\mP^2 \mZ^2$ & & & 4\\
0 & -3& & & & $\mP^3$ &&&& 1\\ 
\hline
0 & $\Sigma$ & 1 & 4 & 10 & 20 & 10 & 4& 1 & 50\\
\hline
\hline
 1&  $\frac{3}{2}$ & $B_{uuu}$ &  $B_{uuu}\mD$ & $B_{uuu} \mD^2$ & $B_{uuu} \mD^3$ &&&&  4 \\
 1 & $+\frac{1}{2}$ & $B_{uud}$       &  $B_{uud}\,\mU$, $p\,\mD$     & $B_{uud}\,\mZ^2$, $p\,\mD^2$ & $B_{uud} \mZ^2\mD$    &&&&  6 \\
 1 & $-\frac{1}{2}$ & $B_{udd}$       &  $B_{udd}\,\mU$, $n\,\mD$     & $B_{udd}\,\mZ^2$, $n\,\mU^2$ & $B_{udd} \mZ^2\mU$    &&&&  6 \\
 1&  $-\frac{3}{2}$ & $B_{ddd}$ &  $B_{ddd}\mU$ & $B_{ddd} \mU^2$ & $B_{ddd} \mU^3$ &&&&  4 \\
\hline
 1  & $\Sigma$ & 4 & 6 & 6 & 4 & 0 & 0& 0 & 20\\
\hline
\hline
 2& 0 & $p$\,$n$ & & & & & & & 1 \\
\hline
 2 & $\Sigma$ & 1 & 0 & 0 & 0 & 0 & 0& 0 & 1\\
\hline
\hline
$\Sigma$ &  &11 & 16 & 22 & 28 & 10 & 4 & 1 & 92\\    
\hline
\end{tabular}
\caption{All 92 possible quantum states for the $\Nf=2$ Hamiltonian formulation with $\SU(3)$ gauge group. 
The states and their multiplicities are given for the sectors specified baryon number $B$ and isospin number $I$, and meson occupation number $\meson$. Note the mesonic particle-hole symmetry  $\meson \leftrightarrow (\Nf-|B|)\Nc-\meson$ which corresponds to the shift symmetry by $\at$.
}
}
\label{HadronStatesNf2}
\end{table*}
\twocolumngrid
\noindent which simplifies due to flux conservation:
\begin{align}
\kP&=\kM,& \kU+\kD+\kP+\kM&=\Nc.
\end{align}
Just as for $\Nf=1$, we can define vertices in the same way as in Eq.~(\ref{Vertices}).
 This allows to compose them into line segments between spatial dimer emission in terms of alternating dimers of $\kU$, $\kD$, orientied fluxes $\kP$, $\kM$, $B_{uuu}$, $\bar{B}_{uuu}$, etc.~or combinations thereof. 
%
We note that there are various ways to combine the link states in Eq.~(\ref{LinkStatesNf2}):
in particular, the flavor singlet dimer combinations $\kU\kD$, $\kP\kM$, $\kPPUD$ and $\kUDPP$ mix and have to be resummed. We do so by defining the matrix in the basis of this order to determine what meson states survive: 
\begin{align}
\Pi_0&=\left(
\begin{array}{llll}
\frac{4}{3} & -\frac{2}{3} & -\frac{\sqrt{2}}{3}  & \frac{2\sqrt{2}}{3} \\
-\frac{2}{3} & \frac{4}{3} & \frac{2\sqrt{2}}{3}  & -\frac{\sqrt{2}}{3} \\
\frac{2\sqrt{2}}{3} & -\frac{\sqrt{2}}{3} & -\frac{1}{3}  & \frac{2}{3} \\
-\frac{\sqrt{2}}{3} & \frac{2\sqrt{2}}{3} & \frac{2}{3}  & -\frac{1}{3} \\
\end{array}
\right).
\end{align}
This matrix is a projector with eigenvalues (1,1,0,0), such that it can be diagonalized to the $\Id_{2x2}$ matrix with the basis vectors
\begin{align}
\mPZ&=\frac{1}{3}\lr{\sqrt{2}\kU\kD+\kPPUD}\nn
\mSZ&=\frac{1}{3}\lr{\sqrt{2}\kP\kM+\kUDPP} 
\end{align}
Note that in the strong coupling limit there is no distinction between $\pi_0=\frac{1}{\sqrt{2}}(\bar{u}u-\bar{d}d)$ and $\eta/\eta'=\frac{1}{\sqrt{2}}(\bar{u}u+\bar{d}d)$, due to the lack of topological features. 
All other states do not mix. We will now list the quantum states for the Hamiltonian formulation, classified by the baryonic sectors $n_B\in \{-\Nf,\ldots \Nf\}$ and the isosopin sectors $n_I\in \{-\Nf,\ldots \Nf\}$.
Recall that the states are only distinguishable on the quark level, and there are several possible assignments in terms of hadrons:
\begin{align}
 \mZ^2&\equiv \mU\mD=\mP\mM,\nn
 \mZ^6&=B_{uud}\bar{B}_{udd}= B_{udd}\bar{B}_{uud}= B_{ddd}\bar{B}_{uuu}=(\mZ^2)^3,\nn
 \bar{p}\,\bar{n}&\equiv\bar{B}_{uud}\bar{B}_{udd}=\bar{B}_{uuu}\bar{B}_{ddd},\nn
  p\,n &\equiv B_{uud}B_{udd}=B_{uuu}B_{ddd}.
\end{align}
The final 92 quantum states are given in Tab.~\ref{HadronStatesNf2}. 
The 50 purely mesonic states can be further classified by the set of charges
$(Q_{\pi_0},Q_{\pi^+})$, resulting in the Hamiltonian
\begin{align}
\hat{\calH}&=\frac{1}{2}\sum_{
\langle\vec{x},\vec{y}\rangle}
\left(
\hat{J}^{+}_{\pi_0,\vec{x}} \hat{J}^{-}_{\pi_0,\vec{y}}+
\hat{J}^{+}_{\bar{\pi}_0,\vec{x}} \hat{J}^{-}_{\bar{\pi}_0,\vec{y}}\right.\nn
&\left.\hspace{15mm}
+\hat{J}^{+}_{\pi^+,\vec{x}} \hat{J}^{-}_{\pi^+,\vec{y}}+
\hat{J}^{+}_{\pi^-,\vec{x}} \hat{J}^{-}_{\pi^-,\vec{y}}+h.c.
\right)
\label{ParFuncHamNf2}
\end{align}
with the occupation number raising and lowering operators defined for each conserved charge.
The full $\Nf=2$ partition function including the baryonic states and flavored observables will be discussed in a forthcoming publication.
We also derived the number of quantum states for arbitrary $\Nc$ and $\Nf$, resulting in the 1-dim.~partition function:
\begin{align}
Z_{\Nf}(\mu_B/T)&=\sum_{B=-\Nf}^\Nf \prod_{a=0}^{\Nc-1} \frac{a!(2\Nf+a)!\;e^{B\mu_B/T}}{(\Nf+a-B)!(\Nf+a+B)!}.
 \end{align}
For $B=0$ the multiplicities are given in Tab.~\ref{MultiplicitiesNf}.
An important application of the $\Nf=2$ partition function is to determine the QCD phase diagram with both finite baryon and isospin chemical potential. 
Our formulation is still sign-problem free in the continuous time limit. As we have not yet performed dynamical simulations, we can only provide analytic results on the static limit, correpsonding to 1-dim.~QCD. For $\Nc=3$:
\begin{align}
Z\lr{\frac{\mu_B}{T},\frac{\mu_I}{T}}= & 2\cosh\frac{3\mu_{I}}{T}+8\cosh\frac{2\mu_{I}}{T}+20\cosh\frac{\mu_{I}}{T}+20\nn
&+ 2\cosh\frac{\mu_{B}}{T}\left( 8\cosh{\frac{\frac{3}{2}\mu_{I}}{T}}+12\cosh\frac{\frac{1}{2}\mu_{I}}{T}\right) \nn
&+2\cosh\frac{2 \mu_{B}}{T}
\label{ZNf2Nc2}.
\end{align}
\begin{table}
\begin{tabular}{|r|rrrrr|}
\hline
\diagbox{$\Nc$}{$\Nf$} & 0 & 1 & 2 & 3 & 4\\
\hline
1 & 1 & 2 & 3 & 4 & 5 \\
2 & 1 & 6 & 20 & 50 & 105 \\
3 & 1 & 20 & 275 & 1430 & 7007 \\
4 & 1 & 170 & 5814 & 94692 & 980628 \\
\hline
\end{tabular}
\caption{Multiplicities of quantum states (static lines) in the mesonic sector, i.e.~from $\U(\Nc)$ integrals.}
\label{MultiplicitiesNf}
\end{table}

\begin{figure}
 \includegraphics[width=0.49\textwidth]{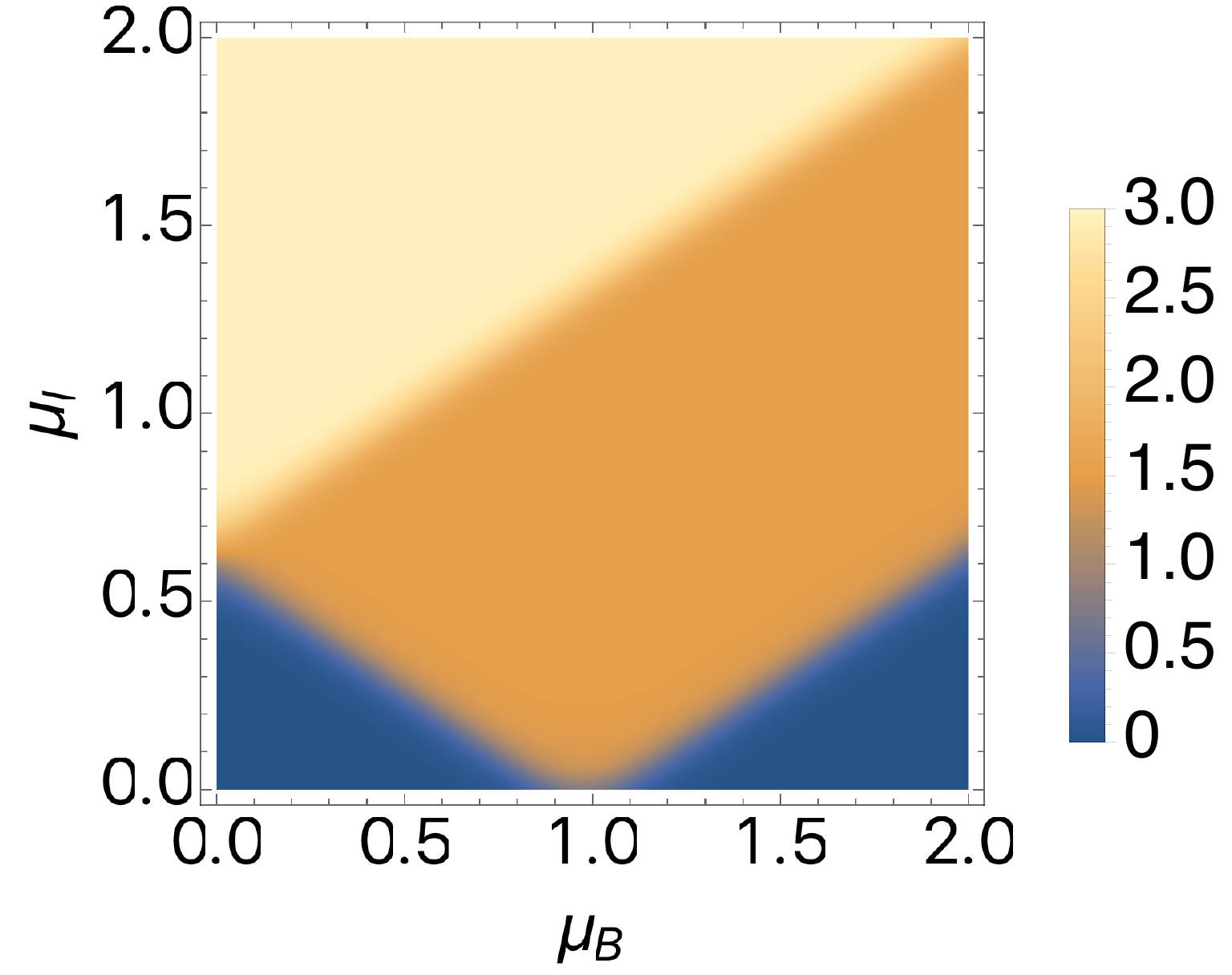}
\caption{Isospin density in the $\mu_B - \mu_I$ plane, based on the limit $T\rightarrow 0$ of Eq.~(\ref{ZNf2Nc2}), showing strong phase boundaries form pion condensation.
}
 \label{muBmuIPhaseDiag}
\end{figure}
Even though interactions will be crucial at low temperatures, we can plot the zero-temperature limit of Eq.~(\ref{ZNf2Nc2}) to obtain a naive picture of the phases in the $\mu_B-\mu_I$ plane, shown in Fig.~\ref{muBmuIPhaseDiag}.\\

\subsection{Finite Quark Mass}
\label{FiniteMass}
The chiral limit is the most interesting regime when studying the chiral transition, but we need to extend the derivation of the continuous time partition function to finite quark mass to address the quark mass dependence of zero and finite temperature observables. Only then it is possible to study the $p$-regime where the pion correlation function fits on the lattice.
Whereas in the chiral limit, the chiral condensate is strictly zero (in a finite volume), already a small quark mass will result in a non-zero chiral condensate.
Likewise the sigma meson becomes much heavier compared to the pion. This can be best understood in the dual representation: 
The number of monomers on even sites always equals the number of monomers on odd sites.
In the pion correlator, the contributions from monomers at even sites have the \emph{opposite sign} from those at odd sites, resulting in a light pion mass.
In the sigma correlator, the contributions from monomers at even sites have the \emph{same sign} as those at odd sites, resulting in a heavy sigma meson.

When attempting to derive the continuous time partition function at finite quark mass in a naive way, i.e.~at fixed quark mass $am_q$, the monomer number will diverge in the limit $\Nt\rightarrow \infty$. We have illustrated in \cite{Bollweg2018} that the continuous time limit is well defined also at finite quark mass, but it turns out that the constant $\kappa$ is now quark mass dependent. This function $\kappa(m_q)$ has been determined non-perturbatively with a condition for keeping the quark mass constant in the limit $\at\rightarrow \infty$. With this knowledge, the continuous time limit can also be derived at finite quark mass, but it is not the bare quark mass $am_q$, but rather the ratio $M_{\pi}/T$ which is the input parameter of the continuous time partition function. We are working on an extension of the CT worm algorithm such that the Poisson process incorporates a finite quark mass.

\bibliography{conttime} 
\bibliographystyle{apsrev4-1}

\end{document}